\documentclass{aa}

\usepackage[varg]{txfonts}
\usepackage{graphicx}
\usepackage{amsmath}
\usepackage{amssymb}
\usepackage{microtype}
\usepackage{booktabs}
\usepackage{multirow}
\usepackage{subcaption}
\usepackage{tablefootnote}
\usepackage{threeparttablex}
\usepackage{hyperref}
\usepackage{diagbox}
\usepackage{pdflscape}
\usepackage{bm}

\makeatletter
\renewcommand*\aa@pageof{, page \thepage{} of \pageref*{LastPage}}
\makeatother

\hypersetup{colorlinks=true, linkcolor=blue, citecolor=blue}

\begin{document}

    \title{Fine-tuning the complex organic molecule formation: sulfur and CO ice as regulators of surface chemistry}
    
    \titlerunning{Sulfur and CO ice as regulators of surface chemistry}
    \author{D. Navarro-Almaida
          \inst{1}\fnmsep\thanks{\email{dnavarro@cab.inta-csic.es}}
          \and
          A. Taillard\inst{1}
          \and
          A. Fuente\inst{1}
          \and
          P. Caselli\inst{2}
          \and
          R. Mart\'in-Dom\'enech\inst{1}
          \and
          J. J. Miranzo-Pastor\inst{1}
          }

    \institute{
        Centro de Astrobiolog\'ia (CAB), CSIC-INTA, Ctra. de Torrej\'on a Ajalvir km 4, 28850, Torrej\'on de Ardoz, Madrid, Spain
        \and
        Max-Planck-Institut f\"ur extraterrestrische Physik, Gie{\ss}enbachstrasse 1, D-85748 Garching bei M\"unchen, Germany
        }
    \date{}
    
    \abstract{Grain-surface chemistry plays a crucial role in the formation of molecules of astrobiological interest, including H$_{2}$S and complex organic molecules (COMs). These are commonly observed in the gas phase toward star-forming regions, but their detection in ices remains limited. Combining gas-phase observations with chemical modeling is therefore essential for advancing our understanding of their chemistry.}
    {The goal is to investigate the factors that promote or hinder molecular complexity combining gas-phase observations of CH$_{3}$OH, H$_{2}$S, OCS, N$_{2}$H$^{+}$, and C$^{18}$O with chemical modeling in two prototypical dense cores: Barnard-1b and IC348.}
    {We observed millimeter emission lines of CH$_{3}$OH, H$_{2}$S, OCS, N$_{2}$H$^{+}$, and C$^{18}$O along strips using the IRAM 30m and Yebes 40m telescopes. We used the gas-grain chemical model \texttt{Nautilus} to reproduce the observed abundance profiles adjusting parameters such as initial sulfur abundances and binding energies.}
    {H$_{2}$S, N$_{2}$H$^{+}$ and C$^{18}$O gas-phase abundances vary up to one order of magnitude towards the extinction peak. CH$_{3}$OH abundance remains quite uniform. Our chemical modeling revealed that these abundances can only be reproduced assuming a decreasing sulfur budget, which lowers H$_{2}$S and enhances CH$_{3}$OH abundances. Decreasing binding energies, which are expected in CO-rich apolar ices, are also required. The sulfur depletion required to explain H$_2$S is generally higher than that required by CH$_3$OH, suggesting an unknown sulfur sink. These findings highlight the intricate relationship between sulfur chemistry and COM formation, driven by the competition between sulfur and CO for hydrogen atoms.}
    {The formation of COMs begins in the low-density envelopes of molecular clouds. The growth of CO ice and the progressive sequestration of hydrogen atoms are critical in determining whether chemical complexity can develop. Our study emphasizes that molecular complexity is tightly linked to the sulfur chemistry within dense cores, providing key insights into the early stages of star and planet formation.}


	\maketitle

	\section{Introduction}
    	
    	More than 300 different molecules of all shapes and sizes have been identified in the interstellar medium (ISM). Among them, those carbon-bearing molecules composed by more than 6 atoms are referred to as complex organic molecules (COMs, \citet{Herbst2009}). These molecules, already detected in the earliest stages of star formation, follow the star and planet formation, potentially seeding the origin of life in the newly formed planets.
    	
    	The main challenges currently present in the study of COMs are where and how these molecules form. These molecules were first detected in the so-called hot cores \citep[see, e.g.,][]{Cummins1986, Blake1987}, that is, compact, warm, and dense regions surrounding high-mass protostars. Back then, COMs were thought to form by gas-phase processes collectively known as warm gas-phase chemistry: upon heating from the high-mass protostar, the interstellar ice content is sublimated and its constituents, called parent molecules, are injected in the gas-phase to then engage in ion-molecule or neutral-neutral reactions that form COMs. In order to explain the observations toward OMC-1 \citep{Blake1987} and Sgr B2 \citep{Cummins1986}, several chemical models were developed following this idea \citep{Millar1991, Charnley1992, Caselli1993, Viti1999, Garrod2008}. The warm gas-phase formation scenario was, however, unable to explain the high abundances observed \citep{Cazaux2003, RequenaTorres2006} and it was challenged by experimental results where these pathways were found to be much less efficient than previously thought \citep[see, for instance,][]{Horn2004, Geppert2006}. Mechanisms involving ultraviolet (UV) radiation and cosmic-ray induced secondary UV photons \citep{Prasad1983} were also proposed to describe the formation of COMs by ice photochemistry, in which the parent molecules, radicals, are produced on grain surfaces by these sources of radiation. If these radicals are sufficiently mobile on the grain surface they undergo into reactions generally lacking of energy barriers \citep{Garrod2006, Garrod2008}, although there are some exceptions \citep{EnriqueRomero2021, EnriqueRomero2022, EnriqueRomero2023}. The COMs produced in these reactions would be then thermally desorbed as star formation proceeds \citep[see, e.g.,][]{Oberg2016}. 
    	
    	This picture of COM formation is, however, not adequate for the earliest stages of star formation. Prestellar and starless cores are dense $\gtrsim 10^{4}$ cm$^{-3}$ \citep{Tafalla2004} and cold $\sim 10$ K regions inside molecular clouds where a great deal of COMs has been detected: methanol \citep[CH$_{3}$OH,][]{Oberg2010, Bizzocchi2014, Vastel2014, JimenezSerra2016, Scibelli2020, JimenezSerra2021, Scibelli2024}, acetaldehyde \citep[CH$_{3}$CHO,][]{Oberg2010,Vastel2014, Bacmann2012, Scibelli2024}, or methyl formate \citep[HCOOCH$_{3}$,][]{Oberg2010, Bacmann2012, Scibelli2024}, among others. These objects are devoid of heating and radiation sources and, consequently, the low temperatures and the absence of UV radiation from the ISM prevent these mechanisms to operate efficiently. Instead, additional phenomena have been proposed to explain the presence of COMs in the gas phase in these objects. On the one hand, cosmic rays, which are able to penetrate deep into dense cores, have been suggested to enhance COM presence in the gas phase via secondary UV photons \citep{Hasegawa1993, Shen2004, Caselli2012, Oberg2016, Sipila2021, Arslan2023} or grain sputtering \citep{Wakelam2021, Dartois2021, Taillard2023}. On the other hand, reactive desorption, a non-thermal desorption mechanism proposed by \citet{Vasyunin2013}, would allow these molecules to leave the grain mantles using the energy released by the exothermic reaction of their formation if it is not absorbed by the grain \citep{Pantaleone2020, Pantaleone2021, Ferrero2023}. In this picture, the diffusion and subsequent reaction of the parent molecules on the surface of interstellar grains lead to the release of the COM to the gas phase if the reaction is exothermic. In cold cores, methanol, the simplest COM, is believed to form almost exclusively on dust grain surfaces via successive hydrogenation of carbon monoxide \citep[see, e.g.,][]{Watanabe2002, JimenezSerra2025}, followed by its release into the gas phase due to cosmic-ray heating desorption \citep{Leger1985}, cosmic-ray sputtering \citep{Taillard2023}, or reactive desorption \citep{Vasyunin2017, Riedel2023}. Its formation is then dependent on the readiness of the parent molecules and the rate at which they diffuse on dust grain surfaces. On the one hand, CO is expected to be highly available in the ices of cold cores in what is known as catastrophic CO depletion \citep{Caselli1999}, a process dependent on dust temperature and gas density. Dust temperatures below $\sim 15$ K have been shown to effectively freeze out CO at the typical densities of cold cores, $\sim 10^{4}$ cm$^{-3}$ \citep{Nakagawa1980, MunozCaro2010}. On the other hand, according to the results in  \citet{Caselli1994}, the presence of atomic hydrogen in the ice matrix, another ingredient needed to synthesize COMs, can vary greatly depending on the abundance of sulfur, as it is sequestered forming sulfur hydrides.
    	
    	The determination of the sulfur abundance is a challenging task due to the so-called sulfur depletion problem. The latter stands for the unknown identity of the main sulfur reservoirs in molecular clouds, where the total amount of detected gas-phase sulfur bearing molecules only accounts for $<1\%$ \citep[see, for instance,][]{Charnley1997, Ruffle1999, Esplugues2014, LeGal2021, Fuente2023} of the cosmic abundance, in contrast with the diffuse medium where the detected sulfur compounds account for the total cosmic abundance S/H $\sim 1.5\times 10^{-5}$ \citep{Asplund2009}. Furthermore, the detection of sulfur-bearing molecules in ices still remains elusive. To date, OCS \citep{Geballe1985, Palumbo1995} and SO$_{2}$ \citep{Boogert1997, Rocha2024}, are the only sulfur-bearing molecules tentatively detected in interstellar ices. The estimation of the sulfur elemental abundance therefore relies on the observation of abundant gas-phase sulfur-bearing species and the use of chemical models to build a coherent view of the sulfur chemistry in molecular clouds. According to chemical models \citep{Vidal2017, Laas2019, NavarroAlmaida2020} and cometary detections \citep{Mumma2011, Calmonte2016}, H$_{2}$S is one of the most abundant sulfur-bearing molecules formed in ices. It has not been, however, detected in interstellar ices, and only upper limits \citep{Escobar2011, McClure2023} are reported. It has been suggested that solid H$_{2}$S may undergo reactions that change the identity of the main sulfur reservoir in ices into sulfur allotropes and polysulfanes \citep{Shingledecker2020, Cazaux2022}. Finally, ammonia salts carrying sulfur NH$_{4}$SH have also been recently proposed as an important reservoir of the missing sulfur from interstellar ices \citep{Slavicinska2024}. All in all, the uncertain outlook of the sulfur chemistry and its connection to the availability of key ingredients for chemical complexity introduce uncertainties in the formation of COMs like methanol.
    	
    	Nevertheless, this picture of COM production, where they are formed by the diffusion of parent molecules that react and are subsequently released into the gas phase by non-thermal desorption mechanisms is, most certainly, still incomplete. For instance, the diffusion process is still not fully understood \citep{Chen2024}. Moreover, the large COMs observed toward cold cores are not expected to form this way due to their low mobility at 10 K. Alternative phenomena collectively known as non-diffusive chemistry emerges in an attempt to address the difficulty of diffusion of heavy species on the ice matrix in the formation of COMs \citep{Jin2020, Chen2024}. Some of these mechanisms have been reported to be responsible for the formation of glycine and methylamine in laboratory water-rich ices under dark cloud conditions \citep{Ioppolo2021}, and seem to provide a good agreement with the observed relative ratios of the structural isomers methyl formate, glycolaldehyde, and acetic acid \citep{Garrod2022}. Moreover, the Eley-Rideal mechanism has also been proved to have a significant role in the production of COMs \citep{Ruaud2015} at low temperatures $\sim 10$ K. Additionally, several mechanisms have been proposed to enhance the diffusion of large species over the surface of the grains. As suggested by \citet{Vasyunin2017}, the efficiency of reactive desorption releasing COMs into the gas-phase possibly depends on the composition of the outermost ice layers. Ice layers in cold cores are expected to contain mostly frozen CO and, due to its low dipole moment compared to that of water, they would increase the reactive desorption efficiency and facilitate the diffusion of species on top of CO ice layers. Likewise, the low dipole moment of CO has also been found responsible for a higher yield of cosmic ray sputtering that would help explain the high gas-phase abundances of methanol observed in dense cores \citep{Taillard2023}. The recent work of \citet{Molpeceres2024} provides a quantitative measure of the decline in binding energies expected in CO-rich ices that increase the efficiency of desorption and the diffusion rate of species in the ice matrix. Recent temperature programmed desorption (TPD) laboratory experiments and theoretical quantum chemistry computations highlight the need to consider variations of the usually tabulated binding energies. Several works, such as \citet{Chaabouni2018}, \citet{Tinacci2022}, \citet{Perrero2024}, \citet{Bariosco2024}, and \citet{Bariosco2025}, revealed the importance of shifting from the single-valued binding energy assumption to the more complex binding energy distribution description. These binding energy distributions naturally incorporate the variation in binding energies that are expected at different adsorption/desorption sites in realistic grain and ice surfaces. This phenomenon also helps enhance the production of COMs in the ice and their release into the gas-phase \citep{Bariosco2025}.
    	
    	In this paper we analyze the presence of methanol, the simplest COM, in two star-forming regions and investigate the link between its formation, the sulfur abundance, and the binding energies. As commented above, the formation of COMs greatly depends on the composition of the surface and the readiness of reactants, mediated by sulfur. We present and analyze millimeter observations of C$^{18}$O, H$_{2}$S, N$_{2}$H$^{+}$, OCS, and CH$_{3}$OH toward the sources Barnard 1b and IC 348, located in the Perseus molecular cloud. C$^{18}$O and N$_{2}$H$^{+}$ were selected to supply information about CO depletion and growth of CO ices that potentially weaken the binding of molecules to the ice surface and enhance COM formation. At the same time, H$_{2}$S and OCS provide an estimate of the sulfur abundance that sets the availability of reactive species to form COMs. Finally, these two factors are modeled and compared to the observed CH$_{3}$OH to have a complete picture of the chemical processes that regulate the production of complex organic molecules.
    	
	\section{Observations}
		
		\subsection{Source sample}
		
			We observed toward two star-forming regions inside the Perseus molecular cloud: Barnard 1b and IC348.
		
			\subsubsection{Barnard 1b}
			
				Barnard 1 (B1), at a distance of $\sim 300$ pc \citep{Zucker2018}, is a young, intermediate-mass, moderately active star forming cloud located in the south-eastern sector of the molecular cloud complex Perseus. Barnard 1b (B1-b, left side of Fig. \ref{fig:maps}) is the most prominent core in Barnard 1, and it is known to host a multiple system of three young stellar objects (YSOs), B1b-N, B1b-S, and B1b-W \citep[see, e.g.,][]{Hirano2014, Gerin2015}. This source has been mapped in many molecular tracers such as CS, NH$_3$, $^{13}$CO \citep{Bachiller1990}, N$_2$H$^+$ \citep{Huang2013}, H$^{13}$CO$^+$ \citep{Hirano1999}, or CH$_3$OH \citep{Hiramatsu2010, Oberg2010}, and is characterized by a rich chemistry. For instance, interferometric observations with ALMA have shown signatures of  hot corino chemistry in B1b-S \citep{Marcelino2018}, rich in COMs. Barnard 1b also displays a high degree of deuterium fractionation \citep{Lis2002, Marcelino2005} and an enhanced sulfur chemistry when compared to cold cores \citep{Fuente2016}.
			
			\subsubsection{IC 348}
				
\begin{figure*}
					\centering
					\includegraphics[width=\textwidth]{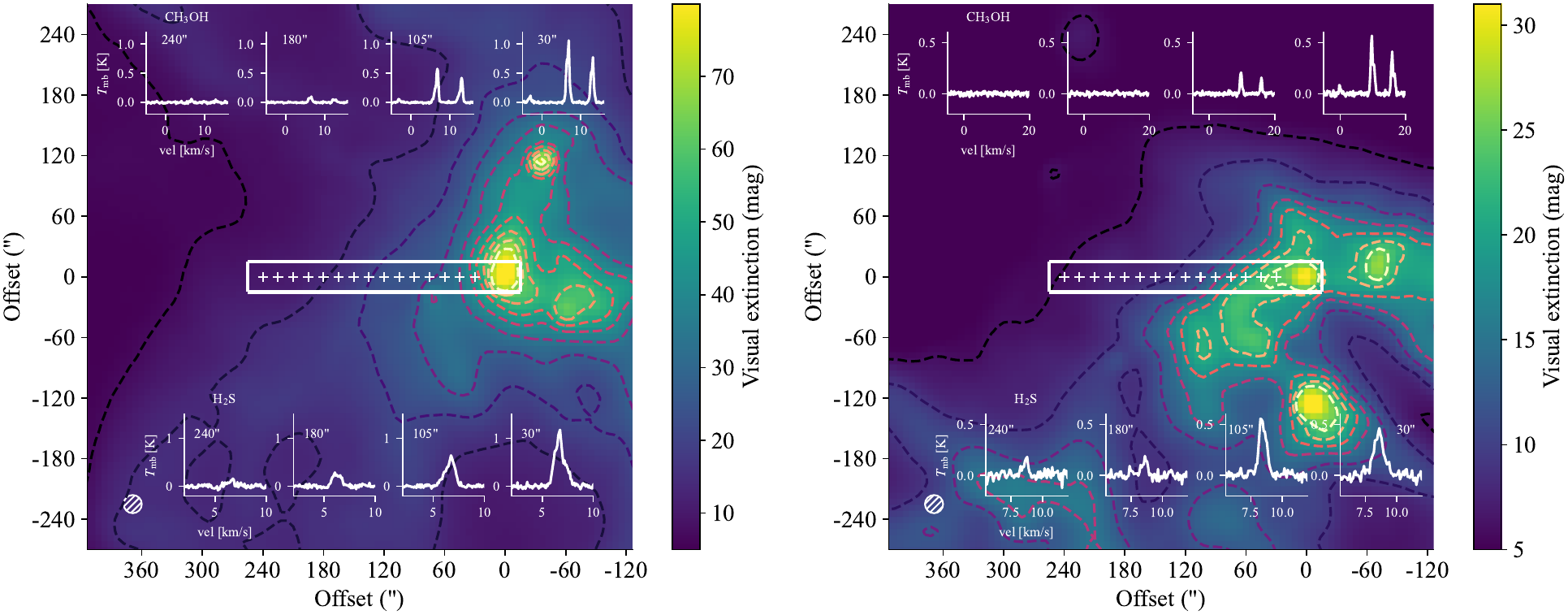}
					\caption{Visual extinction maps of the observed regions. Contours in the Barnard 1b map (left) correspond to $[7, 14, 21, 28, 35, 42, 49, 56, 63, 70]$ mag levels. The contours in the IC 348 map (right) correspond to $[6, 9, 12, 15, 18, 21, 24]$ mag levels. These maps were obtained from the $\tau_{850}$ optical depth maps in \citet{Zari2016} applying the K-band to V-band extinction ratio of $\sim 1/16$ \citep{Nishiyama2008}. The crosses mark the pointings from where data was taken. Selected spectra of CH$_{3}$OH lines at 90 GHz (Table \ref{tab:lines}) and H$_{2}$S $1_{1,0}\rightarrow 1_{0,1}$ line observed at different offsets are included in the maps.}
					\label{fig:maps}
		\end{figure*}

				We observed the horizontal strip across the dense core IC 348-SW1 shown in the right side of Fig. \ref{fig:maps}, located in the south-western part of the IC 348 cluster. From now on, this dense core is simply referred to as IC 348. This cluster, at a distance of $\sim 337$ pc \citep{Poggio2021}, is a nearby and young $\sim 2-3$ Myr cluster with a total mass of $\sim 90$ M$_{\odot}$ \citep{Herbst2008}. It has been the target of several studies about its age and disk population \citep{Lada2006, Muench2007}. Evidences of its star-forming activity and outflows have been reported in \citet{Tafalla2006}, displaying hints of CO depletion at high densities. It exhibits a rich chemistry in carbon chains and, in particular, polycyclic aromatic hydrocarbons (PAHs) and fullerenes \citep{Iglesias-Groth2019, Iglesias-Groth2023}.
		
		\subsection{Single-dish millimeter observations}
		
		The observations toward Barnard 1b and IC 348 (Table \ref{tab:sources}) presented here were taken using IRAM 30m and Yebes 40m telescopes. The data were calibrated and reduced using the \texttt{GILDAS-CLASS}\footnote{\url{https://www.iram.fr/IRAMFR/GILDAS}} software.
		
		\begin{table}
			\resizebox{0.49\textwidth}{!}{
			\begin{threeparttable}
   				\caption{Target coordinates and properties.}\label{tab:sources}
   				\centering
   				\begin{tabular}{lcccc}
     				\toprule
		        	Source &  RA (J2000) &  Dec (J2000) & $v_{\rm lsr}$ (km s$^{-1}$) & Distance (pc) \\ \midrule 
		        	Barnard 1b & 03:33:20.90 & +31:07:34.0 & 6.5 & 301\tnote{(1)} \\
		        	IC 348-SW1\tnote{(2)} & 03:43:57.00	 & +32:02:54.8 & 9.2 & 337\tnote{(3)} \\
    		    	\bottomrule 
     			\end{tabular}
     			\begin{tablenotes}
       				\item [(1)] \citet{Zucker2018}.
       				\item [(2)] Name convention as in \citet{Tafalla2006}.
       				\item [(3)] \citet{Poggio2021}.
     			\end{tablenotes}
  			\end{threeparttable}
  			}
    	\end{table}
    	
    	The IRAM 30m observations were carried out in two schedule blocks: from the 20th to the 22nd of December 2022, and the 26th and 27th of February 2023. The Yebes 40m telescope data were observed from the 2nd to the 9th of July 2018. The IRAM 30m data were taken in position-switching mode along an horizontal strip from offsets $+30''$ to $+240''$, spaced by $+15''$, with respect to the target coordinates shown in Table \ref{tab:sources} and Fig. \ref{fig:maps}. The off-source position was located at an offset of $+500''$ with respect to the targets (Table \ref{tab:sources}). CH$_{3}$OH lines at 96 GHz were also observed at offsets $0''$ and $+15''$ toward Barnard 1b. A deep integration over the off-source position was performed in frequency-switching mode with a frequency throw of 6 MHz, which was then added to the products of the position-switching procedure to ensure that no line was subtracted from the resulting data. The IRAM 30m observations used the EMIR receivers in combination with the Fast Fourier Transform Spectrometers (FFTS) as backends, yielding a spectral resolution of 50 kHz ($\sim 0.17$ km s$^{-1}$ at 90 GHz). The intensity scale of the data is the main-beam temperature $T_{\rm mb}$, which is related to the antenna temperature $T_{\rm A}$ via the main-beam efficiency $\eta_{\rm mb}$ in such a way that $T_{\rm mb}\ \eta_{\rm mb} = T_{\rm A}$. The main-beam efficiencies $\eta_{\rm mb}$ and half power beam widths (HPBWs) $\theta_{\rm HPBW}$ are shown in Table \ref{tab:lines}.
    	
    	\begin{table}
    		\resizebox{0.49\textwidth}{!}{
			\begin{threeparttable}
   				\caption{Parameters of the observed lines.}\label{tab:lines}
   				\centering
   				\begin{tabular}{lcccc}
     				\toprule
		        	Molecule &  Transition &  Frequency (MHz) & $\theta_{\rm HPBW}$ ($''$)\tnote{(1)} & $\eta_{\rm mb}$\tnote{(1)} \\ \midrule
		        	C$^{18}$O & $1\rightarrow 0$ & 109782.1734 & 22 & 0.79 \\
		        	o-H$_{2}$S & $1_{1,0}\rightarrow 1_{0,1}$ & 168762.7624 & 15 & 0.70\\
		        	N$_{2}$H$^{+}$ & $1\rightarrow 0$ & 93173.3977 & 26 & 0.81 \\
		        	%OCS & $8\rightarrow 7$ & 97301.2085 & 25 & 0.80 \\
		        	OCS & $9\rightarrow 8$ & 109463.0630 & 22 & 0.79 \\
		        	CH$_{3}$OH\tnote{(2)} & A$^{+}$\ \ $1_{0,1}\rightarrow 0_{0,0}$ & 48372.4600 & 36 & 0.49 \\
		        	CH$_{3}$OH & E$_2$\ \ $2_{1,2}\rightarrow 1_{1,1}$ & 96739.3580 & 25 & 0.80 \\
		        	CH$_{3}$OH & A$^{+}$\ \ $2_{0,2}\rightarrow 1_{0,1}$ & 96741.3710 & 25 & 0.80 \\
		        	CH$_{3}$OH & E$_1$\ \ $2_{0,2}\rightarrow 1_{0,1}$ & 96744.5450 & 25 & 0.80 \\
    		    	\bottomrule 
     			\end{tabular}
     			\begin{tablenotes}
     				\item [(1)] Yebes 40m and IRAM 30m telescope efficiencies taken from the \href{https://rt40m.oan.es/rt40m_en.php}{Yebes 40m technical information webpage} (measurements taken in 2020) and \href{https://publicwiki.iram.es/Iram30mEfficiencies}{IRAM 30m efficiencies wiki}, respectively.
       				\item [(2)] Observed with the 40m Yebes telescope only toward Barnard 1b. The rest of the lines were observed with the IRAM 30m telescope.
     			\end{tablenotes}
  			\end{threeparttable}
  			}
    	\end{table}
    	
    	The Yebes 40m data were taken in position-switching mode to provide additional methanol lines (Table \ref{tab:lines}) in those positions toward Barnard 1b where methanol lines at 90 GHz were no longer detected, that is, at offsets from $+150''$ to $+240''$. The line A$^{+}$\ \ $1_{0,1}\rightarrow 0_{0,0}$ was observed using the HEMT receiver in the Q band $(31.5-50$ GHz), covering a bandwidth of $\sim 9$ GHz with a spectral resolution of $\sim 38$ kHz.
	\section{Results}
	
		Observations were performed along horizontal strips toward Barnard 1b and IC 348 as described in the previous section. These strips are shown in Fig. \ref{fig:maps}. We targeted and detected the lines listed in Table \ref{tab:lines} in most parts of the two regions. Line spectra toward Barnard 1b and IC 348 are shown in Appendixes \ref{sec:spectraBarnard1b} and \ref{sec:spectraIC348}, respectively. The resulting main-beam temperatures, line widths, and integrated intensities are listed in Appendixes \ref{sec:linePropBarnard1b} and \ref{sec:linePropIC348}, respectively. We used the integrated intensities and line widths of CH$_{3}$OH listed in Tables \ref{tab:linePropertiesCH3OHB1b}, \ref{tab:linePropertiesCH3OHB1b40m}, and \ref{tab:linePropertiesCH3OHIC348} to derive first the gas density and CH$_{3}$OH column density, and then, assuming this gas density, the column density of the remaining species.
		
		\subsection{Derivation of gas density and methanol column density toward Barnard 1b and IC 348}\label{sec:methanolDens}
			
\begin{figure*}
				\centering
				\includegraphics[width=\textwidth]{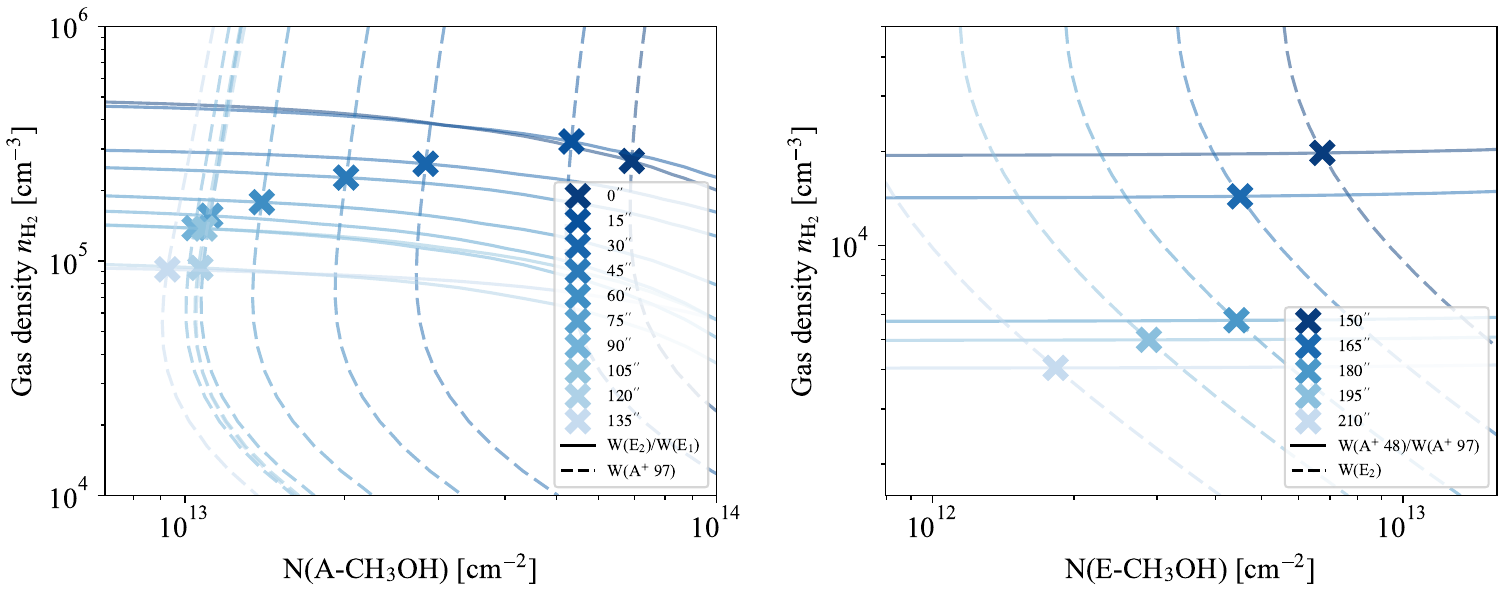}
				\caption{Contours of integrated intensities and integrated intensity ratios to estimate gas density and methanol column density. \emph{Left}: the contours of the integrated intensity $W({\rm A^{+}}\ 2_{0,2}\rightarrow 1_{0,1})$ (vertical dashed lines) and the contours of the integrated intensity ratio $W({\rm E_{2}})/W({\rm E_{1}})$ (horizontal dashed lines) intersect on a grid of \texttt{RADEX} models to yield gas density $n_{\rm H_{2}}$ and N($A$-CH$_{3}$OH) in offsets from $0''$ to $+135''$ in Barnard 1b. \emph{Right}: contours of the integrated intensity $W({\rm E_{2}})$ (vertical dashed lines) and the contours of the integrated intensity ratio $W({\rm A^{+}}\ 2_{0,2}\rightarrow 1_{0,1})/W({\rm A^{+}}\ 1_{0,1}\rightarrow 0_{0,0})$ (horizontal dashed lines) intersect on a grid of \texttt{RADEX} models to yield gas density $n_{\rm H_{2}}$ and N($E$-CH$_{3}$OH) in offsets from $+150''$ to $+210''$ in Barnard 1b.}
				\label{fig:gridB1b}
			\end{figure*}

\begin{figure}
				\centering
				\includegraphics[width=0.49\textwidth]{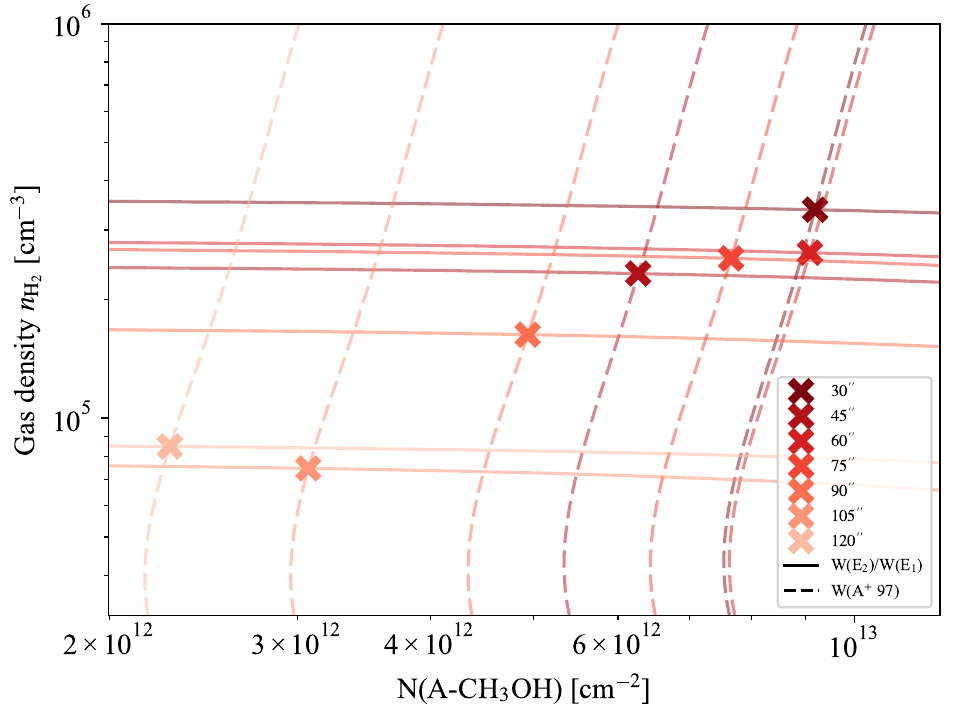}
				\caption{Contours of the integrated intensity $W({\rm A^{+}}\ 2_{0,2}\rightarrow 1_{0,1})$ (vertical dashed lines) and the contours of the integrated intensity ratio $W({\rm E_{2}})/W({\rm E_{1}})$ (horizontal dashed lines) intersect on a grid of \texttt{RADEX} models to yield gas density $n_{\rm H_{2}}$ and N($A$-CH$_{3}$OH) in offsets from $+30''$ to $+120''$ in IC 348.}
				\label{fig:gridIC348}
			\end{figure}

			The molecular hydrogen number density $n_{\rm H_{2}}$ and the column density of CH$_{3}$OH were derived using the non-LTE molecular radiative transfer code \texttt{RADEX} \citep{vanDerTak2007} in the escape probability approach over a uniform sphere and the collisional rates of the two methanol spin isomers $E$-CH$_{3}$OH and $A$-CH$_{3}$OH reported in \citet{Rabli2010}. Gas temperature was assumed to be equal to the dust temperature, which was taken from the dust temperature maps of Perseus presented in \citet{Zari2016}. This approximation is adequate for densities $n_{\rm H_{2}} > 10^{4}$ cm$^{-3}$ \citep{Fuente2019} and fair for lower densities inside molecular clouds, where the kinetic temperature derived with ammonia inversion lines is $\sim 1-2$ K lower than the dust temperature \citep{Friesen2017}. The uncertainties in our observations account for these discrepancies. The process of computing the gas density and the CH$_{3}$OH column density is as follows. When the three lines of CH$_{3}$OH at $\sim 90$ GHz are detected, we ran a grid of \texttt{RADEX} models to estimate the line integrated intensity ratio $W({\rm E_{2}})/W({\rm E_{1}})$, where $W({\rm E_{2}})$ and $W({\rm E_{1}})$ stand for the integrated intensity of the CH$_{3}$OH E$_2$\ $2_{1,2}\rightarrow 1_{1,1}$ line and the integrated intensity of CH$_{3}$OH E$_1$\ $2_{0,2}\rightarrow 1_{0,1}$ line, respectively, for varying collider densities and CH$_{3}$OH column densities in the ranges $n_{\rm H_{2}} = 10^{2} - 10^{8}$ cm$^{-3}$ and $N({\rm CH_{3}OH}) = 10^{10}-10^{16}$ cm$^{-2}$, respectively. The assumed linewidths in the \texttt{RADEX} models are the average of the observed linewidths of the lines involved (see Tables \ref{tab:linePropertiesCH3OHB1b}, \ref{tab:linePropertiesCH3OHB1b40m}, and \ref{tab:linePropertiesCH3OHIC348}). In this case, models for the $W({\rm A^{+}})$ integrated intensity assume the observed A$^{+}$\ $2_{0,2}\rightarrow 1_{0,1}$ linewidth while models for the $W({\rm E_{2}})/W({\rm E_{1}})$ integrated intensity ratio assume the average linewidth of the observed E$_1$\ $2_{0,2}\rightarrow 1_{0,1}$ and E$_2$\ $2_{1,2}\rightarrow 1_{1,1}$ linewidths. The set of (N, $n_{{\rm H}_{2}}$) points whose predicted $W({\rm E_{2}})/W({\rm E_{1}})$ integrated intensity ratio with \texttt{RADEX} is equal to the observed one at a given offset appears as horizontal lines in this plane, providing a good estimate of the density (see, for instance, Fig. \ref{fig:gridB1b}). We then ran a similar computation of the integrated intensity of the CH$_{3}$OH A$^{+}$\ $2_{0,2}\rightarrow 1_{0,1}$ line, $W({\rm A^{+}})$, for each point of the N-$n_{{\rm H}_{2}}$ grid with \texttt{RADEX}. Again, the set of (N, $n_{{\rm H}_{2}}$) pairs whose predicted $W({\rm A^{+}})$ integrated intensity matches the observations appear as vertical lines. The gas density $n_{\rm H_{2}}$ and the column density N($A$-CH$_{3}$OH) at a given offset are provided by the intersection of these two sets of level curves (see Fig. \ref{fig:gridB1b} for the results of Barnard 1b and Fig. \ref{fig:gridIC348} for those of IC 348). Finally, we obtained the column density N($E$-CH$_{3}$OH) by fitting the integrated intensity of the line ${\rm E_{2}}$ using the gas density value calculated with the grid.
			
			From offsets $+150''$ to $+210''$ in Barnard 1b, where the CH$_{3}$OH line E$_{1}$ $2_{0,2}\rightarrow 1_{0,1}$ was not detected, but ${\rm E_{2}}$ $2_{1,2}\rightarrow 1_{1,1}$ was, we used the $A^{+}\ \ 1_{0,1}\rightarrow 0_{0,0}$ line observed with the Yebes 40m telescope to compute gas densities and CH$_{3}$OH column densities. This time, the gas density and N($E$-CH$_{3}$OH) column density were given by the intersection of the integrated intensity ratio $W({\rm A^{+}}\ 2_{0,2}\rightarrow 1_{0,1})/W({\rm A^{+}}\ 1_{0,1}\rightarrow 0_{0,0})$ level curves with those of the integrated intensity of the ${\rm A^{+}}$ $2_{0,2}\rightarrow 1_{0,1}$ line. We then obtained the column density N($A$-CH$_{3}$OH) by fitting the integrated intensity of the line ${\rm A^{+}}$ $2_{0,2}\rightarrow 1_{0,1}$ using the gas density value calculated with the grid (see right panel of Fig. \ref{fig:gridB1b}). Finally, at offsets $+225''$ and $+240''$ in Barnard 1b, we assumed the gas density obtained in \citet{NavarroAlmaida2020}. This allowed us to compute the column density of N($A$-CH$_{3}$OH) at offset $+225''$ and provide an upper bound for it at offset $+240''$. The densities and methanol column densities obtained toward Barnard 1b with this method are summarized in Table \ref{tab:methanolColDensB1b}. The uncertainty in their estimation is calculated as the quadratic sum of the flux calibration uncertainty and the relative uncertainty in the observed integrated intensities of methanol. The flux calibration uncertainty is tipically assumed to be $\sim 20\%$, higher than the tabulated $\sim 10\%$ \footnote{\href{https://publicwiki.iram.es/Iram30mEfficiencies}{IRAM 30m efficiencies wiki}} that is achieved with excellent weather conditions and no anomalous refraction.		
			\begin{table*}
				\resizebox{\textwidth}{!}{
				\begin{threeparttable}
   					\caption{Hydrogen number density and CH$_{3}$OH column density in Barnard 1b.}
   					\label{tab:methanolColDensB1b}
   					\centering
   					\begin{tabular}{rccc|ccccc}
     					\toprule
		        		Offset & $A_{\rm v}$ & $T_{\rm dust}$ & Gas density $n_{\rm H_{2}}$ &  N($A$-CH$_{3}$OH) & N($E$-CH$_{3}$OH) & N(CH$_{3}$OH) & \multirow{2}{*}{N(CH$_{3}$OH)/N(H)} & \multirow{2}{*}{$E/A$ Ratio} \\ 
		        		$('')$ & (mag)$^{(1)}$ & (K)$^{(2)}$ & (cm$^{-3}$) & (cm$^{-2}$) & (cm$^{-2}$) & (cm$^{-2}$) \\ \midrule
		        		0 & 72 & 10.0 & $(2.7\pm 0.6)\times 10^{5}$ & $(7\pm 1)\times 10^{13}$ & $(6\pm 1)\times 10^{13}$ & $(1.3\pm 0.2)\times 10^{14}$ & $(1.0\pm 0.1)\times 10^{-9}$ & ${0.9\pm 0.3}$\\
		        		15 & 53 & 10.5 & $(3.2\pm 0.7)\times 10^{5}$ & $(5\pm 1)\times 10^{13}$ & $(4.1\pm 0.9)\times 10^{13}$ & $(9\pm 1)\times 10^{13}$ & $(1.0\pm 0.1)\times 10^{-9}$ & $0.8\pm 0.2$\\
		        		30 & 38.7 & 11.7 & $(2.6\pm 0.9)\times 10^{5}$ & $(2.8\pm 0.6)\times 10^{13}$ & $(2.3\pm 0.5)\times 10^{13}$ & $(5.1\pm 0.8)\times 10^{13}$ & $(7\pm 1)\times 10^{-10}$ & $0.8\pm 0.2$\\
		        		45 & 24.2 & 12.4 & $(2.3\pm 0.5)\times 10^{5}$ & $(2.0\pm 0.4)\times 10^{13}$ & $(1.8\pm 0.4)\times 10^{13}$ & $(3.8\pm 0.6)\times 10^{13}$ & $(8\pm 1)\times 10^{-10}$ & $0.9\pm 0.3$\\
		        		60 & 20.0 & 12.8 & $(1.9\pm 0.4)\times 10^{5}$ & $(1.4\pm 0.3)\times 10^{13}$ & $(1.3\pm 0.6)\times 10^{13}$ & $(2.7\pm 0.6)\times 10^{13}$ & $(7\pm 2)\times 10^{-10}$ & $0.9\pm 0.4$\\
		        		75 & 17.8 & 13.3 & $(1.6\pm 0.4)\times 10^{5}$ & $(1.1\pm 0.2)\times 10^{13}$ & $(1.0\pm 0.2)\times 10^{13}$ & $(2.1\pm 0.3)\times 10^{13}$ & $(6\pm 1)\times 10^{-10}$ & $0.9\pm 0.3$\\
		        		90 & 16.2 & 13.7 & $(1.4\pm 0.4)\times 10^{5}$ & $(1.0\pm 0.2)\times 10^{13}$ & $(1.0\pm 0.2)\times 10^{13}$ & $(2.0\pm 0.3)\times 10^{13}$ & $(7\pm 1)\times 10^{-10}$ & $0.9\pm 0.3$\\
		        		105 & 14.9 & 14.1 & $(1.4\pm 0.4)\times 10^{5}$ & $(1.1\pm 0.2)\times 10^{13}$ & $(1.0\pm 0.2)\times 10^{13}$ & $(2.1\pm 0.3)\times 10^{13}$ & $(7\pm 1)\times 10^{-10}$ & $0.9\pm 0.3$\\
		        		120 & 13.6 & 14.4 & $(9\pm 2)\times 10^{4}$ & $(1.1\pm 0.2)\times 10^{13}$ & $(1.0\pm 0.2)\times 10^{13}$ & $(2.1\pm 0.3)\times 10^{13}$ & $(8\pm 1)\times 10^{-10}$ & $0.9\pm 0.3$\\
		        		135 & 12.3 & 14.7 & $(9\pm 3)\times 10^{4}$ & $(9\pm 2)\times 10^{12}$ & $(9\pm 6)\times 10^{12}$ & $(1.8\pm 0.6)\times 10^{13}$ & $(8\pm 2)\times 10^{-10}$ & $0.9\pm 0.6$\\
		        		150 & 11.0 & 15.0 & $(1.8\pm 0.5)\times 10^{4}$ & $(7\pm 1)\times 10^{12}$ & $(7\pm 3)\times 10^{12}$ & $(1.3\pm 0.3)\times 10^{13}$ & $(6\pm 2)\times 10^{-10}$ & $1.0\pm 0.5$\\
		        		165 & 9.8 & 15.3 & $(1.2\pm 0.3)\times 10^{4}$ & $(5\pm 1)\times 10^{12}$ & $(4.4\pm 0.9)\times 10^{12}$ & $(1.0\pm 0.1)\times 10^{13}$ & $(5.3\pm 0.8)\times 10^{-10}$ & $0.8\pm 0.2$\\
		        		180 & 8.6 & 15.7 & $(5\pm 2)\times 10^{3}$ & $(4.2\pm 0.9)\times 10^{12}$ & $(4.4\pm 0.9)\times 10^{12}$ & $(9\pm 1)\times 10^{12}$ & $(5.3\pm 0.8)\times 10^{-10}$ & $1.0\pm 0.3$\\
		        		195 & 8.0 & 16.0 & $(5\pm 2)\times 10^{3}$ & $(3.5\pm 0.8)\times 10^{12}$ & $(2.7\pm 0.6)\times 10^{12}$ & $(6\pm 1)\times 10^{12}$ & $(4.2\pm 0.7)\times 10^{-10}$ & $0.8\pm 0.3$\\
		        		210 & 7.4 & 16.3 & $(4\pm 2)\times 10^{3}$ & $(3.3\pm 0.7)\times 10^{12}$ & $(1.6\pm 0.4)\times 10^{12}$ & $(4.9\pm 0.8)\times 10^{12}$ & $(3.5\pm 0.6)\times 10^{-10}$ & $0.5\pm 0.2$\\
		        		225 & 6.3 & 16.5 & $(1.9\pm 0.9)\times 10^{3}$ $^{(a)}$ & $(2.7\pm 0.6)\times 10^{12}$ & $(2.4\pm 0.7)\times 10^{12}$ & $(5\pm 1)\times 10^{12}$ & $(4.3\pm 0.8)\times 10^{-10}$ & $0.9$ $^{(b)}$\\
		        		240 & 6.2 & 16.7 & $(2\pm 1)\times 10^{3}$ $^{(a)}$ & $< 4\times 10^{12}$ & $< 4\times 10^{12}$ & $< 8\times 10^{12}$ & $< 7\times 10^{-10}$ & $-$\\
    		    		\bottomrule 
     				\end{tabular}
					\begin{tablenotes}
     					\item [(1)] The uncertainty in the visual extinction across the strip is $\sim 2\%$.
       					\item [(2)] The uncertainty in the dust temperature across the strip is $\pm\ 0.1$ K.
       					\item [(a)] The density at this offset is taken from \citet{NavarroAlmaida2020}.
       					\item [(b)] This $E/A$ ratio at this offset is set to be equal to the theoretical one.
     				\end{tablenotes}
  				\end{threeparttable}
  				}
    		\end{table*}
			
\begin{figure*}
				\centering
				\includegraphics[width=\textwidth]{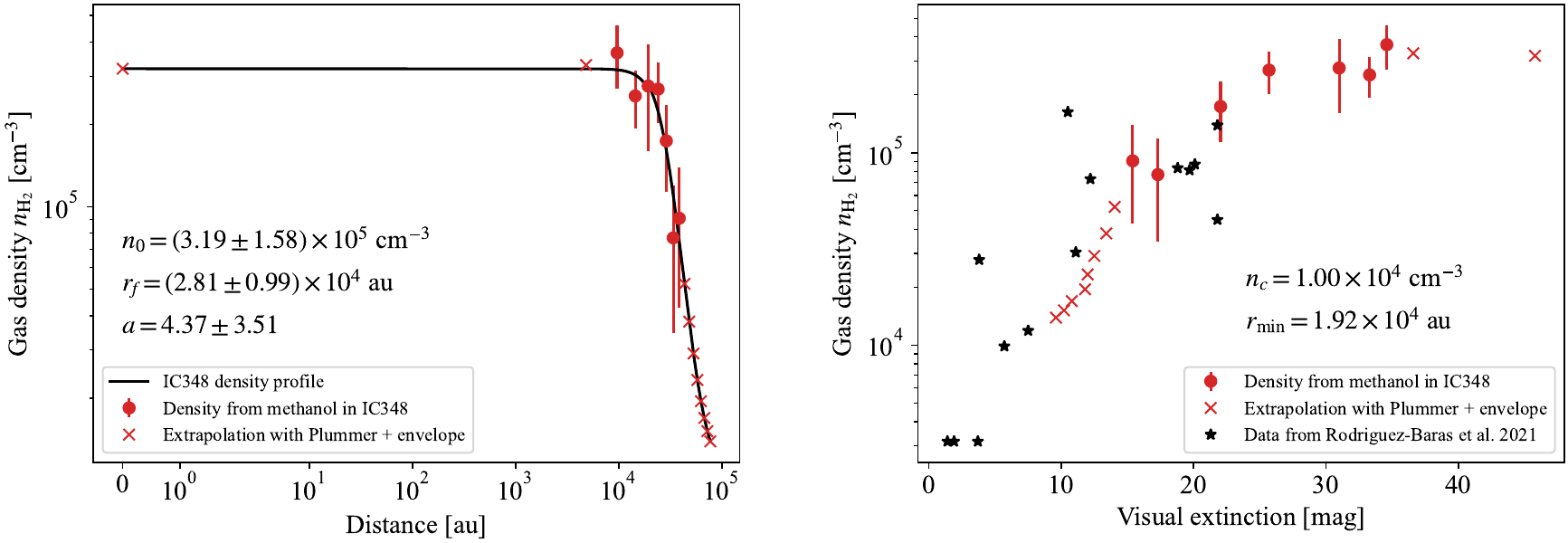}
				\caption{Density across the observed strip in IC 348. \emph{Left:} Plummer-like profile and envelope (black solid line) that best fit the density derived from methanol integrated intensities (red dots). The red crosses mark the density that would correspond to the points where an estimation using methanol was not possible. \emph{Right:} the parameters of the envelope were chosen so that the density at the outermost areas (red dots and crosses) matches the values of density from \citet{RodriguezBaras2021} (black stars) found at points of similar extinction in the same region.}
				\label{fig:densityProfileIC348}
			\end{figure*}

			In IC 348 we did not detect the three methanol lines at 96 GHz (Table \ref{tab:lines}) at offsets $+135''$ to $+240''$. At these positions, we modeled the gas density toward IC 348 by fitting the density values that were possible to calculate with the method described above to a Plummer-like profile, an analytical approximation to the Bonnor-Ebert sphere that is commonly used to describe the density distribution of dense cores \citep{Tafalla2004}. The Plummer-like profile, as a function of the radius $r$, has the following analytical expression:
			\begin{equation}\label{eq:plummer}
				n_{\rm H_{2}}(r) = \frac{n_{0}}{1+\left(\frac{r}{r_{f}}\right)^{a}},
			\end{equation}
			where $n_{0}$ is the central gas density, $r_{f}$ is the flat radius, and $a$ is the asymptotic power index. The values of the parameters in Eq. \ref{eq:plummer} that fit the densities derived from methanol data toward IC 348 are shown in Fig. \ref{fig:densityProfileIC348}. We then added an envelope to the Plummer-like profile so that density does not fall sharply at longer radii where methanol is not detected. Such envelope $n_{\rm env}$ is proportional to a Heaviside step function
			\begin{equation}\label{eq:envelope}
				n_{\rm env}(r) = \frac{n_{c}}{1+\exp(-k(r-r_{\rm min}))},
			\end{equation}
			where $n_{c}$ is the constant density of the envelope (the height of the step), $r_{\rm min}$ is the minimum radial extent of the envelope, and the parameter $k$ determines how sharp is the transition between the envelope and the Plummer-like density profile when $r=r_{\rm min}$. In this case, $k$ was chosen sufficiently large so that the envelope does not alter the density of the Plummer profile for radii $r<r_{\rm min}$. Both $n_{c}$ and $r_{\rm min}$ were chosen so that the gas density at a given visual extinction is in agreement with values found in the literature. Since we know the visual extinction at different radii (Fig. \ref{fig:maps}), we modified $n_{c}$ and $r_{\rm min}$ in order to find a good agreement with the results published in \citet{RodriguezBaras2021} (see Fig. \ref{fig:densityProfileIC348}). The resulting density profile is shown in Fig. \ref{fig:densityProfileIC348}.
			Once gas density is calculated at all offsets toward IC 348, we obtained N($A$-CH$_{3}$OH) and N($E$-CH$_{3}$OH) independently at offsets from $+135''$ to $+180''$ by matching, respectively, the integrated intensity of CH$_{3}$OH $A^{+}\ 2_{0,2}\rightarrow 1_{0,1}$ and CH$_{3}$OH $E_2\ 2_{1,2}\rightarrow 1_{1,1}$ lines with \texttt{RADEX}. At offsets $+195''$ and $+210''$, we could only determine N($A$-CH$_{3}$OH) with \texttt{RADEX} using its observed integrated intensity. The gas densities and methanol column densities obtained toward IC 348 are summarized in Table \ref{tab:methanolColDensIC348}.
			
			\begin{table*}
				\resizebox{\textwidth}{!}{
				\begin{threeparttable}
   					\caption{Hydrogen number density and CH$_{3}$OH column density in IC 348.}
   					\label{tab:methanolColDensIC348}
   					\centering
   					\begin{tabular}{rccc|ccccc}
     					\toprule
		        		Offset & $A_{\rm v}$ & $T_{\rm dust}$ & Gas density $n_{\rm H_{2}}$ &  N($A$-CH$_{3}$OH) & N($E$-CH$_{3}$OH) & N(CH$_{3}$OH) & \multirow{2}{*}{N(CH$_{3}$OH)/N(H)} & \multirow{2}{*}{$E/A$ Ratio} \\ 
		        		$('')$ & (mag)$^{(1)}$ & (K)$^{(2)}$ & (cm$^{-3}$) & (cm$^{-2}$) & (cm$^{-2}$) & (cm$^{-2}$) \\ \midrule
		        		30 & 24.0 & 17.5 & $(4 \pm 1)\times 10^{5}$ & $(9\pm 2)\times 10^{12}$ & $	(9\pm 2)\times 10^{12}$ & $(1.9\pm 0.3)\times 10^{13}$ & $(4.2\pm 0.7)\times 10^{-10}$ & $1.0\pm 0.3$\\
		        		45 & 23.1 & 17.6 & $(2.5\pm 0.6)\times 10^{5}$ & $(6\pm 1)\times 10^{12}$ & $(8\pm 2)\times 10^{12}$ & $(1.4\pm 0.2)\times 10^{13}$ & $(3.3\pm 0.5)\times 10^{-10}$ & $1.3\pm 0.4$\\
		        		60 & 20.0 & 17.7 & $(3\pm 1)\times 10^{5}$ & $(9\pm 2)\times 10^{12}$ & $(9\pm 4)\times 10^{12}$ & $(1.8\pm 0.4)\times 10^{13}$ & $(5\pm 1)\times 10^{-10}$ & $1.0\pm 0.5$\\
		        		75 & 17.6 & 17.9 & $(2.7\pm 0.7)\times 10^{5}$ & $(8\pm 2)\times 10^{12}$ & $(7\pm 2)\times 10^{12}$ & $(1.5\pm 0.2)\times 10^{13}$ & $(4.5\pm 0.7)\times 10^{-10}$ & $0.9\pm 0.3$\\
		        		90 & 14.0 & 18.0 & $(1.7\pm 0.6)\times 10^{5}$ & $(5\pm 1)\times 10^{12}$ & $(4\pm 1)\times 10^{12}$ & $(9\pm 2)\times 10^{12}$ & $(3.5\pm 0.6)\times 10^{-10}$ & $0.9\pm 0.3$\\
		        		105 & 11.2 & 18.1 & $(8\pm 4)\times 10^{4}$ & $(3.1\pm 0.6)\times 10^{12}$ & $(2\pm 1)\times 10^{12}$ & $(6\pm 1)\times 10^{12}$ & $(2.6\pm 0.7)\times 10^{-10}$ & $0.8\pm 0.5$\\
		        		120 & 10.1 & 18.2 & $(9\pm 5)\times 10^{4}$ & $(2.3\pm 0.5)\times 10^{12}$ & $(1.8\pm 0.9)\times 10^{12}$ & $(4\pm 1)\times 10^{12}$ & $(2.2\pm 0.5)\times 10^{-10}$ & $0.8\pm 0.4$\\
		        		135 & 9.4 & 18.3 & $(5\pm 2)\times 10^{4}$ & $(1.4\pm 0.3)\times 10^{12}$ & $(1.1\pm 0.3)\times 10^{12}$ & $(2.6\pm 0.4)\times 10^{12}$ & $(1.4\pm 0.2)\times 10^{-10}$ & $0.8\pm 0.3$\\
		        		150 & 8.7 & 18.6 & $(4\pm 2)\times 10^{4}$ & $(7\pm 2)\times 10^{11}$ & $(5\pm 1)\times 10^{11}$ & $(1.2\pm 0.2)\times 10^{12}$ & $(8\pm 1)\times 10^{-11}$ & $0.7\pm 0.3$\\
		        		165 & 8.3 & 18.8 & $(3\pm 1)\times 10^{4}$ & $(3\pm 1)\times 10^{11}$ & $(3\pm 1)\times 10^{11}$ & $(6\pm 2)\times 10^{11}$ & $(4\pm 1)\times 10^{-11}$ & $0.8\pm 0.4$\\
		        		180 & 8.1 & 18.9 & $(2\pm 1)\times 10^{4}$ & $(5\pm 2)\times 10^{11}$ & $(4\pm 2)\times 10^{11}$ & $(9\pm 2)\times 10^{11}$ & $(6\pm 2)\times 10^{-11}$ & $0.9\pm 0.5$\\
		        		195 & 8.1 & 19.1 & $(1.7\pm 0.8)\times 10^{4}$ & $(3\pm 1)\times 10^{11}$ & $(3\pm 1)\times 10^{11}$ & $(6\pm 2)\times 10^{11}$ & $(4\pm 1)\times 10^{-11}$ & $0.9$\\
		        		210 & 7.5 & 19.2 & $(1.4\pm 0.7)\times 10^{4}$ & $(3\pm 1)\times 10^{11}$ & $(2\pm 1)\times 10^{11}$ & $(5\pm 1)\times 10^{11}$ & $(4\pm 1)\times 10^{-11}$ & $0.9$\\
		        		225 & 7.1 & 19.2 & $(1.2\pm 0.6)\times 10^{4}$ & $<7\times 10^{11}$ & $<6\times 10^{11}$ & $<1\times 10^{12}$ & $<1\times 10^{-10}$ & $-$\\
		        		240 & 6.7 & 19.3 & $(1.1\pm 0.5)\times 10^{4}$ & $<1\times 10^{12}$ & $<8\times 10^{11}$ & $<2\times 10^{12}$ & $<2\times 10^{-10}$ & $-$\\
    		    		\bottomrule 
     				\end{tabular}
					\begin{tablenotes}
     					\item [(1)] The uncertainty in the visual extinction across the strip is $\sim 2\%$.
       					\item [(2)] The uncertainty in the dust temperature across the strip is $\pm\ 0.1$ K.
     				\end{tablenotes}
  				\end{threeparttable}
  				}
    		\end{table*}
			
\begin{figure*}
				\centering
				\includegraphics[width=0.52\textwidth]{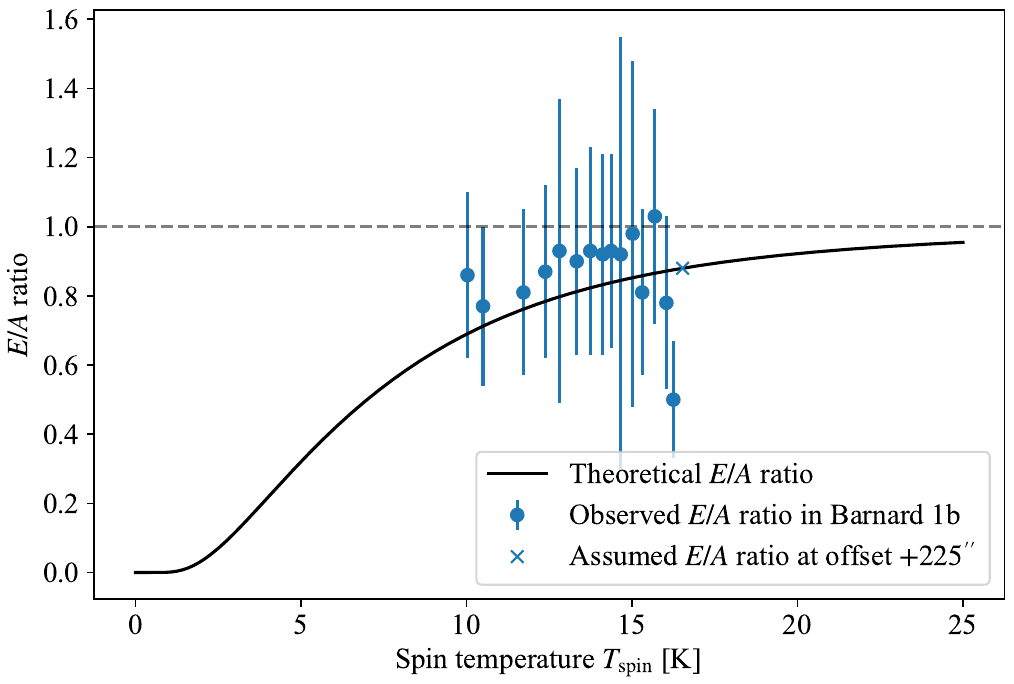}
				\includegraphics[width=0.4675\textwidth]{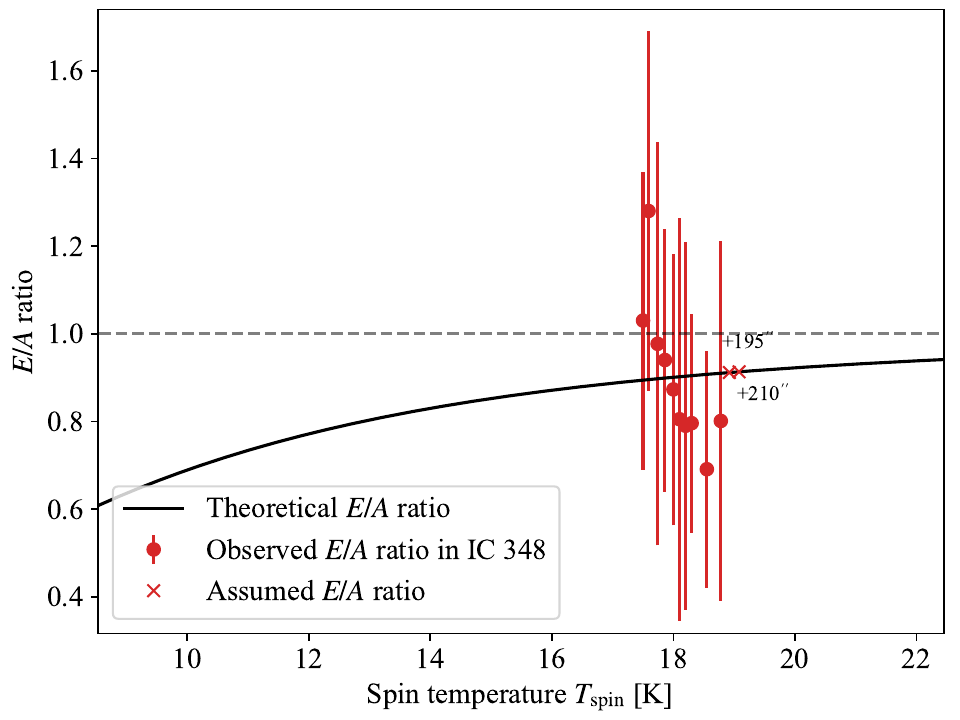}
				\caption{Theoretical methanol $E/A$ ratios (black solid line) compared to those observed in Barnard 1b (blue dots, left panel) and IC 348 (red dots, right panel). The blue and red crosses indicate that the $E/A$ ratio used to estimate the total methanol column density is set to be equal to the theoretical one. The horizontal dashed line corresponds to the statistical 1:1 ratio.}
				\label{fig:spinTemp}
			\end{figure*}

			When the column density of only one of the isomers could be calculated, that is, in offset $+225''$ toward Barnard 1b and offsets $+195''$ and $+210''$ toward IC 348, we used theoretical considerations to obtain the column density of its isomer counterpart by assuming that the relative population of the methanol spin isomers is governed by Maxwell-Boltzmann statistics with the so-called nuclear spin temperature $T_{\rm spin}$. The N($E$-CH$_{3}$OH)/N($A$-CH$_{3}$OH) ratio, $E/A$ ratio for short, is therefore given by:
			\begin{equation}\label{eq:spinTemp}
				\frac{E}{A} = \frac{\displaystyle\sum\limits_{i\in E}g_{i}\exp\left(-\frac{E_{i}^{E}}{kT_{\rm spin}}\right)}{\displaystyle\sum\limits_{j\in A}g_{j}\exp\left(-\frac{E_{j}^{A}}{kT_{\rm spin}}\right)},
			\end{equation}
			where $i$ and $j$ are summed over the statistical weights $g_{i}$, $g_{j}$, and the energy levels $E_{i}^{E}$ and $E_{j}^{A}$ of the species $E$-CH$_{3}$OH and $A$-CH$_{3}$OH, respectively. Transitions connecting the two spin states $E$-CH$_{3}$OH and $A$-CH$_{3}$OH are strongly forbidden \citep{Kawakita2009}, so it is customarily assumed to be negligible in the interstellar medium (ISM) and thus the $E/A$ ratio is set at the time of formation of this molecule \citep{Wirstrom2011}. If methanol is thermalized at the time of formation, spin temperature $T_{\rm spin}$ is an approximation of the temperature of the formation environment. Since methanol is formed most efficiently on the surface of interstellar grains, we tested whether $T_{\rm dust}$ is a good estimate of $T_{\rm spin}$ \citep{Holdship2019}. To do so, we compared the observed $E/A$ ratio (see Tables \ref{tab:methanolColDensB1b} and \ref{tab:methanolColDensIC348}) with the theoretical result (Eq. \ref{eq:spinTemp} with $T_{\rm spin}=T_{\rm dust}$) in Fig. \ref{fig:spinTemp}. Our estimates of the methanol column density are compatible with a formation in an environment with a temperature similar to that of the dust, although they are also consistent with the statistical 1:1 ratio given the uncertainties in the dust temperatures and column densities. Thus the choice of the $E/A$ ratios assuming $T_{\rm spin}=T_{\rm dust}$ is not significantly different from the statistical value.
			
\begin{figure*}
				\centering
				\includegraphics[width=\textwidth]{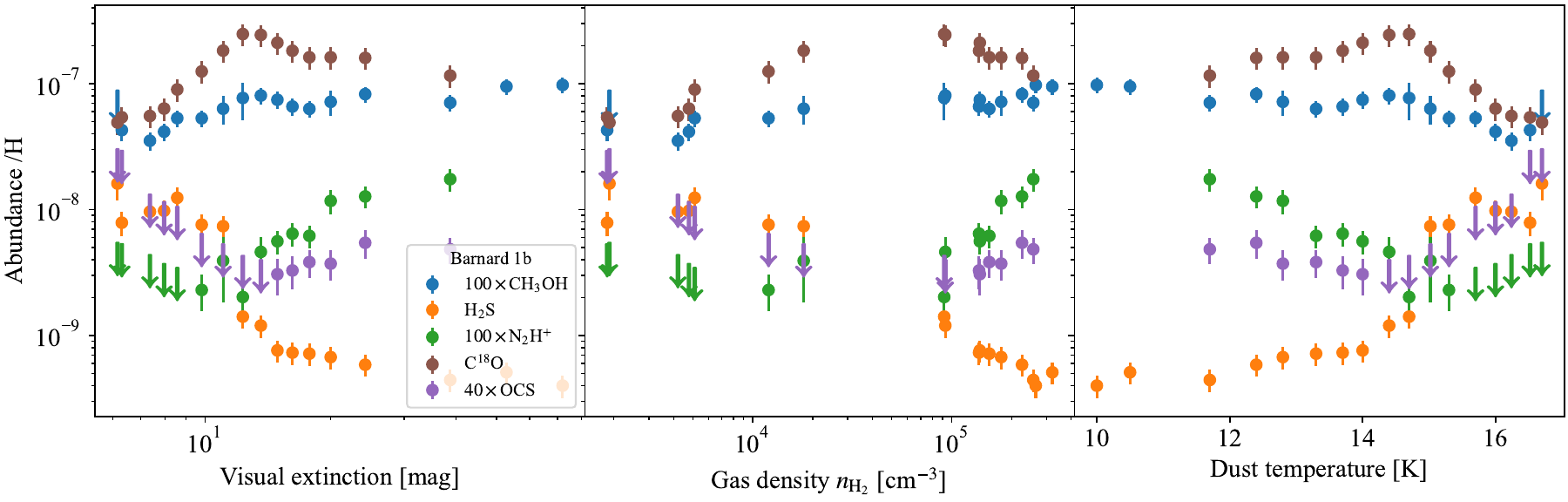}
				\caption{Scaled gas-phase abundances of methanol (blue dots), H$_{2}$S (orange dots), N$_{2}$H$^{+}$ (green dots), C$^{18}$O (brown dots), and OCS (purple dots) in Barnard 1b as a function of the visual extinction (left panel), gas density $n_{\rm H_{2}}$ (middle panel), and dust temperature (right panel).}
					\label{fig:allAbundancesB1b}
			\end{figure*}

\begin{figure*}
				\centering
				\includegraphics[width=\textwidth]{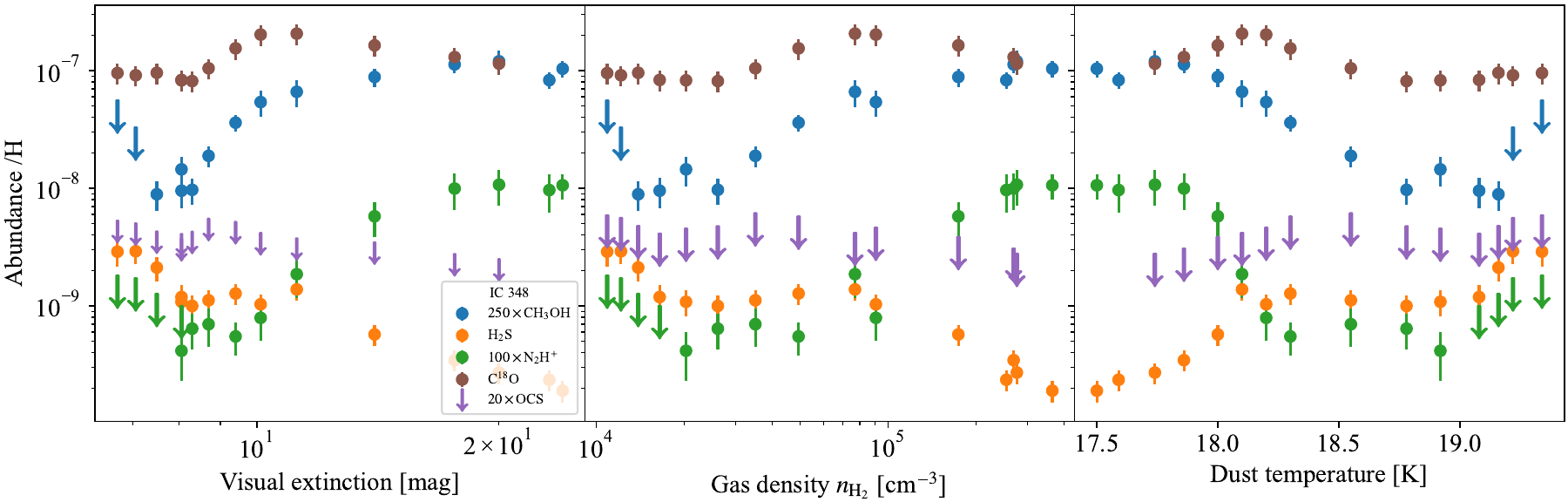}
				\caption{Scaled gas-phase abundances of methanol (blue dots), H$_{2}$S (orange dots), N$_{2}$H$^{+}$ (green dots), C$^{18}$O (brown dots), and OCS (purple dots) in IC348 as a function of the visual extinction (left panel), gas density $n_{\rm H_{2}}$ (middle panel), and dust temperature (right panel).}
					\label{fig:allAbundancesIC348}
			\end{figure*}

			With the resulting column densities of methanol N(CH$_{3}$OH), computed as N(CH$_{3}$OH) $=$ N($E$-CH$_{3}$OH) $+$ N($A$-CH$_{3}$OH), we calculated its abundance N(CH$_{3}$OH)/N(H), with N(H) being the total column density of H nuclei. N(H) is related to the visual extinction so that N(H) = $1.88\times 10^{21}\ A_{\rm v}$ \citep{Bohlin1978}, with $A_{\rm v}$ given by the extinction maps in Fig. \ref{fig:maps}. In the range of physical conditions across the different pointings, methanol abundance spans one order of magnitude from the most abundant areas at the highest extinction to the warmer and less dense ones where it is less abundant. In Fig. \ref{fig:allAbundancesB1b}, we plot the abundances of methanol toward Barnard 1b as a function of the visual extinction, gas density, and dust temperature. Methanol abundance is correlated with the visual extinction, with the highest abundances found at the extinction peak. This is also found for the gas density, as the methanol abundance reaches its highest value on the densest positions. Finally, methanol abundance is anti-correlated with the dust temperature. The decreasing abundances of methanol beyond $\sim 15$ K are due to the lower availability of the reactants needed to form methanol on the surface of dust grains. Indeed, hydrogenation may become ineffective due to the low abundance of H over grain surfaces at these temperatures \citep{Caselli1993}. It has also been shown \citep[see, e.g.,][]{Fuchs2009, MunozCaro2010} that thermal desorption of CO is triggered when $T_{\rm dust} \gtrsim 15$ K, having important implications for the composition of ices and the formation of molecules like methanol on the surface of interstellar grains. Methanol abundances toward IC 348, shown in Fig. \ref{fig:allAbundancesIC348}, behave in a similar fashion to Barnard 1b, spanning one order of magnitude while being generally lower by a factor of $\sim 2$. The peak of methanol abundance follows the same trend, being located in the coldest, most far ultraviolet (FUV)-shielded positions. Thus, methanol abundance is again correlated with the visual extinction and gas density and anti-correlated with the dust temperature. Dust temperature is higher in IC 348 and this is apparent in the absence of a methanol abundance plateau at the lowest temperatures.
			
		\subsection{H\textsubscript{2}S, N\textsubscript{2}H\textsuperscript{+}, C\textsuperscript{18}O, and OCS abundances}
			
			We derived the abundances for the rest of species listed in Table \ref{tab:lines} using the gas density estimation. The uncertainties in the column densities were calculated following the same procedure as explained above for methanol. The results are shown in Fig. \ref{fig:allAbundancesB1b}.
				
			We obtained the column density of H$_{2}$S toward Barnard 1b and IC 348 by assuming the statistical ortho-to-para ratio (3:1) and fitting the integrated intensity of the o-H$_{2}$S $1_{1,0}\rightarrow 1_{0,1}$ line (Table \ref{tab:lines}) with \texttt{RADEX}. We used the collisional rates of H$_{2}$S with molecular hydrogen presented in \citet{Dagdigian2020}. The results are listed in Table \ref{tab:b1bAbundances} and shown in Fig. \ref{fig:allAbundancesB1b}. H$_{2}$S abundances observed in IC 348 span about one order of magnitude and are generally lower than those found in Barnard 1b, where they span two orders of magnitude. Its peak in both regions is found toward the outermost positions. Compared to the abundances reported in \citet{NavarroAlmaida2020}, ours are in general lower, but compatible within a factor of two. This is a consequence of gas densities we derived with methanol, which are higher than the ones they obtained with CS lines. CS lines are expected to trace more external layers of gas in the pointings of the strip observed toward Barnard 1b since it is a molecule prone to freeze out. The resulting abundances are summarized in Tables \ref{tab:b1bAbundances} and \ref{tab:IC348Abundances} and shown in \ref{fig:allAbundancesB1b} and \ref{fig:allAbundancesIC348}.
				
					\begin{table*}
						\resizebox{\textwidth}{!}{
						\begin{threeparttable}
   							\caption{Molecular abundances observed toward Barnard 1b.}
   							\label{tab:b1bAbundances}
   							\centering
   							\begin{tabular}{r|cccccccc}
     							\toprule
		        				Offset & N(o-H$_{2}$S) & \multirow{2}{*}{N(H$_{2}$S)/N(H)} & N(N$_{2}$H$^{+}$) & \multirow{2}{*}{N(N$_{2}$H$^{+}$)/N(H)} & N(C$^{18}$O) & \multirow{2}{*}{N(C$^{18}$O)/N(H)} & N(OCS) & \multirow{2}{*}{N(OCS)/N(H)} \\ 
		        				$('')$ & (cm$^{-2}$) & & (cm$^{-2}$) &  & (cm$^{-2}$) & & (cm$^{-2}$) & \\ \midrule
		        				0 & $(4.1\pm 0.8)\times 10^{13}$ & $(4.0\pm 0.8)\times 10^{-10}$ & $-$ & $-$ & $-$ & $-$ & $-$ & $-$ \\
		        				15 & $(3.8\pm 0.8)\times 10^{13}$ & $(5\pm 1)\times 10^{-10}$ & $-$ & $-$ & $-$ & $-$ & $-$ & $-$ \\
		        				30 & $(2.4\pm 0.5)\times 10^{13}$ & $(4.4\pm 0.9)\times 10^{-10}$ & $(1.3\pm 0.3)\times 10^{13}$ & $(1.8\pm 0.4)\times 10^{-10}$ & $(8\pm 2)\times 10^{15}$ & $(1.2\pm 0.2)\times 10^{-7}$ & $(9\pm 2)\times 10^{12}$ & $(1.2\pm 0.3)\times 10^{-10}$\\
		        				45 & $(2.0\pm 0.4)\times 10^{13}$ & $(6\pm 1)\times 10^{-10}$ & $(6\pm 1)\times 10^{12}$ & $(1.3\pm 0.3)\times 10^{-10}$ & $(7\pm 1)\times 10^{15}$ & $(1.6\pm 0.3)\times 10^{-7}$ & $(6\pm 2)\times 10^{12}$ & $(1.4\pm 0.4)\times 10^{-10}$ \\
		        				60 & $(1.9\pm 0.4)\times 10^{13}$ & $(7\pm 1)\times 10^{-10}$ & $(4\pm 1)\times 10^{12}$ & $(1.2\pm 0.3)\times 10^{-10}$ & $(6\pm 1)\times 10^{15}$ & $(1.6\pm 0.3)\times 10^{-7}$ & $(4\pm 1)\times 10^{12}$ & $(9\pm 3)\times 10^{-11}$ \\
		        				75 & $(1.8\pm 0.4)\times 10^{13}$ & $(7\pm 1)\times 10^{-10}$ & $(2.1\pm 0.4)\times 10^{12}$ & $(6\pm 1)\times 10^{-11}$ & $(5\pm 1)\times 10^{15}$ & $(1.6\pm 0.3)\times 10^{-7}$ & $(3.2\pm 0.8)\times 10^{12}$ & $(1.0\pm 0.2)\times 10^{-10}$ \\
		        				90 & $(1.7\pm 0.3)\times 10^{13}$ & $(7\pm 1)\times 10^{-10}$ & $(2.0\pm 0.4)\times 10^{12}$ & $(6\pm 1)\times 10^{-11}$ & $(6\pm 1)\times 10^{15}$ & $(1.8\pm 0.3)\times 10^{-7}$ & $(2.5\pm 0.7)\times 10^{12}$ & $(8\pm 2)\times 10^{-11}$ \\
		        				105 & $(1.6\pm 0.3)\times 10^{13}$ & $(8\pm 2)\times 10^{-10}$ & $(1.6\pm 0.3)\times 10^{12}$ & $(6\pm 1)\times 10^{-11}$ & $(6\pm 1)\times 10^{15}$ & $(2.1\pm 0.4)\times 10^{-7}$ & $(2.2\pm 0.7)\times 10^{12}$ & $(8\pm 2)\times 10^{-11}$ \\
		        				120 & $(2.3\pm 0.5)\times 10^{13}$ & $(1.2\pm 0.2)\times 10^{-9}$ & $(1.2\pm 0.4)\times 10^{12}$ & $(5\pm 1)\times 10^{-11}$ & $(6\pm 1)\times 10^{15}$ & $(2.4\pm 0.5)\times 10^{-7}$ & $ < 2\times 10^{12}$ & $<7\times 10^{-11}$ \\
		        				135 & $(2.5\pm 0.5)\times 10^{13}$ & $(1.4\pm 0.3)\times 10^{-9}$ & $(5\pm 2)\times 10^{11}$ & $(2.0\pm 0.7)\times 10^{-11}$ & $(6\pm 1)\times 10^{15}$ & $(2.5\pm 0.5)\times 10^{-7}$ & $<2\times 10^{12}$ & $<8\times 10^{-11}$ \\
		        				150 & $(1.2\pm 0.2)\times 10^{14}$ & $(7\pm 1)\times 10^{-9}$ & $(8\pm 4)\times 10^{11}$ & $(4\pm 2)\times 10^{-11}$ & $(3.8\pm 0.8)\times 10^{15}$ & $(1.8\pm 0.4)\times 10^{-7}$ & $<2\times 10^{12}$ & $<1\times 10^{-10}$ \\
		        				165 & $(1.1\pm 0.2)\times 10^{14}$ & $(6\pm 2)\times 10^{-9}$ & $(4\pm 1)\times 10^{11}$ & $(2.3\pm 0.7)\times 10^{-11}$ & $(2.3\pm 0.5)\times 10^{15}$ & $(1.3\pm 0.3)\times 10^{-7}$ & $<2\times 10^{12}$ & $<1\times 10^{-10}$ \\
		        				180 & $(1.5\pm 0.3)\times 10^{14}$ & $(1.2\pm 0.3)\times 10^{-8}$ & $<4\times 10^{11}$ & $<3\times 10^{-11}$ & $(1.5\pm 0.3)\times 10^{15}$ & $(9\pm 2)\times 10^{-8}$ & $<3\times 10^{12}$ & $<2\times 10^{-10}$ \\
		        				195 & $(1.1\pm 0.2)\times 10^{14}$ & $(1.0\pm 0.2)\times 10^{-8}$ & $<4\times 10^{11}$ & $<3\times 10^{-11}$ & $(1\pm 0.2)\times 10^{15}$ & $(6\pm 1)\times 10^{-8}$ & $<3\times 10^{12}$ & $<2\times 10^{-10}$ \\
		        				210 & $(1.0\pm 0.2)\times 10^{14}$ & $(1.0\pm 0.2)\times 10^{-8}$ & $<5\times 10^{11}$ & $<3\times 10^{-11}$ & $(8\pm 2)\times 10^{14}$ & $(6\pm 1)\times 10^{-8}$ & $<3\times 10^{12}$ & $<2\times 10^{-10}$ \\
		        				225 & $(7\pm 2)\times 10^{13}$ & $(8\pm 2)\times 10^{-9}$ & $<5\times 10^{11}$ & $<4\times 10^{-11}$ & $(6\pm 1)\times 10^{14}$ & $(5\pm 1)\times 10^{-8}$ & $<6\times 10^{12}$ & $<5\times 10^{-10}$ \\
		        				240 & $(1.4\pm 0.4)\times 10^{14}$ & $(1.6\pm 0.4)\times 10^{-8}$ & $<5\times 10^{11}$ & $<4\times 10^{-11}$ & $(6\pm 1)\times 10^{14}$ & $(5\pm 1)\times 10^{-8}$ & $<7\times 10^{12}$ & $<6\times 10^{-10}$\\
    		    				\bottomrule 
     						\end{tabular}
%     						\begin{tablenotes}
%     						\end{tablenotes}
  						\end{threeparttable}
  						}
    				\end{table*}
    				
    				\begin{table*}
					\resizebox{\textwidth}{!}{
					\begin{threeparttable}
   						\caption{Molecular abundances observed toward IC348.}
   						\label{tab:IC348Abundances}
   						\centering
   						\begin{tabular}{r|cccccccc}
     						\toprule
		        			Offset & N(o-H$_{2}$S) & \multirow{2}{*}{N(H$_{2}$S)/N(H)} & N(N$_{2}$H$^{+}$) & \multirow{2}{*}{N(N$_{2}$H$^{+}$)/N(H)} & N(C$^{18}$O) & \multirow{2}{*}{N(C$^{18}$O)/N(H)} & N(OCS) & \multirow{2}{*}{N(OCS)/N(H)} \\ 
		        				$('')$ & (cm$^{-2}$) & & (cm$^{-2}$) &  & (cm$^{-2}$) & & (cm$^{-2}$) & \\ \midrule
		        			30 & $(6\pm 1)\times 10^{12}$ & $(1.9\pm 0.4)\times 10^{-10}$ & $(5\pm 1)\times 10^{12}$ & $(1.1\pm 0.3)\times 10^{-10}$ & $(3.7\pm 0.7)\times 10^{15}$ & $(8.1\pm 2)\times 10^{-8}$ & $<4\times 10^{12}$ & $<8\times 10^{-11}$ \\
		        			45 & $(8\pm 2)\times 10^{12}$ & $(2.4\pm 0.5)\times 10^{-10}$ & $(4\pm 2)\times 10^{12}$ & $(1.0\pm 0.3)\times 10^{-10}$ & $(3.9\pm 0.8)\times 10^{15}$ & $(9\pm 2)\times 10^{-8}$ & $<4\times 10^{12}$ & $<9\times 10^{-11}$ \\
		        			60 & $(8\pm 2)\times 10^{12}$ & $(2.7\pm 0.6)\times 10^{-10}$ & $(4\pm 1)\times 10^{12}$ & $(1.1\pm 0.4)\times 10^{-10}$ & $(4.3\pm 0.9)\times 10^{15}$ & $(1.2\pm 0.2)\times 10^{-7}$ & $<4\times 10^{12}$ & $<1\times 10^{-10}$ \\
		        			75 & $(9\pm 2)\times 10^{12}$ & $(3.5\pm 0.7)\times 10^{-10}$ & $(3\pm 1)\times 10^{12}$ & $(1.0\pm 0.3)\times 10^{-10}$ & $(4.3\pm 0.9)\times 10^{15}$ & $(1.3\pm 0.3)\times 10^{-7}$ & $<4\times 10^{12}$ & $<1\times 10^{-10}$ \\
		       				90 & $(1.1\pm 0.2)\times 10^{13}$ & $(6\pm 1)\times 10^{-10}$ & $(1.5\pm 0.5)\times 10^{12}$ & $(6\pm 2)\times 10^{-11}$ & $(4.3\pm 0.9)\times 10^{15}$ & $(1.6\pm 0.3)\times 10^{-7}$ & $<4\times 10^{12}$ & $<1\times 10^{-10}$ \\
		       				105 & $(2.2\pm 0.4)\times 10^{13}$ & $(1.4\pm 0.3)\times 10^{-9}$ & $(4\pm 1)\times 10^{11}$ & $(1.9\pm 0.7)\times 10^{-11}$ & $(4.3\pm 0.9)\times 10^{15}$ & $(2.1\pm 0.4)\times 10^{-7}$ & $<3\times 10^{12}$ & $<2\times 10^{-10}$ \\
		       				120 & $(1.5\pm 0.3)\times 10^{13}$ & $(1.0\pm 0.2)\times 10^{-9}$ & $(1.5\pm 0.5)\times 10^{11}$ & $(8\pm 3)\times 10^{-12}$ & $(3.8\pm 0.8)\times 10^{15}$ & $(2.0\pm 0.4)\times 10^{-7}$ & $<3\times 10^{12}$ & $<1.7\times 10^{-10}$ \\
		       				135 & $(1.7\pm 0.4)\times 10^{13}$ & $(1.3\pm 0.2)\times 10^{-9}$ & $(1.0\pm 0.3)\times 10^{11}$ & $(6\pm 2)\times 10^{-12}$ & $(2.7\pm 0.6)\times 10^{15}$ & $(1.5\pm 0.3)\times 10^{-7}$ & $<4\times 10^{12}$ & $<2\times 10^{-10}$ \\
		       				150 & $(1.4\pm 0.3)\times 10^{13}$ & $(1.1\pm 0.3)\times 10^{-9}$ & $(1.1\pm 0.4)\times 10^{11}$ & $(7\pm 3)\times 10^{-12}$ & $(1.7\pm 0.3)\times 10^{15}$ & $(1.1\pm 0.2)\times 10^{-7}$ & $<4\times 10^{12}$ & $<2\times 10^{-10}$ \\
		       				165 & $(1.2\pm 0.3)\times 10^{13}$ & $(1.0\pm 0.2)\times 10^{-9}$ & $(1.0\pm 0.3)\times 10^{11}$ & $(6\pm 2)\times 10^{-12}$ & $(1.3\pm 0.3)\times 10^{15}$ & $(8\pm 2)\times 10^{-7}$ & $<3\times 10^{12}$ & $<2\times 10^{-10}$ \\
		       				180 & $(1.2\pm 0.3)\times 10^{13}$ & $(1.1\pm 0.3)\times 10^{-9}$ & $(6\pm 3)\times 10^{10}$ & $(4\pm 2)\times 10^{-12}$ & $(1.2\pm 0.3)\times 10^{15}$ & $(8\pm 2)\times 10^{-8}$ & $<3\times 10^{12}$ & $<2\times 10^{-10}$ \\
		       				195 & $(1.4\pm 0.4)\times 10^{13}$ & $(1.2\pm 0.3)\times 10^{-9}$ & $<1\times 10^{11}$ & $<7\times 10^{-12}$ & $(1.3\pm 0.2)\times 10^{15}$ & $(8\pm 2)\times 10^{-8}$ & $<2\times 10^{12}$ & $<2\times 10^{-10}$ \\
	        				210 & $(2.2\pm 0.5)\times 10^{13}$ & $(2.1\pm 0.5)\times 10^{-9}$ & $<1\times 10^{11}$ & $<9\times 10^{-12}$ & $(1.4\pm 0.3)\times 10^{15}$ & $(1.0\pm 0.2)\times 10^{-7}$ & $<2\times 10^{12}$ & $<2\times 10^{-10}$ \\
	        				225 & $(2.9\pm 0.7)\times 10^{13}$ & $(2.9\pm 0.7)\times 10^{-9}$ & $<2\times 10^{11}$ & $<1\times 10^{-11}$ & $(1.2\pm 0.2)\times 10^{15}$ & $(9\pm 2)\times 10^{-8}$ & $<3\times 10^{12}$ & $<2\times 10^{-10}$ \\
		        			240 & $(2.7\pm 0.7)\times 10^{13}$ & $(2.9\pm 0.7)\times 10^{-9}$ & $<2\times 10^{11}$ & $<1\times 10^{-11}$ & $(1.2\pm 0.3)\times 10^{15}$ & $(1.0\pm 0.2)\times 10^{-7}$ & $<3\times 10^{12}$ & $<2\times 10^{-10}$ \\
    		    				\bottomrule 
     					\end{tabular}
%     						\begin{tablenotes}
%     						\end{tablenotes}
  					\end{threeparttable}
  					}
    			\end{table*}
    				
    			The N$_{2}$H$^{+}$ $1\rightarrow 0$ line presents hyperfine splitting. The relative intensities of the hyperfine components provided us with a measurement of the opacity of this line. We found moderate opacities $\tau_{\rm main}\leq 1$ across the horizontal strip we observed. To avoid opacity effects, we fitted the integrated intensity of the weakest hyperfine component using the collisional rates without hyperfine splitting reported in \citet{Schoier2005}. Then, the resulting column density is scaled by the spectroscopic ratio between the integrated intensity of the total N$_{2}$H$^{+}$ $1\rightarrow 0$ line and that of the weakest hyperfine component we chose. The results are listed in Tables \ref{tab:b1bAbundances} and \ref{tab:IC348Abundances} and shown in Figs. \ref{fig:allAbundancesB1b} and \ref{fig:allAbundancesIC348}. They exhibit a trend of decreasing abundance at lower extinctions. It is when dust temperatures go beyond $T_{\rm dust}\gtrsim 15-18$ K and densities go below $n_{\rm H}\sim 10^{4}$ cm$^{-3}$ when N$_{2}$H$^{+}$ is no longer detected \citep{Priestley2023}. This is consistent with the decline in methanol abundance and our previous discussion about the lower availability of CO over grain surfaces at this temperature, since it is well-known that N$_{2}$H$^{+}$ is easily destroyed by CO \citep[see, e.g.,][]{Jorgensen2004, Lee2004}.
											
				The C$^{18}$O column density was estimated by fitting the integrated intensity of the C$^{18}$O $1\rightarrow 0$ line with \texttt{RADEX}. Its abundance spans one order of magnitude in Barnard 1b, with its peak at offset $+135''$, where dust temperature reaches $T_{\rm dust}\sim 15$ K (see Fig. \ref{fig:allAbundancesB1b}). This is again in agreement with our findings about CH$_{3}$OH and N$_{2}$H$^{+}$. In IC 348 we found a shallower abundance profile compared to Barnard 1b, possibly due to dust temperatures being consistently above $\sim 15$ K (see Fig. \ref{fig:allAbundancesIC348}).
					
				Finally, we obtained the OCS column density and abundance by fitting the OCS $9\rightarrow 8$ integrated intensities detected toward Barnard 1b using the collisional rates presented in \citet{Schoier2005}. These were detected only in a few positions, from offsets $+30''$ to $+105''$. We compared our column densities with those calculated in \citet{RodriguezBaras2021}, and we found a good agreement within a factor of two. OCS abundance spans a factor of two across the points where its abundance was estimated (see Fig. \ref{fig:allAbundancesB1b}). It shows a slightly increasing trend at higher extinctions, mimicking CH$_{3}$OH, right when C$^{18}$O shows signs of depletion and N$_{2}$H$^{+}$ abundance is enhanced. This is compatible with the production of OCS by grain surface reactions \citep{Garrod2007, elAkel2022} in molecular clouds. Unlike Barnard 1b, OCS was not detected toward IC 348 and only upper limits could be obtained (Table \ref{tab:IC348Abundances}).			
			
	\section{Chemical modeling}
		
		In this section we carried out the astrochemical modeling of the two regions, Barnard 1b and IC348, in order to interpret the observations, understand the variations of molecular abundances in different environments, and investigate the possible interplay between them. 
		
		\subsection{Fiducial chemical model of Barnard 1b}\label{sec:fiducialB1b}
		
			We used the gas-grain chemical code \texttt{Nautilus} \citep{Ruaud2016}, a chemical model that takes into account the chemical processes and interactions in the gas-phase, the surface of icy grains, and ice mantles. \texttt{Nautilus} uses the rate equation approach to compute the chemical composition through time given an initial set of physical and chemical conditions. We used the most up-to-date version \texttt{Nautilus} 2.0.0 with its corresponding updated chemical network, described in \citet{Wakelam2024}.
			
\begin{figure*}
				\centering
				\includegraphics[width=\textwidth]{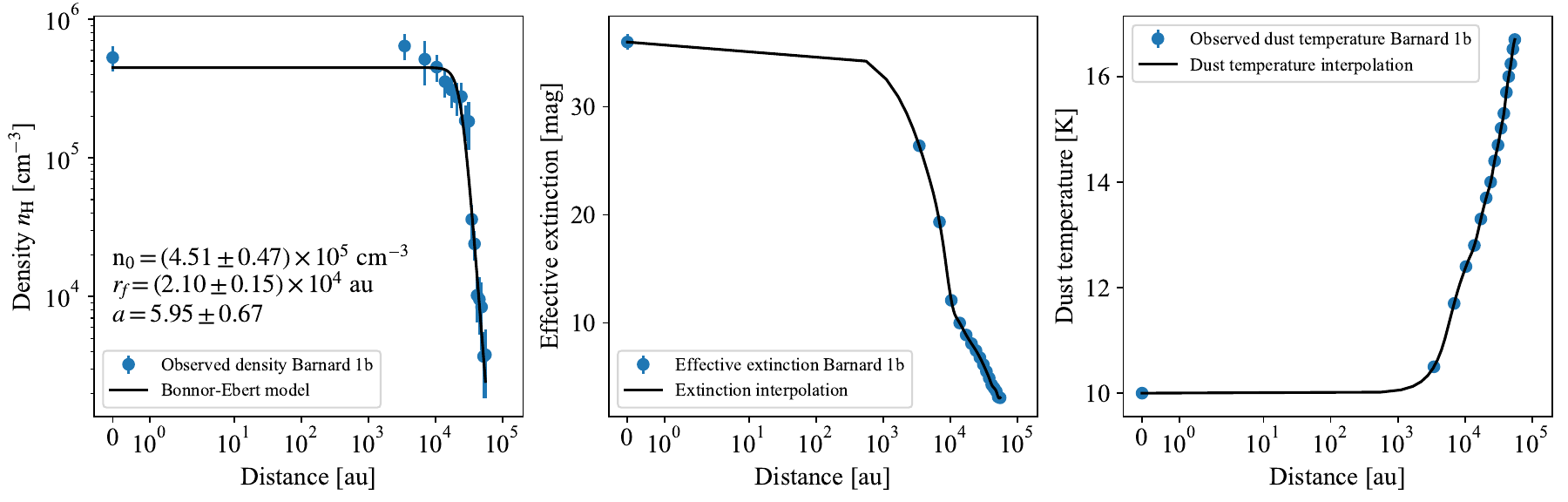}
				\caption{Physical model of Barnard 1b. \emph{Left:} density structure fitted to the Plummer-like profile in Eq. \ref{eq:plummer} with the parameters that best fit the observations. \emph{Middle:} interpolation of the effective extinction toward Barnard 1b. \emph{Right:} interpolation of the dust temperature in Barnard 1b.}
					\label{fig:modelDensB1b}
			\end{figure*}

			We modeled the density structure of Barnard 1b as a static Bonnor-Ebert sphere with the Plummer-like profile (Eq. \ref{eq:plummer}) that best fits the densities derived from methanol. The best fit to the observed densities is shown in Fig. \ref{fig:modelDensB1b}. We assumed a spherical model with isotropic illumination from the interstellar FUV field. Therefore, the extinction at each point in the model is half of the total extinction (we call it effective extinction from now on) shown in Fig. \ref{fig:maps}, since the extinction in these maps is measured as a projection along the total line of sight. The effective extinction is interpolated to cover all the points in the model (Fig. \ref{fig:modelDensB1b}). Similarly, we interpolated dust temperature to all the points in the model (Fig. \ref{fig:modelDensB1b}) and assumed it is equal to the gas temperature. The strength of the ambient FUV field is taken from \citet{NavarroAlmaida2020} as $\chi\sim$ 24 in Draine field units \citep{Draine1978}. Finally, the chosen cosmic-ray ionization rate $\zeta_{\rm H_{2}}$ is $\zeta_{\rm H_{2}} = 6.5\times 10^{-17}$ s$^{-1}$, as in \citet{NavarroAlmaida2020}.
			
			\begin{table}
				\centering
				\begin{threeparttable}
   					\caption{Initial chemical abundances for the fiducial model of Barnard 1b.}\label{tab:initChemAbModelB1b}
   					\begin{tabular}{ll}
     				\toprule
		        	Species & Initial abundance (/H) \\ \midrule 
		        	He & $9.0\times 10^{-2}$ \\
		        	N & $6.2\times 10^{-5}$	 \\
		        	O & $2.4\times 10^{-4}$	 \\
		        	H$_{2}$ & $5.0\times 10^{-1}$ \\
		        	C$^{+}$ & $1.7\times 10^{-4}$	 \\
		        	S$^{+}$ & $7.0\times 10^{-6}$ \tnote{(1)}	 \\
		        	Si$^{+}$ & $8.0\times 10^{-9}$	 \\
		        	Fe$^{+}$ & $3.0\times 10^{-9}$	 \\
		        	Na$^{+}$ & $2.0\times 10^{-9}$	 \\
		        	Mg$^{+}$ & $7.0\times 10^{-9}$	 \\
		        	P$^{+}$ & $2.0\times 10^{-10}$	 \\
		        	Cl$^{+}$ & $1.0\times 10^{-9}$	 \\
		        	F & $6.7\times 10^{-9}$	 \\
    		    	\bottomrule 
     				\end{tabular}
     				\begin{tablenotes}
       					\item [(1)] \citet{Fuente2023}.
     				\end{tablenotes}
  				\end{threeparttable}
    		\end{table}
    	
			The fiducial chemical model of Barnard 1b consists of the physical properties described before and a set of initial chemical abundances. Among the initial chemical abundances we may set, the initial sulfur abundance is one of the most uncertain. \citet{Fuente2023} estimated the average sulfur depletion that better accounts for the chemical abundances of several molecules observed toward this source. Following their results, we set the initial elemental sulfur abundance, that is, the total sulfur budget with respect to H, as ${\rm S/H}\sim 7\times 10^{-6}$. The remaining initial chemical abundances for this model are listed in Table \ref{tab:initChemAbModelB1b}. We then ran \texttt{Nautilus} to extract the chemical abundances of the species in Table \ref{tab:lines} assuming a chemical age of $10^{6}$ years. Finally, for a more accurate comparison with observations, we followed the same methodology as in \citet{NavarroAlmaida2020}, projecting the predicted chemical abundances along the line of sight using the physical model of Barnard 1b. To do so, we computed the average abundance $\overline{X}$ along each offset $r$ of the spherical model, weighted by the density $n_{\rm H}$:

			\begin{equation*}
				\overline{X}(r) = \frac{\sum_{i}\left(l_{i+1}-l_{i}\right)\left(n_{\rm H}(s_{i}){X}(s_{i})+n_{\rm H}(s_{i+1}){X}(s_{i+1})\right)}{\sum_{j}\left(l_{j+1}-l_{j}\right)\left(n_{\rm H}(s_{j})+n_{\rm H}(s_{j+1})\right)},
			\end{equation*}
			where $n_{\rm H}(s_{i})$ and ${X}(s_{i})$ are, respectively, the number density and abundance at a distance $s_{i}=\sqrt{r^2+l_{i}^{2}}$. Here $l_{i}$ is a discretization of the segment along the line of sight at a given offset $r$ such that $l_{\rm max} > \dots > l_{i+1} > l_{i} > \dots > 0$, with $l_{\rm max} = \sqrt{r^2+r_{\rm max}^{2}}$ and $r_{\rm max}$ being the radius of the density profile.
			
			\begin{figure*}
				\centering
				\includegraphics[width=\textwidth]{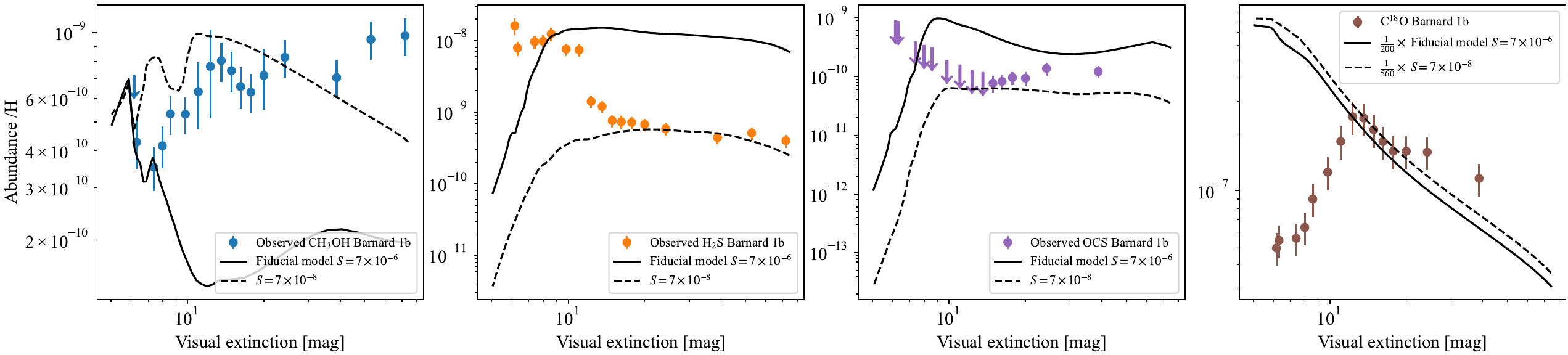}
				\caption{Observed gas-phase abundance toward Barnard 1b of CH$_{3}$OH (blue dots in the first panel), H$_{2}$S (orange dots in the second panel), OCS (purple arrows in the third panel), and C$^{18}$O (brown dots in the fourth panel) compared to the respective abundances predicted by the fiducial model (solid black line) and those of an alternative model with higher sulfur depletion (dashed black line). Model predictions of CO were scaled by a factor of $\frac{1}{200}$ to account for the observed C$^{18}$O abundances.}
					\label{fig:fiducialB1bComp}
			\end{figure*}
			
			The performance of the fiducial model fitting the CH$_{3}$OH, H$_{2}$S, and OCS abundances in Barnard 1b is shown in Fig. \ref{fig:fiducialB1bComp}. Fiducial model (${\rm S/H}=7\times 10^{-6}$) predictions of CH$_{3}$OH and H$_{2}$S abundances are in agreement only in the external part of the model ($A_{v}\leq 10$ mag). At higher extinctions, the predicted CH$_{3}$OH abundance is lower than our observations, while H$_{2}$S is highly overestimated. To improve the performance of the model in the case of H$_{2}$S, we modified the fiducial model with the introduction of an alternative high sulfur depletion model, meaning that it has a lower initial elemental sulfur budget by two orders of magnitude: ${\rm S/H}=7\times 10^{-8}$. In this case, the H$_{2}$S abundance in the innermost areas is well reproduced by the model. Interestingly, in this depleted model the predicted CH$_{3}$OH abundance changes significantly, offering a better fitting at intermediate extinctions ($10\leq A_{v}\leq 40$ mag). However, it still underestimates CH$_{3}$OH at the highest extinctions ($A_{v}\geq 40$ mag). The model with low sulfur budget offers a reasonable agreement with the observed OCS abundances (Fig. \ref{fig:fiducialB1bComp}). The initial sulfur abundance required to reproduce OCS is therefore compatible with that required by H$_{2}$S.

			Our chemical network does not currently include isotopologues for CO, and so the comparison between the model and observations is uncertain. We found the best agreement between the model and the observations if we assume the isotopic ratio $^{16}{\rm O}/^{18}{\rm O} \sim 200$ (Fig. \ref{fig:fiducialB1bComp}), lower than the local ISM \citep[$557\pm 30$,][]{Wilson1999} or the solar system \citep[500,][]{Lodders2003} estimates, but in good agreement with the isotopic ratio observed in dense molecular clouds by \cite{Loison2019} using ${\rm S}^{16}{\rm O}$ and ${\rm S}^{18}{\rm O}$. As seen in Fig. \ref{fig:fiducialB1bComp}, the initial sulfur budget does not have a great impact in the gas-phase C$^{18}$O abundance. Both models provide a good agreement at moderate extinctions ($A_{\rm v}\geq 12$ mag) while it overestimates the abundance of C$^{18}$O by a factor of $\sim 10$ at lower extinctions. The fixed isotopic ratio we assumed and other physical processes occurring at the edge of the cloud like selective photodissociation are the main potential sources of uncertainty in this comparison that could explain the discrepancies between models and observations.
			
		\subsection{Fiducial chemical model of IC 348}\label{sec:fiducialIC348}
			
			\begin{figure*}
				\centering
				\includegraphics[width=\textwidth]{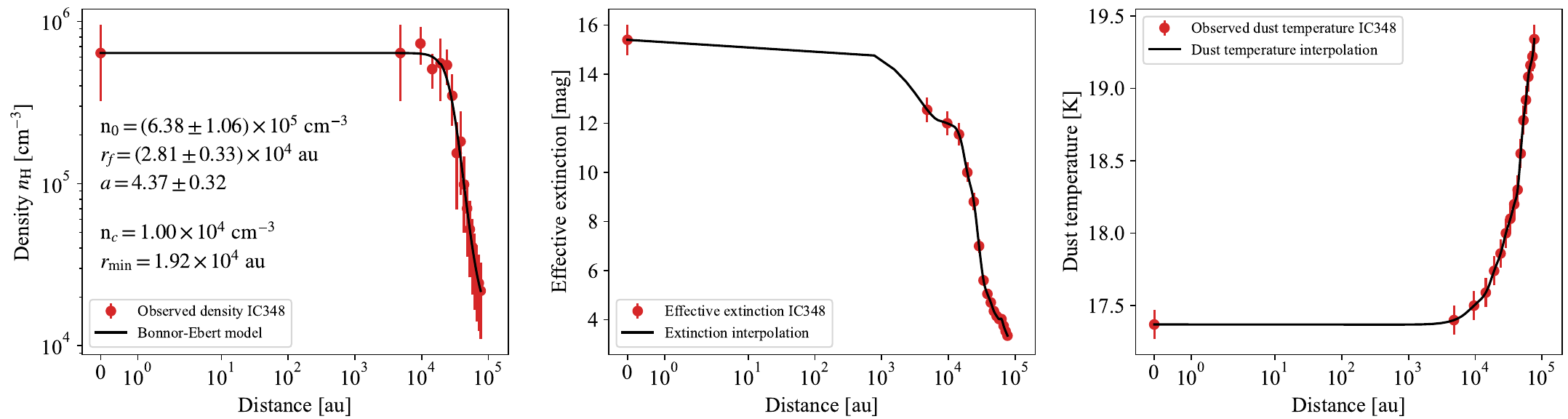}
				\caption{Physical model of IC 348. \emph{Left panel:} density structure fitted to the Plummer-like and envelope profiles in Eqs. \ref{eq:plummer} and \ref{eq:envelope}, respectively, with the parameters that best fit the observations. \emph{Middle panel:} interpolation of the effective extinction toward IC 348. \emph{Right panel:} interpolation of dust temperature in IC 348.}
					\label{fig:modelDensIC348}
			\end{figure*}
			
\begin{figure*}
				\centering
				\includegraphics[width=\textwidth]{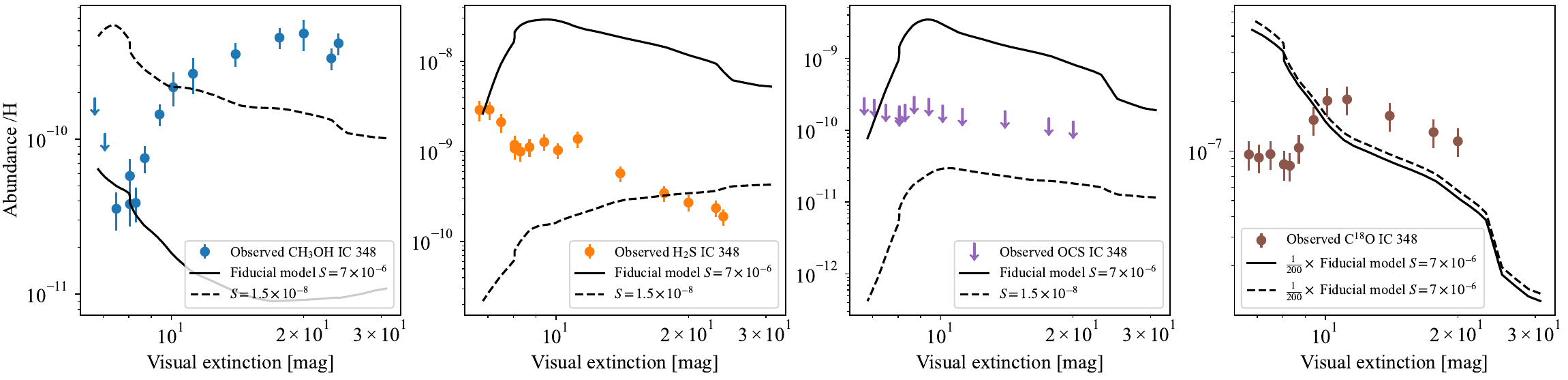}
				\caption{Observed gas-phase abundance toward IC 348 of CH$_{3}$OH (blue dots in the first panel), H$_{2}$S (orange dots in the second panel), OCS (purple arrows in the third panel), and C$^{18}$O (brown dots in the fourth panel) compared to the respective abundances predicted by the fiducial model (solid black line) and those of an alternative model with higher sulfur depletion (dashed black line). Model predictions of CO were scaled by a factor of $\frac{1}{200}$ to account for the observed C$^{18}$O abundances.}
					\label{fig:fiducialIC348Comp}
			\end{figure*}

			The density structure of IC 348 was previously modeled in Sect. \ref{sec:methanolDens} (see Fig. \ref{fig:densityProfileIC348}). Similar to the Barnard 1b model, the effective visual extinction and dust temperature are interpolated to all the points of the model (Fig. \ref{fig:modelDensIC348}). Gas temperature was assumed to be equal to dust temperature. The cosmic-ray ionization rate was chosen to be $\zeta_{\rm H_{2}} = 5\times 10^{-16}$ s$^{-1}$ \citep[see, e.g.,][]{Luo2023}. We computed the FUV field strength using the observed dust temperature and extinction in the parameterization derived by \citet{Hocuk2017}. The Draine field strength $\chi_{\rm FUV}$ that best fits IC 348 was found to be $\chi_{\rm FUV}=88\pm13$. The initial chemical abundances were set as in Table \ref{tab:initChemAbModelB1b} and, like Barnard 1b, we considered an alternative model with a lower sulfur budget. The results of these models are shown in Fig. \ref{fig:fiducialIC348Comp}.
			
			Similar to how the fiducial model described the observations toward Barnard 1b, here we noted that CH$_{3}$OH toward IC 348 is also well predicted by this model only at the outermost points (Fig. \ref{fig:fiducialIC348Comp}). The total sulfur budget impacts the predicted abundances of methanol significantly, enhancing them as the total sulfur budget decreases. Despite this enhancement, the low sulfur budget model is unable to reach the highest abundances of methanol measured toward the extinction peak (Fig. \ref{fig:fiducialIC348Comp}). The sulfur depletion needed for this model is even higher than that in Barnard 1b in order to fit the gas-phase H$_{2}$S abundances observed toward the extinction peak of IC 348. The fiducial model is only able to fit the external gas-phase abundances of H$_{2}$S we observed. The upper limits to the gas-phase abundance of OCS indicate that, in agreement with CH$_{3}$OH and H$_{2}$S, the fiducial model only provides reasonable predictions at the edge of the region we observed, while a higher sulfur depletion is needed in order to be compatible with the upper limits of OCS derived at the highest extinctions. Finally, both the fiducial and the sulfur depleted model provide similar results regarding the CO abundance. Applying the same isotopic ratio as in Barnard 1b, we concluded that both models produce acceptable results beyond the edge of IC 348 ($A_{\rm v}\geq 10$ mag), while they overestimate C$^{18}$O by a factor of $\sim 5$ at lower extinctions. The fixed isotopic ratio we chose and physical processes not taken into account by the model such as selective photodissociation are again a great source of uncertainty in this comparison.
		
		\subsection{Enhancing the formation of COMs: sulfur abundance, CO freeze out, and binding energies.}
		
			The scenarios of formation of methanol and other saturated COMs in cold cores involve diffusion, reaction on grain surfaces, and non-thermal desorption processes \citep[see, e.g.,][]{Watanabe2002, Minissale2016, Taillard2023}. Since large and heavy molecules do not diffuse on grain surfaces easily, one possible explanation to the observed high abundance of COMs, suggested by, for instance, \citet{Vasyunin2017} and \citet{Molpeceres2024}, is the lower binding energy of adsorbates when the ice contains a significant portion of CO. Due to the low value of the dipole moment of the CO molecule, compared to that of water, the intermolecular forces that trap adsorbed molecules onto the surface are much lower. The effect on COMs and other molecules is then two-fold: CO-rich ices lead to weaker bonds between large molecules and grain surfaces, allowing them to move and react more easily, and raises non-thermal desorption efficiencies. The combination of these two effects leads to higher COM abundances in the gas-phase.
			
			\begin{table*}
				\centering
				\begin{threeparttable}
					\centering
   					\caption{List of chemical models with decreasing binding energies.}
   					\label{tab:molpeceresModels}
   					\begin{tabular}{lccccccc}
     					\toprule
     					\multirow{2}[2]{*}{Model} & \multicolumn{7}{c}{Binding Energy (K)} \\ \cmidrule{2-8}
		        		& CH$_{3}$OH & H$_{2}$CO & OCS & O & CO & HCO & H$_{2}$S \\ \midrule
		        		1\,\tnote{(a)} & 5000.00 & 4500.00 & 2400.00 & 1660.00 & 1300.00 & 2400.00 & 2700.00 \\
		        		2 & 4910.53 & 4419.47 & 2357.05 & 1605.83 & 1276.74 & 2322.95 & 2651.68 \\
		        		3 & 4821.05 & 4338.95 & 2314.11 & 1551.66 & 1253.47 & 2245.89 & 2603.37 \\
		        		4 & 4731.58 & 4258.42 & 2271.16 & 1497.49 & 1230.21 & 2168.84 & 2555.05 \\ \midrule
		        		5 & 4642.11 & 4177.90 & 2228.21 & 1443.33 & 1206.95 & 2091.79 & 2506.74 \\
		        		6 & 4552.63 & 4097.37 & 2185.26 & 1389.16 & 1183.68 & 2014.74 & 2458.42 \\
		        		7 & 4463.16 & 4016.84 & 2142.32 & 1334.99 & 1160.42 & 1937.68 & 2410.11 \\
		        		8 & 4373.68 & 3936.32 & 2099.37 & 1280.82 & 1137.16 & 1860.63 & 2361.79 \\
		        		\midrule
		        		9 & 4284.21 & 3855.79 & 2056.42 & 1226.65 & 1113.89 & 1783.58 & 2313.47 \\
		        		10 & 4194.74 & 3775.27 & 2013.47 & 1172.48 & 1090.63 & 1706.53 & 2265.16 \\
		        		11 & 4105.26 & 3694.74 & 1970.53 & 1118.32 & 1067.37 & 1629.47 & 2216.84 \\
		        		12 & 4015.79 & 3614.21 & 1927.58 & 1064.15 & 1044.11 & 1552.42 & 2168.53 \\
		        		\midrule
						13 & 3926.32 & 3533.69 & 1884.63 & 1009.98 & 1020.84 & 1475.37 & 2120.21 \\
						14 & 3836.84 & 3453.16 & 1841.68 & 955.81 & 997.58 & 1398.32 & 2071.89 \\
						15 & 3747.37 & 3372.64 & 1798.74 & 901.64 & 974.32 & 1321.26 & 2023.58 \\
						16\,\tnote{(b)} & 3300.00 & 2970.00 & 1584.00 & 631.00 & 858.00 & 936.00 & 1782.00  \\
    		    		\bottomrule 
     				\end{tabular}
					\begin{tablenotes}
       					\item [(a)] The binding energies in Model 1 are taken from the chemical network presented in \citet{Wakelam2024}, with no scaling applied.
       					\item [(b)] Model 16 uses the scaling of binding energies proposed in \citet{Molpeceres2024}.
     				\end{tablenotes}
  				\end{threeparttable}
    		\end{table*}
    		
    		As shown in Sects. \ref{sec:fiducialB1b} and \ref{sec:fiducialIC348}, the fiducial chemical models of Barnard 1b and IC 348 were unable to reproduce the high abundances of methanol observed toward the extinction peak of each dense core. While a high sulfur depletion seems to enhance them, there is a fraction of this molecule that is still missing. In this section, we investigate if there is a coherent picture that consistently describes the observed CH$_{3}$OH, H$_{2}$S, OCS, and CO abundances, while also accounts for sulfur depletion, the progressive build-up of CO in ices, and the consequent decline in binding energies. To do so, we followed the definition of binding energies on CO presented in \citet{Molpeceres2024}. In Table \ref{tab:molpeceresModels}, we list 16 models with decreasing binding energies for CH$_{3}$OH, H$_{2}$CO, OCS, O, CO, HCO, and H$_{2}$S, varying from the reference values of the chemical network (Model 1) to those after the scaling described in \citet{Molpeceres2024} is applied (Model 16). This scaling is $\varepsilon_{\rm CO}$, following the notation introduced in \citet{Molpeceres2024}. The binding energies of the remaining molecules present in the chemical network were not modified. 
    		
    		\begin{table}
				\centering
				\begin{threeparttable}
					\centering
   					\caption{List of initial sulfur abundances.}
   					\label{tab:initialSulfurAb}
   					\begin{tabular}{lc}
     					\toprule
     					Model & Initial sulfur abundance \\ \midrule
		        		a\,\tnote{(1)} & $7.0\times 10^{-6}$ \\
		        		b & $5.0\times 10^{-6}$ \\
		        		c & $3.0\times 10^{-6}$ \\
		        		d & $1.5\times 10^{-6}$ \\ \midrule
		        		e & $7.0\times 10^{-7}$ \\
		        		f & $5.0\times 10^{-7}$ \\
		        		g & $3.0\times 10^{-7}$ \\
		        		h & $1.5\times 10^{-7}$ \\ \midrule
		        		i & $7.0\times 10^{-8}$ \\
		        		j & $5.0\times 10^{-8}$ \\
		        		k & $3.0\times 10^{-8}$ \\
		        		l & $1.5\times 10^{-8}$ \\ \bottomrule 
     				\end{tabular}
					\begin{tablenotes}
       					\item [(1)] The highest initial sulfur abundance was chosen to be compatible with the average sulfur depletion estimated toward Barnard 1b and IC 348 in \citet{Fuente2023}.
     				\end{tablenotes}
  				\end{threeparttable}
    		\end{table}
    		
    		\subsubsection{H\textsubscript{2}S and initial sulfur abundance}\label{sec:initialSAb}
    		
\begin{figure*}
					\centering
					\includegraphics[width=\textwidth]{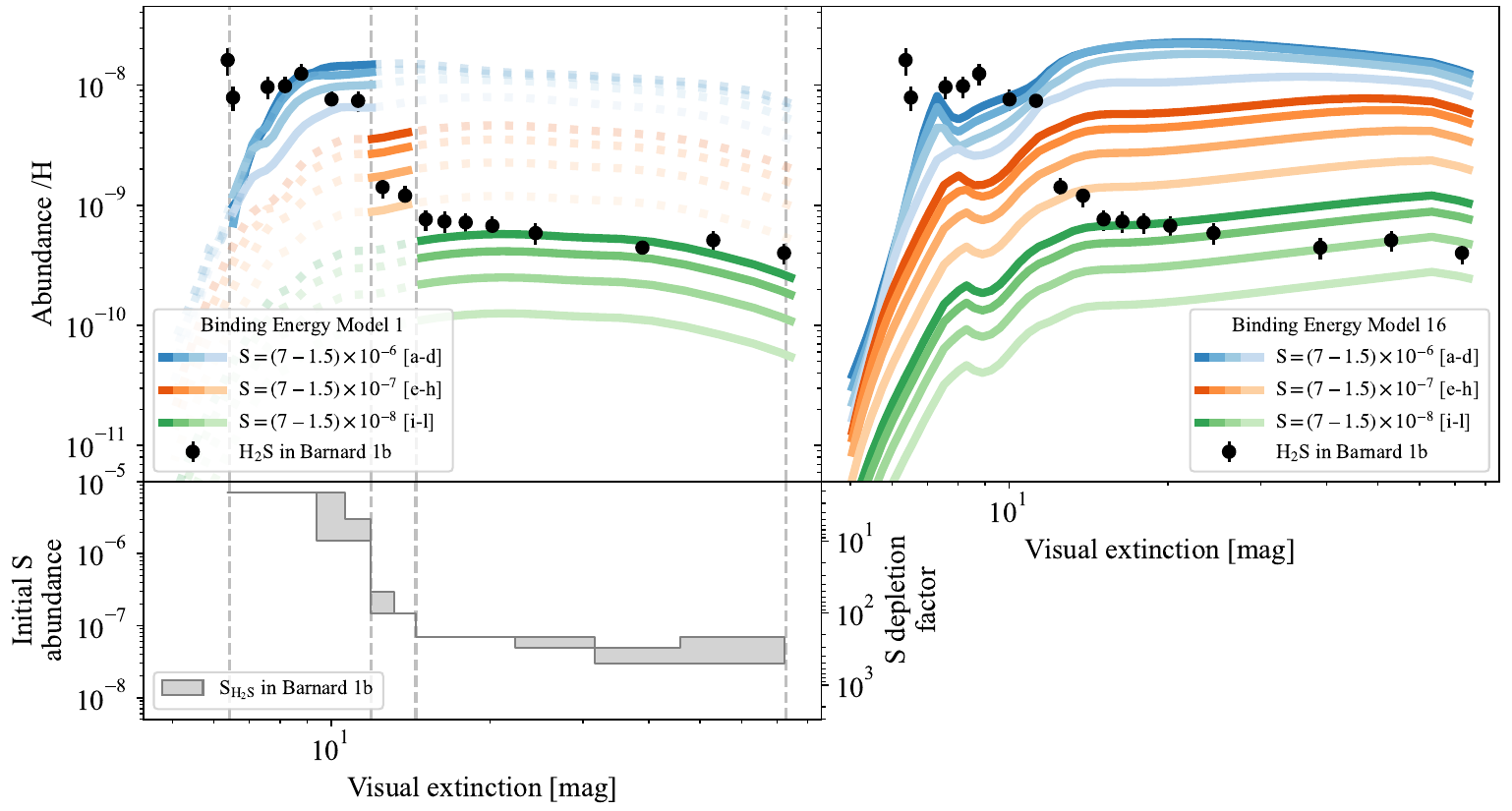}
					\caption{Comparison between the observed abundance of H$_{2}$S in Barnard 1b (black dots) and chemical models. In the left panel of the top row, the observed abundances are compared to models with different initial sulfur abundances and the binding energies of Model 1 from Table \ref{tab:molpeceresModels}. The sets of models that best fit the observations are shown with solid lines. The right panel of the top row shows the same comparison, but with the binding energies of Model 16 from Table \ref{tab:molpeceresModels}, evidencing the low impact of lower binding energies in the predicted gas-phase abundances of H$_{2}$S. The dependence between the sulfur abundance and the extinction is shown in the left panel of the bottom row. The filled area S$_{{\rm H}_{2}{\rm S}}$ represents the appropriate ranges of sulfur abundances that models with other sets of binding energies apart from Model 1 (Table \ref{tab:molpeceresModels}) require.}
					\label{fig:modelsB1bH2S}
				\end{figure*}

\begin{figure*}
					\centering
					\includegraphics[width=\textwidth]{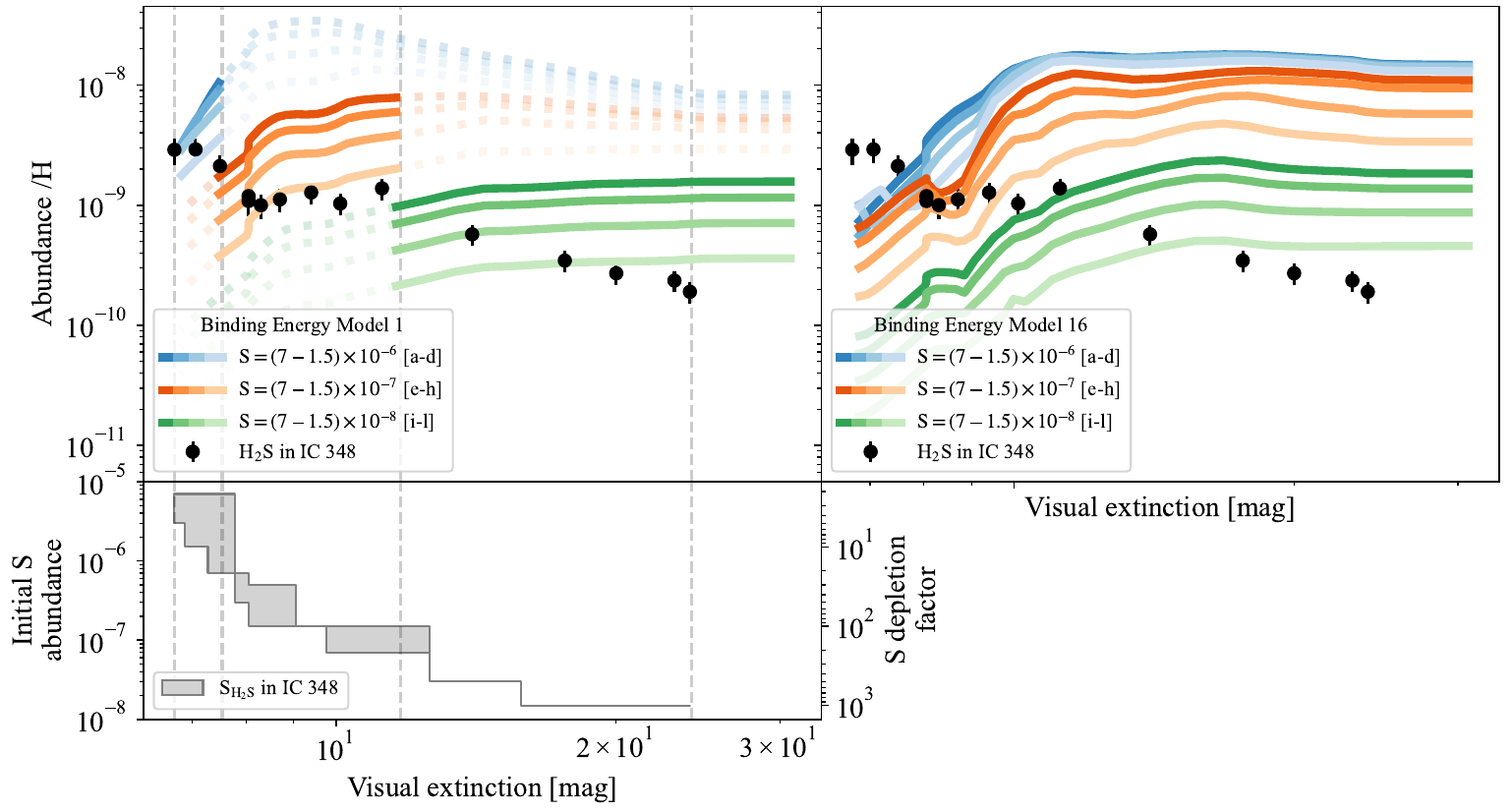}
					\caption{Comparison between the observed abundance of H$_{2}$S in IC 348 (black dots) and chemical models. In the top row, the left panel shows the observed abundances compared to models with different initial sulfur abundances and the binding energies of Model 1 from Table \ref{tab:molpeceresModels}. The sets of models that best fit the observations are shown with solid lines. The right panel of the same row shows a similar comparison, but with the binding energies of Model 16 from Table \ref{tab:molpeceresModels}, evidencing again the low impact of declining binding energies in the predicted gas-phase abundances of H$_{2}$S. The dependence between the sulfur abundance and the extinction is shown in the left panel of the bottom row. The filled area S$_{{\rm H}_{2}{\rm S}}$ represents the appropriate ranges of sulfur abundances that models with other sets of binding energies apart from Model 1 (Table \ref{tab:molpeceresModels}) require.}
					\label{fig:modelsIC348H2S}
				\end{figure*}

    			In Sects. \ref{sec:fiducialB1b} and \ref{sec:fiducialIC348}, it became apparent that a great sulfur depletion is needed in order to reproduce the high contrast between the gas-phase H$_{2}$S abundance at the edge of Barnard 1b and IC 348, and its abundance at the extinction peaks. The high sensitivity of H$_{2}$S to the amount of sulfur depletion is ideal to estimate the sulfur depletion needed at a given visual extinction. To do so, for each binding energy model of Table \ref{tab:molpeceresModels}, we tested the set of initial sulfur abundances listed in Table \ref{tab:initialSulfurAb}. The results for the complete set of all combinations of binding energy models and initial sulfur abundances for Barnard 1b and IC 348 are shown in Figs. \ref{fig:allModelsH2SB1b} and \ref{fig:allModelsH2SIC348}, respectively. While the predicted H$_{2}$S abundance is highly sensitive to the initial sulfur abundance of the model, it is not significantly affected by the changing binding energies. Because of that, this molecule breaks the possible degeneracy that could have arisen in the choice of an initial sulfur abundance or binding energy to fit the observed abundances. In Fig. \ref{fig:modelsB1bH2S} we depict how models with different initial sulfur abundances compare to the observations toward Barnard 1b. From this comparison we extracted the required sulfur abundance S$_{{\rm H}_{2}{\rm S}}$ at any given visual extinction, and it clearly indicates a correlation between sulfur depletion and visual extinction, in agreement with the results of \citet{Fuente2019}, regardless of the binding energy model. A similar correlation is seen in Fig. \ref{fig:modelsIC348H2S}, although with a higher sulfur depletion needed to fit the observations toward the extinction peak of IC 348. This is in agreement with the higher sulfur depletion estimated in IC 348 compared to that of Barnard 1b \citep{Fuente2023}.
    			
    		\subsubsection{CO freeze out, CH\textsubscript{3}OH, and binding energies}\label{sec:methanolModelsMolpeceres}
    		
\begin{figure*}
					\centering
					\includegraphics[width=0.495\textwidth]{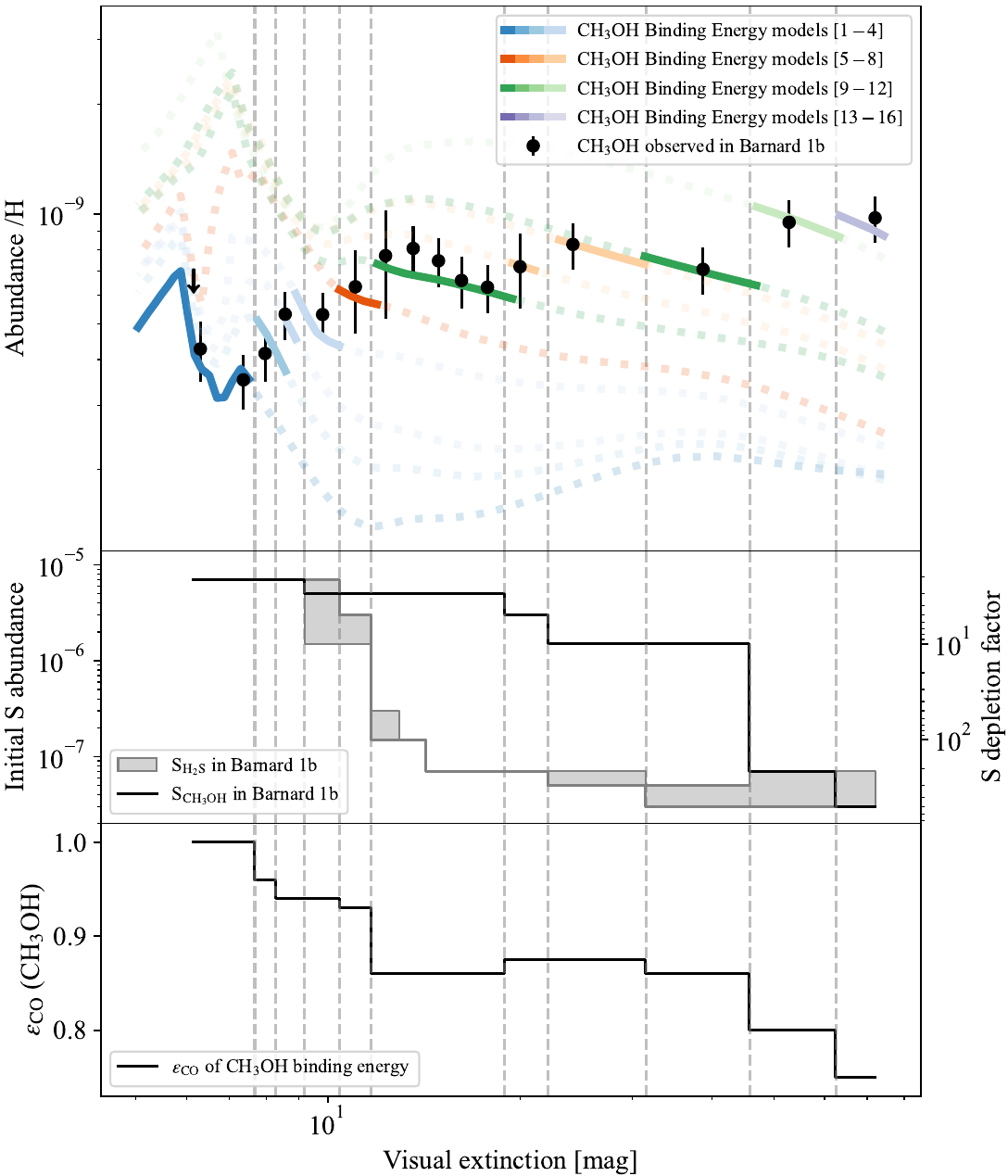}
					\includegraphics[width=0.495\textwidth]{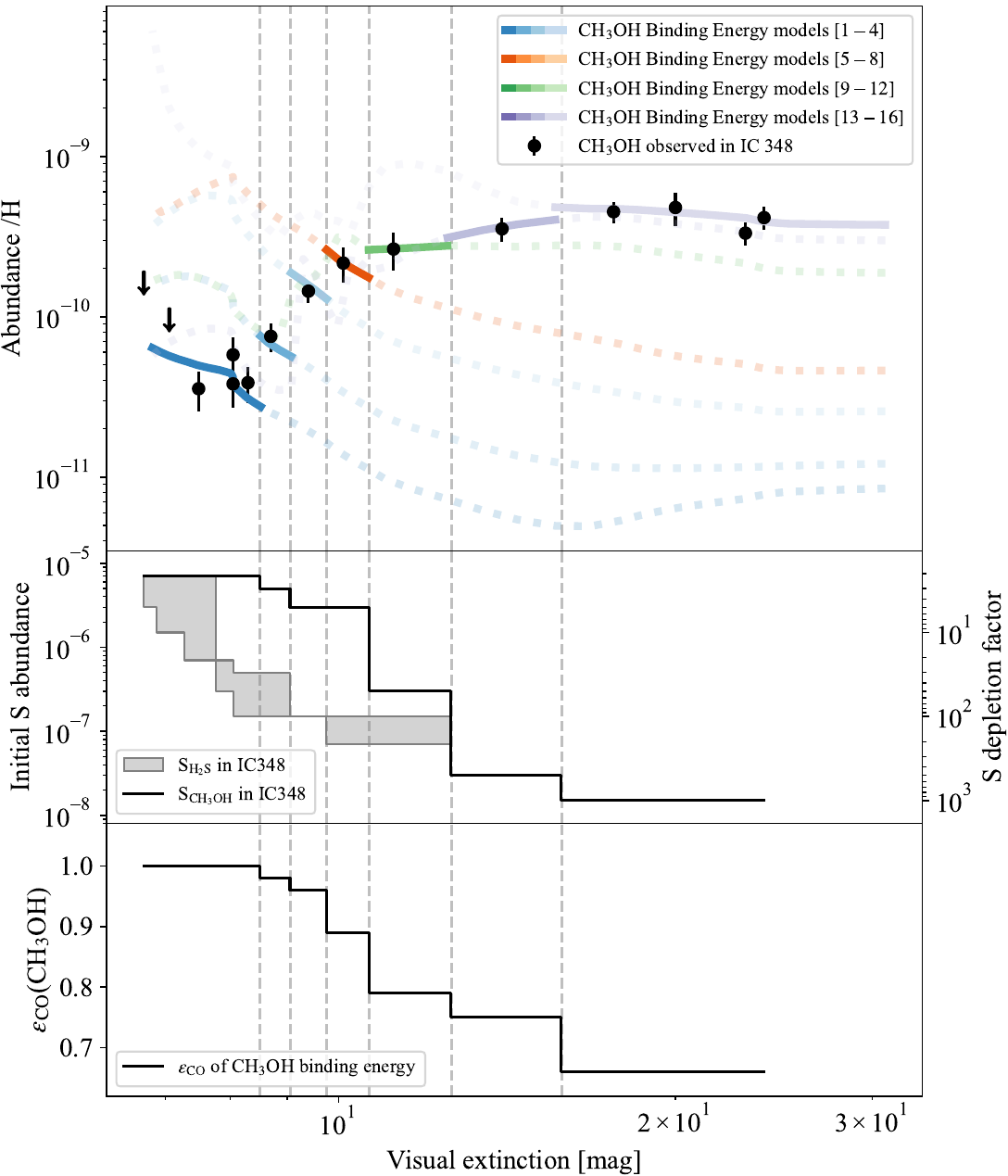}
					\caption{Comparison between the observed abundance of CH$_{3}$OH (black dots) in Barnard 1b (left) and IC 348 (right) with chemical models. There are three panels in each side, with one target per side. In the top panel, the comparison between the observed gas-phase abundance of CH$_{3}$OH and chemical models is made. Each solid segment, delimited by vertical dashed lines, represents the predicted abundance of a model that fits the observed methanol abundance, with a color indicating the set of binding energies of that model. The initial sulfur abundance of that model is displayed in the middle panel of the same column with the black solid curve S$_{{\rm CH}_{3}{\rm OH}}$. The filled area S$_{{\rm H}_{2}{\rm S}}$ from Figs. \ref{fig:modelsB1bH2S} and \ref{fig:modelsIC348H2S} is also shown for comparison. Finally, the scaling applied to the binding energy of CH$_{3}$OH in each segment is depicted in the bottom panel of each column.}
					\label{fig:modelsB1bIC348CH3OH}
				\end{figure*}

    			In Figs. \ref{fig:fiducialB1bComp} and \ref{fig:fiducialIC348Comp} we reported the underestimation of gas-phase CH$_{3}$OH abundance by the fiducial models, even when sulfur depletion is higher. As discussed above, the build up of CO ices could reduce the binding energy of the adsorbates accreted onto grain surfaces, enhancing their gas-phase abundance. Similarly to the analysis of the previous section, we tested all the possible combinations of binding energy models (Table \ref{tab:molpeceresModels}) and initial sulfur abundances (Table \ref{tab:initialSulfurAb}) to find the ones that best fit the observed abundances of methanol. The results of the tests are shown in Figs. \ref{fig:allModelsCH3OHB1b} and \ref{fig:allModelsCH3OHIC348}, and clearly demonstrate the great impact of the total sulfur budget and binding energies on the gas-phase abundance of methanol. In the leftmost panel of the top row of Fig. \ref{fig:allModelsCH3OHB1b} (${\rm S/H}=7\times 10^{-6}$), we first noted that the observed abundances of methanol at the edge of Barnard 1b are only compatible with unmodified or slightly reduced binding energies. This suggests a low effect of CO ice on the binding energies, a reasonable explanation since we expect a low depletion of CO in this area due to its high temperatures and FUV irradiation. At higher extinctions, where sulfur should get even more depleted according to the model predictions of H$_{2}$S, lower binding energies are required in order to obtain a good fit to the data (Fig. \ref{fig:allModelsCH3OHB1b}). Interestingly, even though our CH$_{3}$OH observational data is consistent with an increasing sulfur depletion toward the extinction peak, if we impose the sulfur depletion required for H$_{2}$S, models would overestimate CH$_{3}$OH. In Barnard 1b, a sulfur abundance of ${\rm S/H}\sim 7\times 10^{-8}$ is required at $A_{\rm v}\geq 13$ mag according to the observed H$_{2}$S (Fig. \ref{fig:modelsB1bH2S}). However, this initial sulfur abundance overestimates CH$_{3}$OH around $A_{\rm v}\sim 13-20$ mag (leftmost plot in the bottom row of Fig. \ref{fig:allModelsCH3OHB1b}). It is not possible to assign a unique value for the sulfur depletion that simultaneously reproduces both H$_{2}$S and CH$_{3}$OH observed abundances. We defined S$_{{\rm CH}_{3}{\rm OH}}$ as the appropriate profile of the sulfur abundance that corresponds to methanol (see Fig. \ref{fig:modelsB1bIC348CH3OH}). Nevertheless, it is still possible to find a consistent picture that, on the one hand, incorporates an increasing sulfur depletion toward the extinction peak and, on the other hand, includes increasingly lower binding energies as a result of the building up of CO-ice. This sulfur depletion profile is, in comparison to that obtained with H$_{2}$S, shallower up to $A_{\rm v}\sim 40$ mag. This is summarized in the left side of Fig. \ref{fig:modelsB1bIC348CH3OH}, where we show the combination of sulfur abundance and CH$_{3}$OH binding energy pairs that fits the CH$_{3}$OH observational data.
    			
    			The analysis of models and observational data leads to similar conclusions in IC 348 (Fig. \ref{fig:allModelsCH3OHIC348}). Toward the edge of IC 348, where sulfur abundance is the highest $({\rm S/H}=7\times 10^{-6})$, models with higher binding energies perform better than those with lower binding energies, indicating a little presence of CO ices. At higher extinctions, where, according to H$_{2}$S models, sulfur abundance should be lower, lower binding energies are required in order to fit the CH$_{3}$OH observed abundances. Two different sulfur abundance profiles are needed here as well, the one already derived for H$_{2}$S, and the other that fits CH$_{3}$OH while also retains its decreasing trend toward the extinction peak. The sulfur profile required for methanol is shallower until $A_{\rm v}\sim 10$ mag, where it starts to decrease and match the one derived with H$_{2}$S. These results are summarized in the panels at the right side of Fig. \ref{fig:modelsB1bIC348CH3OH}. It is then possible to build a consistent picture that combines sulfur depletion and lower binding energies to accurately model gas-phase abundances of CH$_{3}$OH toward the two sources observed here.
    			
\begin{figure}
					\centering
					\includegraphics[width=0.495\textwidth]{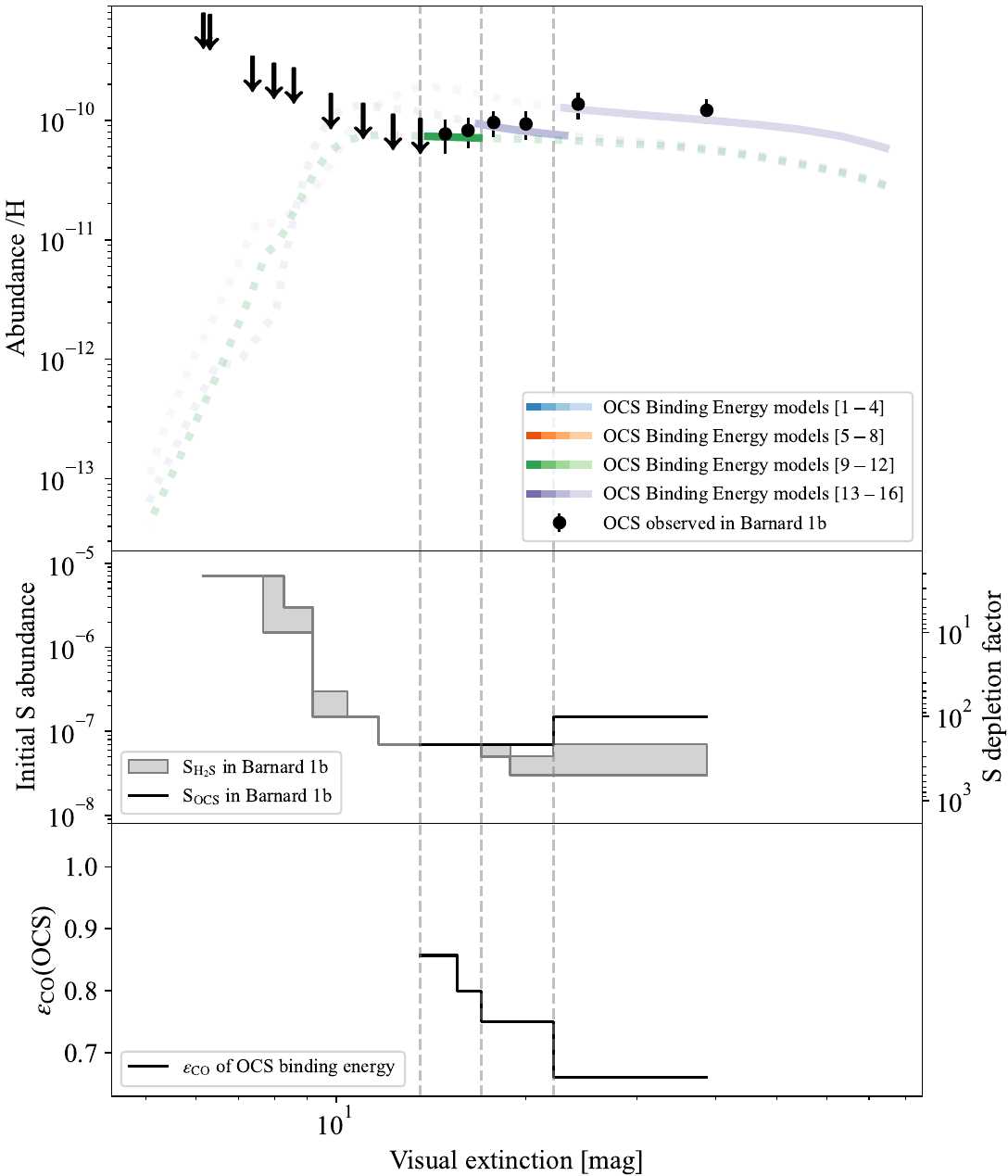}
					\caption{The same analysis as in Fig. \ref{fig:modelsB1bIC348CH3OH}, but with OCS in Barnard 1b. A new sulfur profile, S$_{\rm OCS}$, results from this comparison.}
					\label{fig:modelsB1bOCS}
				\end{figure}

    			OCS line emission was detected toward a few positions in Barnard 1b (Fig. \ref{fig:allAbundancesB1b}), and upper limits to its abundance were derived in IC 348 (Fig. \ref{fig:allAbundancesIC348}). As in the previous section, we tested all possible pairs of initial sulfur abundance and binding energies to find the best fit to the observed abundances of OCS. The results of this test for Barnard 1b are shown in Fig. \ref{fig:allModelsOCSB1b}. In this case, the sulfur depletion profile dictated by H$_{2}$S provides a good fit to the observed abundances of OCS at moderate extinctions, while it must be reduced by a factor of two toward the extinction peak. It is also possible to establish a decreasing trend of binding energies toward the extinction peak as shown in Fig. \ref{fig:modelsB1bOCS}. The paucity of detections of OCS prevents us from extending this trend to lower extinctions. Despite this, it is again possible to establish a consistent pair of profiles for the initial sulfur abundance and the binding energies that describe the declining sulfur abundance and binding energies that are expected at high densities, where CO starts freezing out (Fig. \ref{fig:modelsB1bOCS}). This time, the sulfur profile from OCS, S$_{\rm OCS}$, resembles closely that of H$_{2}$S, S$_{\rm H_{2}S}$.
    			
    			Finally, we tested all possible combinations of sulfur abundances and binding energy models to find the best agreement with the observed abundances of C$^{18}$O. As in Sects. \ref{sec:fiducialB1b} and \ref{sec:fiducialIC348}, we assumed an isotopic ratio of ${\rm C}^{18}{\rm O/CO}\sim 1/200$ \citep{Loison2019}, as our chemical network does not include isotopologues. Processes like selective photodissociation are also not taken into account, so the comparison between models and observations at the edge of the observed region should be taken with care. The performance of the different models reproducing gas-phase C$^{18}$O abundances is plotted in Fig. \ref{fig:allModelsCOB1b}. The gas-phase C$^{18}$O abundance measured at the innermost position, with the highest sulfur depletion, requires the lowest binding energies considered in Table \ref{tab:molpeceresModels}. At lower extinctions ($13 \leq A_{\rm v}\leq 40$ mag), with sulfur becoming more abundant, the remaining gas-phase abundances of C$^{18}$O are in good agreement with models, although there are no significant differences between models with varying binding energies. A similar conclusion is drawn from the comparison in IC 348 (Fig. \ref{fig:allModelsCOIC348}), although here there are notable variations between models with different binding energies. There is however no clear trend in the range $8 \leq A_{\rm v}\leq 10$ mag as to how binding energies should change with extinction to better fit the observations. We conclude that there are several sources of uncertainty in modeling isotopologues like C$^{18}$O and that lower binding energies are needed to reproduce the observed gas-phase C$^{18}$O abundances toward the extinction peak while there is no clear trend at the edge of Barnard 1b and IC 348.
    						
	\section{Discussion}
		
		\subsection{Sulfur abundance and its impact in the chemical complexity}
			
\begin{figure*}
					\centering
					\includegraphics[width=\textwidth]{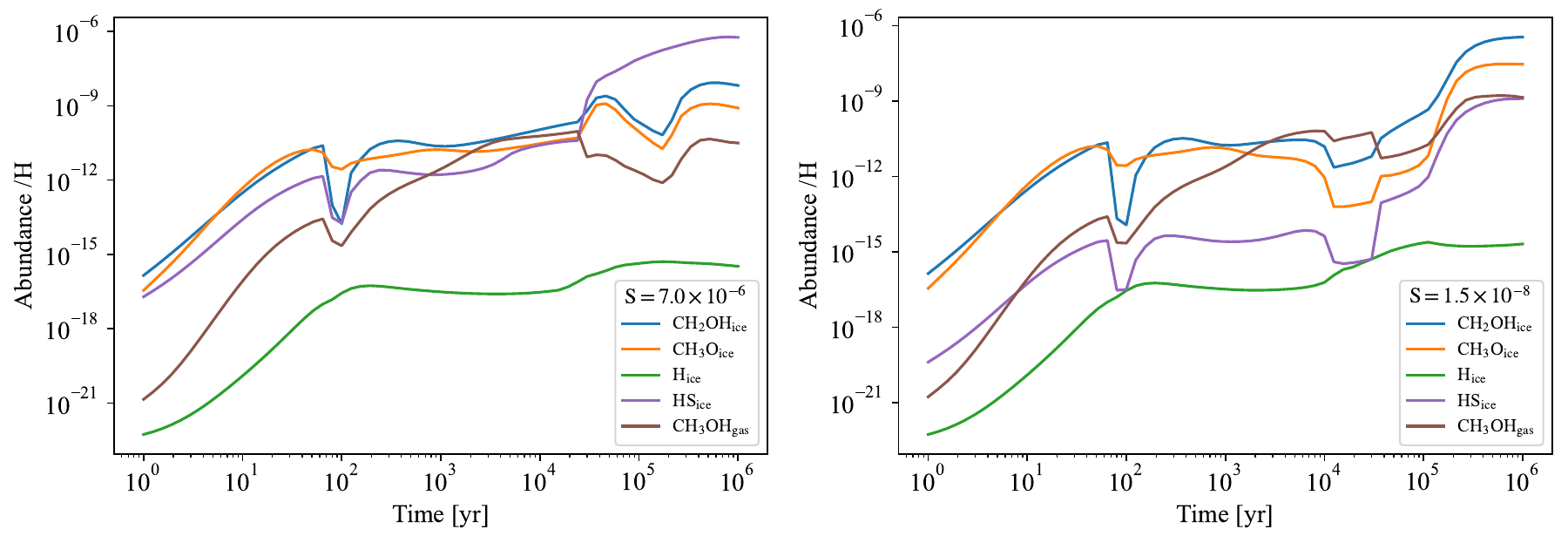}
					\caption{Evolution of gas-phase abundance of CH$_{3}$OH and ice-phase abundance of CH$_{2}$OH, CH$_{3}$O, H, and HS as predicted by the model of Barnard 1b using the binding energies of Model 1 of Table \ref{tab:molpeceresModels} and an initial sulfur abundance of ${\rm S/H}=7.0\times 10^{-6}$ (left) and ${\rm S/H}=7.0\times 10^{-8}$ (right). A significant amount of atomic hydrogen is sequestered when sulfur is highly abundant (left) to form icy HS, leading to a clear reduction of CH$_{2}$OH and CH$_{3}$O radicals in the ice matrix that are involved in the production of methanol.}
					\label{fig:availableH}
			\end{figure*}

			Sulfur-bearing molecules are an important actor in prebiotic chemistry \citep{Leman2004, Chen2019}. Additionally, sulfur impacts the amount of methanol in the gas-phase as shown in Sects. \ref{sec:fiducialB1b}, \ref{sec:fiducialIC348}, and \ref{sec:methanolModelsMolpeceres}. Since methanol is the simplest COM and represents the first step into further chemical complexity, this highlights the importance of an accurate determination of the sulfur elemental abundance in molecular clouds to have a precise knowledge of the formation of COMs. One of the first studies in this regard is the work of \citet{Caselli1994}, who considered the fact that ``low-metal'' initial abundances needed in chemical models to reproduce the abundance of observed species in dark clouds could be the result of freeze-out of ``metals'' on the surface of dust grains. Therefore, in their models, a large amount of atomic sulfur was initially present on grain surfaces, sequestrating the hydrogen found on grain surfaces and thus inhibiting the production of complex organic molecules like CH$_{3}$OH. This was in agreement with the upper limits of solid CH$_{3}$OH along the line of sight of background stars in dark clouds at that time. In our results, regardless of the chosen binding energy model, there is a general decline of predicted gas-phase methanol abundance as the elemental sulfur budget is increased (Figs. \ref{fig:allModelsCH3OHB1b} and \ref{fig:allModelsCH3OHIC348}). To test if the sequestration of atomic hydrogen by sulfur is the reason behind this behavior, we checked the main reactions leading to gas-phase methanol. In both targets, with the binding energies of Model 1 (Table \ref{tab:molpeceresModels}) and at an extinction of $A_{\rm v}= 13$ mag, the main reactions producing methanol are the cosmic ray sputtering, in agreement with the results in \citet{Taillard2023}, and the reactive desorption after hydrogenation of CH$_{2}$OH and CH$_{3}$O on the ice matrix. We then examined the evolution of CH$_{2}$OH$_{\rm ice}$, CH$_{3}$O$_{\rm ice}$, HS$_{\rm ice}$, H$_{\rm ice}$, and CH$_{3}$OH$_{\rm gas}$ abundances predicted by the models of Barnard 1b with two distinct initial sulfur abundances. This is shown in Fig. \ref{fig:availableH}. Up to $\sim 10^{4}$ years, the abundances of CH$_{3}$OH$_{\rm gas}$, CH$_{2}$OH$_{\rm ice}$, CH$_{3}$O$_{\rm ice}$ are quite similar. Notable differences in the abundance of atomic H in the ice appear after $\sim 10^{4}$ years, becoming one order of magnitude more abundant in the case of depleted sulfur. This leads to greater abundances of CH$_{2}$OH$_{\rm ice}$ and CH$_{3}$O$_{\rm ice}$ which, in turn, yield a higher amount of CH$_{3}$OH$_{\rm gas}$. If the initial sulfur budget is enhanced, the ice abundance of HS increases remarkably, suggesting that atomic hydrogen is being indeed captured by the sulfur. Consequently, CH$_{2}$OH$_{\rm ice}$, CH$_{3}$O$_{\rm ice}$, and CH$_{3}$OH do not grow significantly, remaining somewhat stable around $\sim 10^{-9}$, $\sim 10^{-9}$, and $\sim 10^{-11}$, respectively. This amounts to a difference of more than two orders of magnitude between the two scenarios (Fig. \ref{fig:availableH}).
			
			However, one should bear in mind that only diffusive chemistry and cosmic ray sputtering are considered in this paper to be responsible for the formation and release of gas-phase methanol. Non-diffusive chemistry has been invoked to explain the detection of large COMs such as acetaldehyde (CH$_{3}$CHO) or methyl formate (CH$_{3}$OCHO) in very cold environments like dark clouds or prestellar cores \citep[see, e.g.,][]{Matthews1985, Bacmann2012, Vastel2014, JimenezSerra2016, Scibelli2024}, where they are not expected to desorb easily due to their high mass and their low mobility through grain surfaces. Several non-diffusive processes that synthesize large COMs may also affect methanol production \citep{Jin2020}. These include, for instance, the Eley-Rideal mechanism \citep{Ruaud2015}, the recombination of CH$_{n}$O radicals on ices \citep{Fedoseev2015}, or even gas-phase reactions between radicals \citep{Balucani2015}. These processes, in addition to the diffusive chemistry, seem to be required to reproduce the abundances of COMs detected in prestellar cores like L1544 \citep{Jin2020}.
			
    	\subsection{Declining binding energies and the sulfur gap}
		
			Using the state-of-the-art chemical model \texttt{Nautilus}, we investigated whether increasingly lower binding energies, expected when CO is being depleted onto grain mantles, are necessary to explain the observed gas-phase abundances of CH$_{3}$OH, H$_{2}$S, OCS, and C$^{18}$O toward Barnard 1b and IC 348. The scaling of the binding energies proposed by \citet{Molpeceres2024} successfully recreates the gas-phase abundance of methanol at the extinction peak, while the original binding energies of the chemical network \citep{Wakelam2024} are appropriate to its abundance at the edge of the observed regions. The contribution of each declining binding energy to the predicted gas-phase abundances of methanol is not equally distributed among the molecules listed in Table \ref{tab:molpeceresModels}. As discussed in \citet{Molpeceres2024}, CH$_{3}$OH is among the molecules whose ice abundance is not affected by the general decline in binding energies. This is due to its main formation pathways and the very efficient hydrogenation of CO, possible because hydrogen is highly mobile in the ice matrix. However, on the one hand, the lower CH$_{3}$OH binding energy is responsible for its more efficient desorption, as shown by the higher gas-phase CH$_{3}$OH abundances predicted in this paper. The CO binding energy, on the other hand, controls the availability of CO ice over grain surfaces and therefore it has an impact in the total methanol abundance in both gas and ice-phases. In our approach we only considered single-valued binding energies, but as shown by, for instance, \citet{Bariosco2024} and \citet{Bariosco2025}, even lower binding energies than the ones considered here may emerge as a result of the varying nature of adsorption/desorption sites in realistic grain and ice surfaces, even in pure amorphous water ice surfaces. The combined effects of sulfur depletion and lower binding energies help explain the rather flat methanol abundance profile that we observed here and it is also found, for instance, in \citet{Taillard2023}. In this work, however, we do not explore the additional effect of CO ices in the cosmic ray sputtering yield of methanol. The changes we introduced in the binding energies are directly related to the depletion of CO, and thus the physical properties of the two sources analyzed here such as the dust temperature or the gas density should yield distinct rates at which binding energy values drop. Although the observed dust temperature is generally higher toward IC 348, there is a sharper drop of binding energies compared to that calculated for Barnard 1b (Fig. \ref{fig:modelsB1bIC348CH3OH}). The higher gas density toward IC 348 would enhance CO depletion and thus induce such drop. Barnard 1b is also known to host a multiple protostellar system, with three YSOs embedded in the dense core. The activity of the protostars may also be responsible for keeping CO in the gas phase and thus produce a slow decline of binding energies toward the extinction peak of this source. Nevertheless, as discussed previously, the determination of the appropriate scaling for the binding energies is a degenerate problem that relies on the accurate estimation of the sulfur abundance, a parameter that is often poorly constrained. In fact, models with a unique value for the initial sulfur abundance do not necessarily fit the observed gas-phase abundances of all molecules simultaneously. On the one hand, in Sect. \ref{sec:initialSAb} we determined the profile of the initial sulfur abundance ${\rm S}_{\rm H_{2}S}$ needed to fit the observed abundance of H$_{2}$S taking advantage of the small changes that varying binding energies introduce in its predicted gas-phase abundance. We found that, toward the extinction peak, a high sulfur depletion is needed toward Barnard 1b and IC 348. On the other hand, in Sect. \ref{sec:methanolModelsMolpeceres} we found $\epsilon_{\rm CO}-{\rm S}_{\rm CH_{3}OH}$ pairs that fit the gas-phase abundances of methanol and are consistent with the decreasing trends of binding energies and sulfur abundance that are expected at higher extinction. As a result, there is a difference of up to two orders of magnitude between the sulfur profiles ${\rm S}_{\rm H_{2}S}$ and ${\rm S}_{\rm CH_{3}OH}$ in the range $12\lesssim A_{\rm v}\lesssim 19$ mag in Barnard 1b. The difference is still present in the case of ${\rm S}_{\rm OCS}$ and ${\rm S}_{\rm CH_{3}OH}$ (see Fig. \ref{fig:modelsB1bOCS}) toward this source. A similar result was obtained between the sulfur profiles ${\rm S}_{\rm H_{2}S}$ and ${\rm S}_{\rm CH_{3}OH}$ toward IC 348 in the range $9\lesssim A_{\rm v}\lesssim 10$ mag. 
			% These differences may arise as a consequence of the distinct origin of the molecular emission. The rotational emission of H$_{2}$S, CH$_{3}$OH, and OCS could therefore offer a tomographic view of gas with diverse sulfur abundances in dense cores.
			
\begin{figure*}
					\centering
					\includegraphics[width=0.49\textwidth]{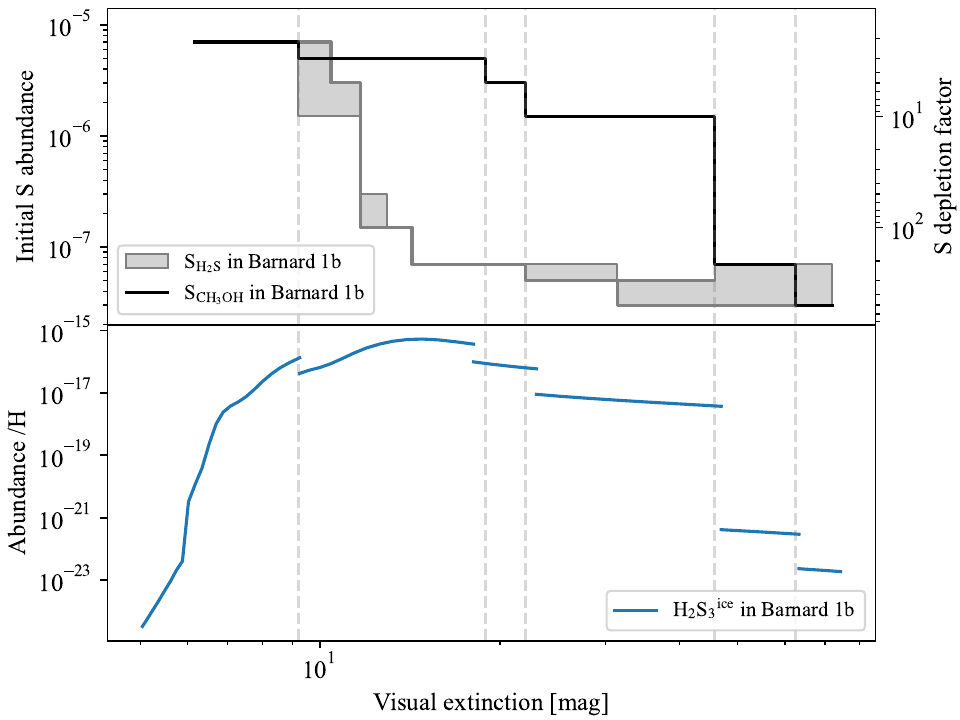}
					\includegraphics[width=0.49\textwidth]{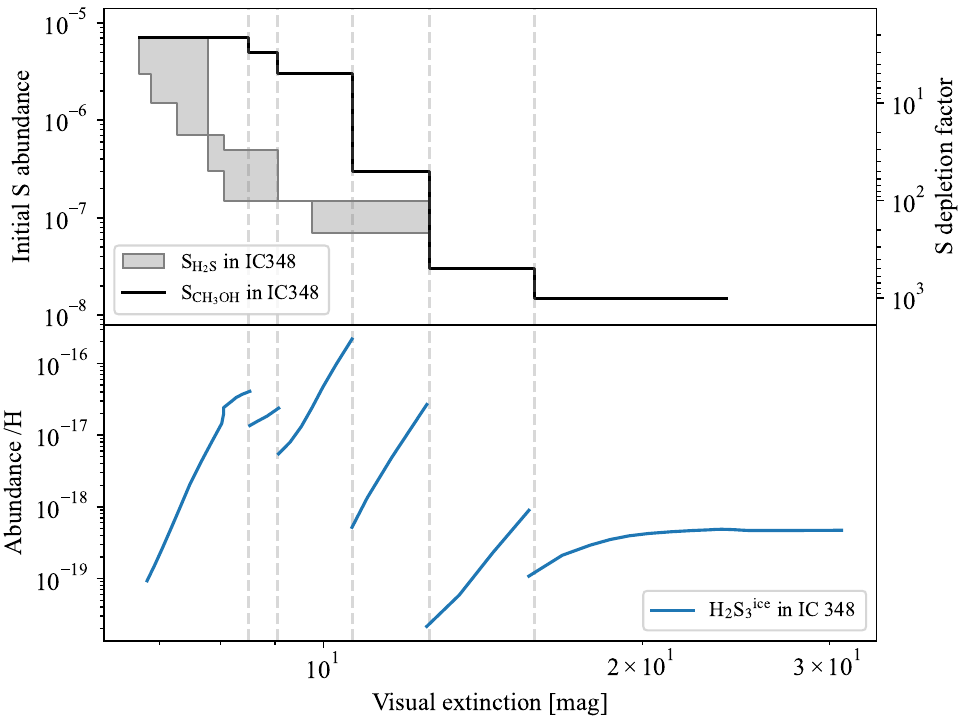}
					\caption{The potential connection between the sulfur gap and the disappearance of H$_{2}$S from interstellar ices with the formation of polysulfanes in Barnard 1b (left side) and IC 348 (right side). In each side, the top panel shows the sulfur gap between the sulfur abundances S$_{{\rm H}_{2}{\rm S}}$ and S$_{{\rm CH}_{3}{\rm OH}}$, required to fit the observations of H$_{2}$S and CH$_{3}$OH, respectively (see also Fig. \ref{fig:modelsB1bIC348CH3OH}). The bottom panel in each side shows the projection along the line of sight of the total icy H$_{2}$S$_{3}$ abundance in the most optimistic case, that is, assuming the sulfur abundance S$_{{\rm CH}_{3}{\rm OH}}$. The chosen binding energies were those of Model 1 in Table \ref{tab:molpeceresModels}.}
					\label{fig:h2s3}
			\end{figure*}

			A possible factor contributing to the differences in the sulfur abundance profiles ${\rm S}_{\rm H_{2}S}$ and ${\rm S}_{\rm OCS}$ when compared to ${\rm S}_{\rm CH_{3}OH}$ is the presence of unknown sulfur sinks. For instance, the lack of observational evidence of H$_{2}$S in interstellar ices, where only a few sulfur-bearing molecules like OCS \citep{Boogert2022, McClure2023} and, tentatively, SO$_{2}$ \citep{Boogert1997, McClure2023, Rocha2024} have been detected, suggests that chemical processing may be taking place in interstellar ices reducing the abundance of H$_{2}$S \citep{Bariosco2024}. The fitting of the gas-phase OCS abundance observed toward Barnard 1b provided us with the sulfur profile S$_{\rm OCS}$ (Fig. \ref{fig:modelsB1bOCS}), which also requires a moderate depletion if compared to S$_{\rm CH_{3}OH}$. Consequently, H$_{2}$S is most likely not being processed into OCS. Polysulfanes (H$_{2}$S$_{x}$) have been proposed first in chemical models \citep{Druard2012} as a product of H$_{2}$S reaction with HS radicals. In particular, they found a large quantity of H$_{2}$S$_{3}$ ice with temperatures $\sim 20$ K that enhance the diffusion of these radicals. Dust temperatures in our observational sample are close to $\sim 20$ K towards the edge, which could help the synthesis of this molecule. However, according to our models, it is found in low quantities. In the most optimistic scenario, that is, assuming the initial sulfur abundance required by methanol ${\rm S}_{\rm CH_{3}OH}$, the predicted total ice abundance along the line of sight peaks at $\sim 10^{-15}$ in Barnard 1b and at $\sim 10^{-16}$ in IC 348 (Fig. \ref{fig:h2s3}). Therefore, despite being most favorably produced at moderate extinctions, where the gap between the sulfur profiles S$_{{\rm H}_{2}{\rm S}}$ and S$_{{\rm CH}_{3}{\rm OH}}$ is at its largest extent, it is not an effective reducer of H$_{2}$S according to our chemical network. Nevertheless, the synthesis of sulfur allotropes and sulfur chains is still poorly constrained, so this should be taken with caution. Additionally, other factors like cosmic rays and secondary photons are also capable of destroying H$_{2}$S to form polysulfanes \citep{Druard2012, Cazaux2022} and sulfur chains \citep{Shingledecker2020, Cazaux2022}. Finally, another potential byproduct of H$_{2}$S reactions in interstellar ices is the ammonium hydrosulfide NH$_{4}$SH as suggested by observations in comets \citep{Altwegg2020}, laboratory work \citep{Vitorino2024, Slavicinska2024}, and chemical modeling \citep{Vitorino2024}. According to the chemical model in \citet{Vitorino2024}, in the evolution of a prestellar core into a protostar, most of the icy H$_{2}$S is processed to form solid NH$_{4}$SH on grain surfaces during the collapse when temperatures are high and reactants are thus mobile enough. While this chemical processing does not take place during the prestellar phase in their model, the enhanced mobility and diffusion rate of molecules in CO-rich ices that we investigated here could potentially favor the conversion of H$_{2}$S into NH$_{4}$SH earlier on in the star formation process. The potential presence of this compound in the interstellar medium has been recently reported in \citet{Slavicinska2024}, where they tentatively identified the NH${_{4}}^{+}\ \nu_{4}$ and SH$^{-}$ libration mode features of the ammonium salt NH$_{4}$SH toward pre- and protostellar objects using the James Webb Space Telescope (JWST). The formation of this compound becomes therefore into another potential culprit in the disappearance of H$_{2}$S from interstellar ices.
			
	\section{Conclusions}
		
		In this paper we presented millimeter observations of rotational lines of H$_{2}$S, CH$_{3}$OH, OCS, C$^{18}$O, and N$_{2}$H$^{+}$ in order to build a consistent picture of the chemical processes that relate these species to each other. Our observations show a widespread detection of CH$_{3}$OH and H$_{2}$S toward Barnard 1b and IC 348. However, OCS was not detected toward IC 348. Methanol exhibits the expected behavior of a molecule mostly produced on grain surfaces. Its abundance starts to decline when CO becomes scarce in the ice matrix due to its thermal desorption at temperatures higher than $\sim 15$ K. At lower temperatures and higher gas densities, the observed gas-phase abundances of C$^{18}$O reveal the catastrophic CO depletion. Is in this setting where methanol and N$_{2}$H$^{+}$ are the most abundant.
		
		We carried out a thorough chemical modeling of these species to reproduce their abundances and learn more about their tight interplay. We found that:
		
		\begin{itemize}
			\item Decreasing sulfur budgets [S] of $\sim 1\times 10^{-7}$ and $\sim 1\times 10^{-8}$ are required to explain the observed gas-phase abundances of H$_{2}$S toward Barnard 1b and IC 348, respectively. The highest sulfur depletion in Barnard 1b, a sulfur rich dense core, is lower than that of IC 348.
			\item The elemental sulfur abundance not only affects the predicted gas-phase abundances of H$_{2}$S, but also those of CH$_{3}$OH. Toward the extinction peak, when sulfur is depleted, CH$_{3}$OH is produced more efficiently, owing to the higher availability of atomic hydrogen in the ice matrix that is required in the hydrogenation of CO into COMs. Therefore, sulfur acts as a regulator promoting or hindering molecular complexity on grain surfaces.
			\item The enhancement of gas-phase methanol abundance due to the depletion of sulfur is not enough to match the observations toward the extinction peak. CO ices in this area are expected to be a significant fraction of the total ice content. To incorporate the weaker bonds of CO ice on the chemical modeling, the binding energies of a set of molecules were progressively lowered in denser areas following the prescription described in \citet{Molpeceres2024}. 
			\item Chemical models with progressively lower sulfur abundances and binding energies successfully predict the observed gas-phase abundances of H$_{2}$S, CH$_{3}$OH, C$^{18}$O, and OCS in Barnard 1b and IC 348. However, the amount of sulfur that must be depleted in order to fit the  molecules in our sample is different: the amount of sulfur to be depleted that provides the best match to H$_{2}$S abundances is higher than that of CH$_{3}$OH or OCS. This suggests a different origin of emission of these molecules or the presence of an unknown sink of H$_{2}$S. Ammonia salts are a potential sulfur sink in this environment. The conditions where these differences between sulfur depletions are higher also appear to be optimal for the synthesis of polysulfanes H$_{2}$S$_{x}$.
		\end{itemize}
		
		These results offer a consistent picture that portrays the formation of methanol, H$_{2}$S, OCS, and C$^{18}$O in the gas phase, while also includes important changes in grain surface chemistry as consequences of CO and sulfur depletion. A more complete picture of COM formation would require additional experimental and theoretical work on the introduction of non-diffusive processes, a more exhaustive description of sulfur chemistry, and the improvement of chemical networks.
		
	\begin{acknowledgements}
		This work is based on observations carried out under project number 121-22 with the IRAM 30m telescope. IRAM is supported by INSU/CNRS (France), MPG (Germany), and IGN (Spain). Based on observations carried out with the Yebes 40 m telescope. The 40 m radio telescope at Yebes Observatory is operated by the Spanish Geographic Institute (IGN; Ministerio de Transportes y Movilidad Sostenible). This work is supported by ERC grant SUL4LIFE, GA No. 101096293. Funded by the European Union. Views and opinions expressed are however those of the authors only and do not necessarily reflect those of the European Union or the European Research Council Executive Agency. Neither the European Union nor the granting authority can be held responsible for them. D.N.-A. also acknowledges funding support from the Fundaci\'on Ram\'on Areces through its international postdoc grant program. A.F. also thanks project PID2022-137980NB-I00 funded by the Spanish Ministry of Science and Innovation/State Agency of Research MCIN/AEI/10.13039/501100011033 and by “ERDF A way of making Europe”. R.M.-D. was supported by a La Caixa Junior Leader grant under agreement LCF/BQ/PI22/11910030.
	\end{acknowledgements}

    \bibliographystyle{aa}
    \bibliography{main.bib}

\begin{thebibliography}{134}
\expandafter\ifx\csname natexlab\endcsname\relax\def\natexlab#1{#1}\fi

\bibitem[{{Altwegg} {et~al.}(2020){Altwegg}, {Balsiger}, {H{\"a}nni}, {Rubin}, {Schuhmann}, {Schroeder}, {S{\'e}mon}, {Wampfler}, {Berthelier}, {Briois}, {Combi}, {Gombosi}, {Cottin}, {De Keyser}, {Dhooghe}, {Fiethe}, \& {Fuselier}}]{Altwegg2020}
{Altwegg}, K., {Balsiger}, H., {H{\"a}nni}, N., {et~al.} 2020, Nature Astronomy, 4, 533

\bibitem[{{Arslan} {et~al.}(2023){Arslan}, {Hocuk}, {Caselli}, \& {K{\"u}{\c{c}}{\"u}k}}]{Arslan2023}
{Arslan}, {\"O}., {Hocuk}, S., {Caselli}, P., \& {K{\"u}{\c{c}}{\"u}k}, {\.I}. 2023, \mnras, 518, 2050

\bibitem[{{Asplund} {et~al.}(2009){Asplund}, {Grevesse}, {Sauval}, \& {Scott}}]{Asplund2009}
{Asplund}, M., {Grevesse}, N., {Sauval}, A.~J., \& {Scott}, P. 2009, \araa, 47, 481

\bibitem[{{Bachiller} {et~al.}(1990){Bachiller}, {Cernicharo}, {Martin-Pintado}, {Tafalla}, \& {Lazareff}}]{Bachiller1990}
{Bachiller}, R., {Cernicharo}, J., {Martin-Pintado}, J., {Tafalla}, M., \& {Lazareff}, B. 1990, \aap, 231, 174

\bibitem[{{Bacmann} {et~al.}(2012){Bacmann}, {Taquet}, {Faure}, {Kahane}, \& {Ceccarelli}}]{Bacmann2012}
{Bacmann}, A., {Taquet}, V., {Faure}, A., {Kahane}, C., \& {Ceccarelli}, C. 2012, \aap, 541, L12

\bibitem[{{Balucani} {et~al.}(2015){Balucani}, {Ceccarelli}, \& {Taquet}}]{Balucani2015}
{Balucani}, N., {Ceccarelli}, C., \& {Taquet}, V. 2015, \mnras, 449, L16

\bibitem[{{Bariosco} {et~al.}(2024){Bariosco}, {Pantaleone}, {Ceccarelli}, {Rimola}, {Balucani}, {Corno}, \& {Ugliengo}}]{Bariosco2024}
{Bariosco}, V., {Pantaleone}, S., {Ceccarelli}, C., {et~al.} 2024, \mnras, 531, 1371

\bibitem[{{Bariosco} {et~al.}(2025){Bariosco}, {Tinacci}, {Pantaleone}, {Ceccarelli}, {Rimola}, \& {Ugliengo}}]{Bariosco2025}
{Bariosco}, V., {Tinacci}, L., {Pantaleone}, S., {et~al.} 2025, \mnras, 539, 82

\bibitem[{{Bizzocchi} {et~al.}(2014){Bizzocchi}, {Caselli}, {Spezzano}, \& {Leonardo}}]{Bizzocchi2014}
{Bizzocchi}, L., {Caselli}, P., {Spezzano}, S., \& {Leonardo}, E. 2014, \aap, 569, A27

\bibitem[{{Blake} {et~al.}(1987){Blake}, {Sutton}, {Masson}, \& {Phillips}}]{Blake1987}
{Blake}, G.~A., {Sutton}, E.~C., {Masson}, C.~R., \& {Phillips}, T.~G. 1987, \apj, 315, 621

\bibitem[{{Bohlin} {et~al.}(1978){Bohlin}, {Savage}, \& {Drake}}]{Bohlin1978}
{Bohlin}, R.~C., {Savage}, B.~D., \& {Drake}, J.~F. 1978, \apj, 224, 132

\bibitem[{{Boogert} {et~al.}(2022){Boogert}, {Brewer}, {Brittain}, \& {Emerson}}]{Boogert2022}
{Boogert}, A.~C.~A., {Brewer}, K., {Brittain}, A., \& {Emerson}, K.~S. 2022, \apj, 941, 32

\bibitem[{{Boogert} {et~al.}(1997){Boogert}, {Schutte}, {Helmich}, {Tielens}, \& {Wooden}}]{Boogert1997}
{Boogert}, A.~C.~A., {Schutte}, W.~A., {Helmich}, F.~P., {Tielens}, A.~G.~G.~M., \& {Wooden}, D.~H. 1997, \aap, 317, 929

\bibitem[{{Calmonte} {et~al.}(2016){Calmonte}, {Altwegg}, {Balsiger}, {Berthelier}, {Bieler}, {Cessateur}, {Dhooghe}, {van Dishoeck}, {Fiethe}, {Fuselier}, {Gasc}, {Gombosi}, {H{\"a}ssig}, {Le Roy}, {Rubin}, {S{\'e}mon}, {Tzou}, \& {Wampfler}}]{Calmonte2016}
{Calmonte}, U., {Altwegg}, K., {Balsiger}, H., {et~al.} 2016, \mnras, 462, S253

\bibitem[{{Caselli} {et~al.}(1993){Caselli}, {Hasegawa}, \& {Herbst}}]{Caselli1993}
{Caselli}, P., {Hasegawa}, T.~I., \& {Herbst}, E. 1993, \apj, 408, 548

\bibitem[{{Caselli} {et~al.}(1994){Caselli}, {Hasegawa}, \& {Herbst}}]{Caselli1994}
{Caselli}, P., {Hasegawa}, T.~I., \& {Herbst}, E. 1994, \apj, 421, 206

\bibitem[{{Caselli} {et~al.}(2012){Caselli}, {Keto}, {Bergin}, {Tafalla}, {Aikawa}, {Douglas}, {Pagani}, {Y{\'\i}ld{\'\i}z}, {van der Tak}, {Walmsley}, {Codella}, {Nisini}, {Kristensen}, \& {van Dishoeck}}]{Caselli2012}
{Caselli}, P., {Keto}, E., {Bergin}, E.~A., {et~al.} 2012, \apjl, 759, L37

\bibitem[{{Caselli} {et~al.}(1999){Caselli}, {Walmsley}, {Tafalla}, {Dore}, \& {Myers}}]{Caselli1999}
{Caselli}, P., {Walmsley}, C.~M., {Tafalla}, M., {Dore}, L., \& {Myers}, P.~C. 1999, \apjl, 523, L165

\bibitem[{{Cazaux} {et~al.}(2022){Cazaux}, {Carrascosa}, {Mu{\~n}oz Caro}, {Caselli}, {Fuente}, {Navarro-Almaida}, \& {Rivi{\'e}re-Marichalar}}]{Cazaux2022}
{Cazaux}, S., {Carrascosa}, H., {Mu{\~n}oz Caro}, G.~M., {et~al.} 2022, \aap, 657, A100

\bibitem[{{Cazaux} {et~al.}(2003){Cazaux}, {Tielens}, {Ceccarelli}, {Castets}, {Wakelam}, {Caux}, {Parise}, \& {Teyssier}}]{Cazaux2003}
{Cazaux}, S., {Tielens}, A.~G.~G.~M., {Ceccarelli}, C., {et~al.} 2003, \apjl, 593, L51

\bibitem[{{Chaabouni} {et~al.}(2018){Chaabouni}, {Diana}, {Nguyen}, \& {Dulieu}}]{Chaabouni2018}
{Chaabouni}, H., {Diana}, S., {Nguyen}, T., \& {Dulieu}, F. 2018, \aap, 612, A47

\bibitem[{{Charnley}(1997)}]{Charnley1997}
{Charnley}, S.~B. 1997, \apj, 481, 396

\bibitem[{{Charnley} {et~al.}(1992){Charnley}, {Tielens}, \& {Millar}}]{Charnley1992}
{Charnley}, S.~B., {Tielens}, A.~G.~G.~M., \& {Millar}, T.~J. 1992, \apjl, 399, L71

\bibitem[{{Chen} {et~al.}(2024){Chen}, {Quan}, {He}, {Wang}, {Li}, \& {Henning}}]{Chen2024}
{Chen}, L.-F., {Quan}, D., {He}, J., {et~al.} 2024, \aap, 685, A55

\bibitem[{Chen \& Yu(2019)}]{Chen2019}
Chen, M. \& Yu, X. 2019

\bibitem[{{Cummins} {et~al.}(1986){Cummins}, {Linke}, \& {Thaddeus}}]{Cummins1986}
{Cummins}, S.~E., {Linke}, R.~A., \& {Thaddeus}, P. 1986, \apjs, 60, 819

\bibitem[{{Dagdigian}(2020)}]{Dagdigian2020}
{Dagdigian}, P.~J. 2020, \mnras, 494, 5239

\bibitem[{{Dartois} {et~al.}(2021){Dartois}, {Chabot}, {Id Barkach}, {Rothard}, {Boduch}, {Aug{\'e}}, \& {Agnihotri}}]{Dartois2021}
{Dartois}, E., {Chabot}, M., {Id Barkach}, T., {et~al.} 2021, \aap, 647, A177

\bibitem[{{Draine}(1978)}]{Draine1978}
{Draine}, B.~T. 1978, \apjs, 36, 595

\bibitem[{{Druard} \& {Wakelam}(2012)}]{Druard2012}
{Druard}, C. \& {Wakelam}, V. 2012, \mnras, 426, 354

\bibitem[{{el Akel} {et~al.}(2022){el Akel}, {Kristensen}, {Le Gal}, {van der Walt}, {Pitts}, \& {Dulieu}}]{elAkel2022}
{el Akel}, M., {Kristensen}, L.~E., {Le Gal}, R., {et~al.} 2022, \aap, 659, A100

\bibitem[{{Enrique-Romero} {et~al.}(2021){Enrique-Romero}, {Ceccarelli}, {Rimola}, {Skouteris}, {Balucani}, \& {Ugliengo}}]{EnriqueRomero2021}
{Enrique-Romero}, J., {Ceccarelli}, C., {Rimola}, A., {et~al.} 2021, \aap, 655, A9

\bibitem[{{Enrique-Romero} {et~al.}(2022){Enrique-Romero}, {Rimola}, {Ceccarelli}, {Ugliengo}, {Balucani}, \& {Skouteris}}]{EnriqueRomero2022}
{Enrique-Romero}, J., {Rimola}, A., {Ceccarelli}, C., {et~al.} 2022, \apjs, 259, 39

\bibitem[{{Enrique-Romero} {et~al.}(2023){Enrique-Romero}, {Rimola}, {Ceccarelli}, {Ugliengo}, {Balucani}, \& {Skouteris}}]{EnriqueRomero2023}
{Enrique-Romero}, J., {Rimola}, A., {Ceccarelli}, C., {et~al.} 2023, in European Conference on Laboratory Astrophysics ECLA2020. The Interplay of Dust, 57--61

\bibitem[{{Esplugues} {et~al.}(2014){Esplugues}, {Viti}, {Goicoechea}, \& {Cernicharo}}]{Esplugues2014}
{Esplugues}, G.~B., {Viti}, S., {Goicoechea}, J.~R., \& {Cernicharo}, J. 2014, \aap, 567, A95

\bibitem[{{Fedoseev} {et~al.}(2015){Fedoseev}, {Cuppen}, {Ioppolo}, {Lamberts}, \& {Linnartz}}]{Fedoseev2015}
{Fedoseev}, G., {Cuppen}, H.~M., {Ioppolo}, S., {Lamberts}, T., \& {Linnartz}, H. 2015, \mnras, 448, 1288

\bibitem[{{Ferrero} {et~al.}(2023){Ferrero}, {Pantaleone}, {Ceccarelli}, {Ugliengo}, {Sodupe}, \& {Rimola}}]{Ferrero2023}
{Ferrero}, S., {Pantaleone}, S., {Ceccarelli}, C., {et~al.} 2023, \apj, 944, 142

\bibitem[{{Friesen} {et~al.}(2017){Friesen}, {Pineda}, {co-PIs}, {Rosolowsky}, {Alves}, {Chac{\'o}n-Tanarro}, {How-Huan Chen}, {Chun-Yuan Chen}, {Di Francesco}, {Keown}, {Kirk}, {Punanova}, {Seo}, {Shirley}, {Ginsburg}, {Hall}, {Offner}, {Singh}, {Arce}, {Caselli}, {Goodman}, {Martin}, {Matzner}, {Myers}, {Redaelli}, \& {GAS Collaboration}}]{Friesen2017}
{Friesen}, R.~K., {Pineda}, J.~E., {co-PIs}, {et~al.} 2017, \apj, 843, 63

\bibitem[{{Fuchs} {et~al.}(2009){Fuchs}, {Cuppen}, {Ioppolo}, {Romanzin}, {Bisschop}, {Andersson}, {van Dishoeck}, \& {Linnartz}}]{Fuchs2009}
{Fuchs}, G.~W., {Cuppen}, H.~M., {Ioppolo}, S., {et~al.} 2009, \aap, 505, 629

\bibitem[{{Fuente} {et~al.}(2016){Fuente}, {Cernicharo}, {Roueff}, {Gerin}, {Pety}, {Marcelino}, {Bachiller}, {Lefloch}, {Roncero}, \& {Aguado}}]{Fuente2016}
{Fuente}, A., {Cernicharo}, J., {Roueff}, E., {et~al.} 2016, \aap, 593, A94

\bibitem[{{Fuente} {et~al.}(2019){Fuente}, {Navarro}, {Caselli}, {Gerin}, {Kramer}, {Roueff}, {Alonso-Albi}, {Bachiller}, {Cazaux}, {Commercon}, {Friesen}, {Garc{\'\i}a-Burillo}, {Giuliano}, {Goicoechea}, {Gratier}, {Hacar}, {Jim{\'e}nez-Serra}, {Kirk}, {Lattanzi}, {Loison}, {Malinen}, {Marcelino}, {Mart{\'\i}n-Dom{\'e}nech}, {Mu{\~n}oz-Caro}, {Pineda}, {Tafalla}, {Tercero}, {Ward-Thompson}, {Trevi{\~n}o-Morales}, {Rivi{\'e}re-Marichalar}, {Roncero}, {Vidal}, \& {Ballester}}]{Fuente2019}
{Fuente}, A., {Navarro}, D.~G., {Caselli}, P., {et~al.} 2019, \aap, 624, A105

\bibitem[{{Fuente} {et~al.}(2023){Fuente}, {Rivi{\`e}re-Marichalar}, {Beitia-Antero}, {Caselli}, {Wakelam}, {Esplugues}, {Rodr{\'\i}guez-Baras}, {Navarro-Almaida}, {Gerin}, {Kramer}, {Bachiller}, {Goicoechea}, {Jim{\'e}nez-Serra}, {Loison}, {Ivlev}, {Mart{\'\i}n-Dom{\'e}nech}, {Spezzano}, {Roncero}, {Mu{\~n}oz-Caro}, {Cazaux}, \& {Marcelino}}]{Fuente2023}
{Fuente}, A., {Rivi{\`e}re-Marichalar}, P., {Beitia-Antero}, L., {et~al.} 2023, \aap, 670, A114

\bibitem[{{Garrod} \& {Herbst}(2006)}]{Garrod2006}
{Garrod}, R.~T. \& {Herbst}, E. 2006, \aap, 457, 927

\bibitem[{{Garrod} {et~al.}(2022){Garrod}, {Jin}, {Matis}, {Jones}, {Willis}, \& {Herbst}}]{Garrod2022}
{Garrod}, R.~T., {Jin}, M., {Matis}, K.~A., {et~al.} 2022, \apjs, 259, 1

\bibitem[{{Garrod} {et~al.}(2007){Garrod}, {Wakelam}, \& {Herbst}}]{Garrod2007}
{Garrod}, R.~T., {Wakelam}, V., \& {Herbst}, E. 2007, \aap, 467, 1103

\bibitem[{{Garrod} {et~al.}(2008){Garrod}, {Widicus Weaver}, \& {Herbst}}]{Garrod2008}
{Garrod}, R.~T., {Widicus Weaver}, S.~L., \& {Herbst}, E. 2008, \apj, 682, 283

\bibitem[{{Geballe} {et~al.}(1985){Geballe}, {Baas}, {Greenberg}, \& {Schutte}}]{Geballe1985}
{Geballe}, T.~R., {Baas}, F., {Greenberg}, J.~M., \& {Schutte}, W. 1985, \aap, 146, L6

\bibitem[{{Geppert} {et~al.}(2006){Geppert}, {Hamberg}, {Thomas}, {{\"O}sterdahl}, {Hellberg}, {Zhaunerchyk}, {Ehlerding}, {Millar}, {Roberts}, {Semaniak}, {Ugglas}, {K{\"a}llberg}, {Simonsson}, {Kaminska}, \& {Larsson}}]{Geppert2006}
{Geppert}, W.~D., {Hamberg}, M., {Thomas}, R.~D., {et~al.} 2006, Faraday Discussions, 133, 177

\bibitem[{{Gerin} {et~al.}(2015){Gerin}, {Pety}, {Fuente}, {Cernicharo}, {Commer{\c{c}}on}, \& {Marcelino}}]{Gerin2015}
{Gerin}, M., {Pety}, J., {Fuente}, A., {et~al.} 2015, \aap, 577, L2

\bibitem[{{Hasegawa} \& {Herbst}(1993)}]{Hasegawa1993}
{Hasegawa}, T.~I. \& {Herbst}, E. 1993, \mnras, 263, 589

\bibitem[{{Herbst} \& {van Dishoeck}(2009)}]{Herbst2009}
{Herbst}, E. \& {van Dishoeck}, E.~F. 2009, \araa, 47, 427

\bibitem[{{Herbst}(2008)}]{Herbst2008}
{Herbst}, W. 2008, in Handbook of Star Forming Regions, Volume I, ed. B.~{Reipurth}, Vol.~4, 372

\bibitem[{{Hiramatsu} {et~al.}(2010){Hiramatsu}, {Hirano}, \& {Takakuwa}}]{Hiramatsu2010}
{Hiramatsu}, M., {Hirano}, N., \& {Takakuwa}, S. 2010, \apj, 712, 778

\bibitem[{{Hirano} {et~al.}(1999){Hirano}, {Kamazaki}, {Mikami}, {Ohashi}, \& {Umemoto}}]{Hirano1999}
{Hirano}, N., {Kamazaki}, T., {Mikami}, H., {Ohashi}, N., \& {Umemoto}, T. 1999, in Star Formation 1999, ed. T.~{Nakamoto}, 181--182

\bibitem[{{Hirano} \& {Liu}(2014)}]{Hirano2014}
{Hirano}, N. \& {Liu}, F.-c. 2014, \apj, 789, 50

\bibitem[{{Hocuk} {et~al.}(2017){Hocuk}, {Sz{\H{u}}cs}, {Caselli}, {Cazaux}, {Spaans}, \& {Esplugues}}]{Hocuk2017}
{Hocuk}, S., {Sz{\H{u}}cs}, L., {Caselli}, P., {et~al.} 2017, \aap, 604, A58

\bibitem[{{Holdship} {et~al.}(2019){Holdship}, {Viti}, {Codella}, {Rawlings}, {Jimenez-Serra}, {Ayalew}, {Curtis}, {Habib}, {Lawrence}, {Warsame}, \& {Horn}}]{Holdship2019}
{Holdship}, J., {Viti}, S., {Codella}, C., {et~al.} 2019, \apj, 880, 138

\bibitem[{{Horn} {et~al.}(2004){Horn}, {M{\o}llendal}, {Sekiguchi}, {Uggerud}, {Roberts}, {Herbst}, {Viggiano}, \& {Fridgen}}]{Horn2004}
{Horn}, A., {M{\o}llendal}, H., {Sekiguchi}, O., {et~al.} 2004, \apj, 611, 605

\bibitem[{{Huang} \& {Hirano}(2013)}]{Huang2013}
{Huang}, Y.-H. \& {Hirano}, N. 2013, \apj, 766, 131

\bibitem[{{Iglesias-Groth}(2019)}]{Iglesias-Groth2019}
{Iglesias-Groth}, S. 2019, \mnras, 489, 1509

\bibitem[{{Iglesias-Groth} \& {Marin-Dobrincic}(2023)}]{Iglesias-Groth2023}
{Iglesias-Groth}, S. \& {Marin-Dobrincic}, M. 2023, \mnras, 521, 2248

\bibitem[{{Ioppolo} {et~al.}(2021){Ioppolo}, {Fedoseev}, {Chuang}, {Cuppen}, {Clements}, {Jin}, {Garrod}, {Qasim}, {Kofman}, {van Dishoeck}, \& {Linnartz}}]{Ioppolo2021}
{Ioppolo}, S., {Fedoseev}, G., {Chuang}, K.~J., {et~al.} 2021, Nature Astronomy, 5, 197

\bibitem[{{Jim{\'e}nez-Escobar} \& {Mu{\~n}oz Caro}(2011)}]{Escobar2011}
{Jim{\'e}nez-Escobar}, A. \& {Mu{\~n}oz Caro}, G.~M. 2011, \aap, 536, A91

\bibitem[{{Jim{\'e}nez-Serra} {et~al.}(2025){Jim{\'e}nez-Serra}, {Meg{\'\i}as}, {Salaris}, {Cuppen}, {Taillard}, {Jin}, {Wakelam}, {Vasyunin}, {Caselli}, {Pendleton}, {Dartois}, {Noble}, {Viti}, {Borshcheva}, {Garrod}, {Lamberts}, {Fraser}, {Melnick}, {McClure}, {Rocha}, {Drozdovskaya}, \& {Lis}}]{JimenezSerra2025}
{Jim{\'e}nez-Serra}, I., {Meg{\'\i}as}, A., {Salaris}, J., {et~al.} 2025, arXiv e-prints, arXiv:2502.10123

\bibitem[{{Jim{\'e}nez-Serra} {et~al.}(2016){Jim{\'e}nez-Serra}, {Vasyunin}, {Caselli}, {Marcelino}, {Billot}, {Viti}, {Testi}, {Vastel}, {Lefloch}, \& {Bachiller}}]{JimenezSerra2016}
{Jim{\'e}nez-Serra}, I., {Vasyunin}, A.~I., {Caselli}, P., {et~al.} 2016, \apjl, 830, L6

\bibitem[{{Jim{\'e}nez-Serra} {et~al.}(2021){Jim{\'e}nez-Serra}, {Vasyunin}, {Spezzano}, {Caselli}, {Cosentino}, \& {Viti}}]{JimenezSerra2021}
{Jim{\'e}nez-Serra}, I., {Vasyunin}, A.~I., {Spezzano}, S., {et~al.} 2021, \apj, 917, 44

\bibitem[{{Jin} \& {Garrod}(2020)}]{Jin2020}
{Jin}, M. \& {Garrod}, R.~T. 2020, \apjs, 249, 26

\bibitem[{{J{\o}rgensen} {et~al.}(2004){J{\o}rgensen}, {Sch{\"o}ier}, \& {van Dishoeck}}]{Jorgensen2004}
{J{\o}rgensen}, J.~K., {Sch{\"o}ier}, F.~L., \& {van Dishoeck}, E.~F. 2004, \aap, 416, 603

\bibitem[{{Kawakita} \& {Kobayashi}(2009)}]{Kawakita2009}
{Kawakita}, H. \& {Kobayashi}, H. 2009, \apj, 693, 388

\bibitem[{{Laas} \& {Caselli}(2019)}]{Laas2019}
{Laas}, J.~C. \& {Caselli}, P. 2019, \aap, 624, A108

\bibitem[{{Lada} {et~al.}(2006){Lada}, {Muench}, {Luhman}, {Allen}, {Hartmann}, {Megeath}, {Myers}, {Fazio}, {Wood}, {Muzerolle}, {Rieke}, {Siegler}, \& {Young}}]{Lada2006}
{Lada}, C.~J., {Muench}, A.~A., {Luhman}, K.~L., {et~al.} 2006, \aj, 131, 1574

\bibitem[{{Le Gal} {et~al.}(2021){Le Gal}, {{\"O}berg}, {Teague}, {Loomis}, {Law}, {Walsh}, {Bergin}, {M{\'e}nard}, {Wilner}, {Andrews}, {Aikawa}, {Booth}, {Cataldi}, {Bergner}, {Bosman}, {Cleeves}, {Czekala}, {Furuya}, {Guzm{\'a}n}, {Huang}, {Ilee}, {Nomura}, {Qi}, {Schwarz}, {Tsukagoshi}, {Yamato}, \& {Zhang}}]{LeGal2021}
{Le Gal}, R., {{\"O}berg}, K.~I., {Teague}, R., {et~al.} 2021, \apjs, 257, 12

\bibitem[{{Lee} {et~al.}(2004){Lee}, {Bergin}, \& {Evans}}]{Lee2004}
{Lee}, J.-E., {Bergin}, E.~A., \& {Evans}, Neal~J., I. 2004, \apj, 617, 360

\bibitem[{{Leger} {et~al.}(1985){Leger}, {Jura}, \& {Omont}}]{Leger1985}
{Leger}, A., {Jura}, M., \& {Omont}, A. 1985, \aap, 144, 147

\bibitem[{Leman {et~al.}(2004)Leman, Orgel, \& Ghadiri}]{Leman2004}
Leman, L.~J., Orgel, L.~E., \& Ghadiri, M.~R. 2004, Science, 306, 283

\bibitem[{{Lis} {et~al.}(2002){Lis}, {Roueff}, {Gerin}, {Phillips}, {Coudert}, {van der Tak}, \& {Schilke}}]{Lis2002}
{Lis}, D.~C., {Roueff}, E., {Gerin}, M., {et~al.} 2002, \apjl, 571, L55

\bibitem[{{Lodders}(2003)}]{Lodders2003}
{Lodders}, K. 2003, \apj, 591, 1220

\bibitem[{{Loison} {et~al.}(2019){Loison}, {Wakelam}, {Gratier}, {Hickson}, {Bacmann}, {Ag{\'u}ndez}, {Marcelino}, {Cernicharo}, {Guzman}, {Gerin}, {Goicoechea}, {Roueff}, {Petit}, {Pety}, {Fuente}, \& {Riviere-Marichalar}}]{Loison2019}
{Loison}, J.-C., {Wakelam}, V., {Gratier}, P., {et~al.} 2019, \mnras, 485, 5777

\bibitem[{{Luo} {et~al.}(2023){Luo}, {Zhang}, {Bisbas}, {Li}, {Tang}, {Wang}, {Zhou}, {Zuo}, {Yue}, {Zhou}, \& {Lin}}]{Luo2023}
{Luo}, G., {Zhang}, Z.-Y., {Bisbas}, T.~G., {et~al.} 2023, \apj, 942, 101

\bibitem[{{Marcelino} {et~al.}(2005){Marcelino}, {Cernicharo}, {Roueff}, {Gerin}, \& {Mauersberger}}]{Marcelino2005}
{Marcelino}, N., {Cernicharo}, J., {Roueff}, E., {Gerin}, M., \& {Mauersberger}, R. 2005, \apj, 620, 308

\bibitem[{{Marcelino} {et~al.}(2018){Marcelino}, {Gerin}, {Cernicharo}, {Fuente}, {Wootten}, {Chapillon}, {Pety}, {Lis}, {Roueff}, {Commer{\c{c}}on}, \& {Ciardi}}]{Marcelino2018}
{Marcelino}, N., {Gerin}, M., {Cernicharo}, J., {et~al.} 2018, \aap, 620, A80

\bibitem[{{Matthews} {et~al.}(1985){Matthews}, {Friberg}, \& {Irvine}}]{Matthews1985}
{Matthews}, H.~E., {Friberg}, P., \& {Irvine}, W.~M. 1985, \apj, 290, 609

\bibitem[{{McClure} {et~al.}(2023){McClure}, {Rocha}, {Pontoppidan}, {Crouzet}, {Chu}, {Dartois}, {Lamberts}, {Noble}, {Pendleton}, {Perotti}, {Qasim}, {Rachid}, {Smith}, {Sun}, {Beck}, {Boogert}, {Brown}, {Caselli}, {Charnley}, {Cuppen}, {Dickinson}, {Drozdovskaya}, {Egami}, {Erkal}, {Fraser}, {Garrod}, {Harsono}, {Ioppolo}, {Jim{\'e}nez-Serra}, {Jin}, {J{\o}rgensen}, {Kristensen}, {Lis}, {McCoustra}, {McGuire}, {Melnick}, {{\~A}-berg}, {Palumbo}, {Shimonishi}, {Sturm}, {van Dishoeck}, \& {Linnartz}}]{McClure2023}
{McClure}, M.~K., {Rocha}, W.~R.~M., {Pontoppidan}, K.~M., {et~al.} 2023, Nature Astronomy, 7, 431

\bibitem[{{Millar} {et~al.}(1991){Millar}, {Herbst}, \& {Charnley}}]{Millar1991}
{Millar}, T.~J., {Herbst}, E., \& {Charnley}, S.~B. 1991, \apj, 369, 147

\bibitem[{{Minissale} {et~al.}(2016){Minissale}, {Dulieu}, {Cazaux}, \& {Hocuk}}]{Minissale2016}
{Minissale}, M., {Dulieu}, F., {Cazaux}, S., \& {Hocuk}, S. 2016, \aap, 585, A24

\bibitem[{{Molpeceres} {et~al.}(2024){Molpeceres}, {Furuya}, \& {Aikawa}}]{Molpeceres2024}
{Molpeceres}, G., {Furuya}, K., \& {Aikawa}, Y. 2024, \aap, 688, A150

\bibitem[{{Mu{\~n}oz Caro} {et~al.}(2010){Mu{\~n}oz Caro}, {Jim{\'e}nez-Escobar}, {Mart{\'\i}n-Gago}, {Rogero}, {Atienza}, {Puertas}, {Sobrado}, \& {Torres-Redondo}}]{MunozCaro2010}
{Mu{\~n}oz Caro}, G.~M., {Jim{\'e}nez-Escobar}, A., {Mart{\'\i}n-Gago}, J.~{\'A}., {et~al.} 2010, \aap, 522, A108

\bibitem[{{Muench} {et~al.}(2007){Muench}, {Lada}, {Luhman}, {Muzerolle}, \& {Young}}]{Muench2007}
{Muench}, A.~A., {Lada}, C.~J., {Luhman}, K.~L., {Muzerolle}, J., \& {Young}, E. 2007, \aj, 134, 411

\bibitem[{{Mumma} \& {Charnley}(2011)}]{Mumma2011}
{Mumma}, M.~J. \& {Charnley}, S.~B. 2011, \araa, 49, 471

\bibitem[{{Nakagawa}(1980)}]{Nakagawa1980}
{Nakagawa}, N. 1980, in IAU Symposium, Vol.~87, Interstellar Molecules, ed. B.~H. {Andrew}, 365

\bibitem[{{Navarro-Almaida} {et~al.}(2020){Navarro-Almaida}, {Le Gal}, {Fuente}, {Rivi{\`e}re-Marichalar}, {Wakelam}, {Cazaux}, {Caselli}, {Laas}, {Alonso-Albi}, {Loison}, {Gerin}, {Kramer}, {Roueff}, {Bachiller}, {Commer{\c{c}}on}, {Friesen}, {Garc{\'\i}a-Burillo}, {Goicoechea}, {Giuliano}, {Jim{\'e}nez-Serra}, {Kirk}, {Lattanzi}, {Malinen}, {Marcelino}, {Mart{\'\i}n-Dom{\`e}nech}, {Mu{\~n}oz Caro}, {Pineda}, {Tercero}, {Trevi{\~n}o-Morales}, {Roncero}, {Hacar}, {Tafalla}, \& {Ward-Thompson}}]{NavarroAlmaida2020}
{Navarro-Almaida}, D., {Le Gal}, R., {Fuente}, A., {et~al.} 2020, \aap, 637, A39

\bibitem[{{Nishiyama} {et~al.}(2008){Nishiyama}, {Nagata}, {Tamura}, {Kandori}, {Hatano}, {Sato}, \& {Sugitani}}]{Nishiyama2008}
{Nishiyama}, S., {Nagata}, T., {Tamura}, M., {et~al.} 2008, \apj, 680, 1174

\bibitem[{{\"O}berg(2016)}]{Oberg2016}
{\"O}berg, K.~I. 2016, Chemical Reviews, 116, 9631, pMID: 27099922

\bibitem[{{{\"O}berg} {et~al.}(2010){{\"O}berg}, {Bottinelli}, {J{\o}rgensen}, \& {van Dishoeck}}]{Oberg2010}
{{\"O}berg}, K.~I., {Bottinelli}, S., {J{\o}rgensen}, J.~K., \& {van Dishoeck}, E.~F. 2010, \apj, 716, 825

\bibitem[{{Palumbo} {et~al.}(1995){Palumbo}, {Tielens}, \& {Tokunaga}}]{Palumbo1995}
{Palumbo}, M.~E., {Tielens}, A.~G.~G.~M., \& {Tokunaga}, A.~T. 1995, \apj, 449, 674

\bibitem[{{Pantaleone} {et~al.}(2021){Pantaleone}, {Enrique-Romero}, {Ceccarelli}, {Ferrero}, {Balucani}, {Rimola}, \& {Ugliengo}}]{Pantaleone2021}
{Pantaleone}, S., {Enrique-Romero}, J., {Ceccarelli}, C., {et~al.} 2021, \apj, 917, 49

\bibitem[{{Pantaleone} {et~al.}(2020){Pantaleone}, {Enrique-Romero}, {Ceccarelli}, {Ugliengo}, {Balucani}, \& {Rimola}}]{Pantaleone2020}
{Pantaleone}, S., {Enrique-Romero}, J., {Ceccarelli}, C., {et~al.} 2020, \apj, 897, 56

\bibitem[{{Perrero} {et~al.}(2024){Perrero}, {Beitia-Antero}, {Fuente}, {Ugliengo}, \& {Rimola}}]{Perrero2024}
{Perrero}, J., {Beitia-Antero}, L., {Fuente}, A., {Ugliengo}, P., \& {Rimola}, A. 2024, \apj, 971, 36

\bibitem[{{Poggio} {et~al.}(2021){Poggio}, {Drimmel}, {Cantat-Gaudin}, {Ramos}, {Ripepi}, {Zari}, {Andrae}, {Blomme}, {Chemin}, {Clementini}, {Figueras}, {Fouesneau}, {Fr{\'e}mat}, {Lobel}, {Marshall}, {Muraveva}, \& {Romero-G{\'o}mez}}]{Poggio2021}
{Poggio}, E., {Drimmel}, R., {Cantat-Gaudin}, T., {et~al.} 2021, \aap, 651, A104

\bibitem[{{Prasad} \& {Tarafdar}(1983)}]{Prasad1983}
{Prasad}, S.~S. \& {Tarafdar}, S.~P. 1983, \apj, 267, 603

\bibitem[{{Priestley} {et~al.}(2023){Priestley}, {Clark}, {Glover}, {Ragan}, {Feh{\'e}r}, {Prole}, \& {Klessen}}]{Priestley2023}
{Priestley}, F.~D., {Clark}, P.~C., {Glover}, S.~C.~O., {et~al.} 2023, \mnras, 526, 4952

\bibitem[{{Rabli} \& {Flower}(2010)}]{Rabli2010}
{Rabli}, D. \& {Flower}, D.~R. 2010, \mnras, 406, 95

\bibitem[{{Requena-Torres} {et~al.}(2006){Requena-Torres}, {Mart{\'\i}n-Pintado}, {Rodr{\'\i}guez-Franco}, {Mart{\'\i}n}, {Rodr{\'\i}guez-Fern{\'a}ndez}, \& {de Vicente}}]{RequenaTorres2006}
{Requena-Torres}, M.~A., {Mart{\'\i}n-Pintado}, J., {Rodr{\'\i}guez-Franco}, A., {et~al.} 2006, \aap, 455, 971

\bibitem[{{Riedel} {et~al.}(2023){Riedel}, {Sipil{\"a}}, {Redaelli}, {Caselli}, {Vasyunin}, {Dulieu}, \& {Watanabe}}]{Riedel2023}
{Riedel}, W., {Sipil{\"a}}, O., {Redaelli}, E., {et~al.} 2023, \aap, 680, A87

\bibitem[{{Rocha} {et~al.}(2024){Rocha}, {van Dishoeck}, {Ressler}, {van Gelder}, {Slavicinska}, {Brunken}, {Linnartz}, {Ray}, {Beuther}, {Caratti o Garatti}, {Geers}, {Kavanagh}, {Klaassen}, {Justtanont}, {Chen}, {Francis}, {Gieser}, {Perotti}, {Tychoniec}, {Barsony}, {Majumdar}, {le Gouellec}, {Chu}, {Lew}, {Henning}, \& {Wright}}]{Rocha2024}
{Rocha}, W.~R.~M., {van Dishoeck}, E.~F., {Ressler}, M.~E., {et~al.} 2024, \aap, 683, A124

\bibitem[{{Rodr{\'\i}guez-Baras} {et~al.}(2021){Rodr{\'\i}guez-Baras}, {Fuente}, {Rivi{\'e}re-Marichalar}, {Navarro-Almaida}, {Caselli}, {Gerin}, {Kramer}, {Roueff}, {Wakelam}, {Esplugues}, {Garc{\'\i}a-Burillo}, {Le Gal}, {Spezzano}, {Alonso-Albi}, {Bachiller}, {Cazaux}, {Commercon}, {Goicoechea}, {Loison}, {Trevi{\~n}o-Morales}, {Roncero}, {Jim{\'e}nez-Serra}, {Laas}, {Hacar}, {Kirk}, {Lattanzi}, {Mart{\'\i}n-Dom{\'e}nech}, {Mu{\~n}oz-Caro}, {Pineda}, {Tercero}, {Ward-Thompson}, {Tafalla}, {Marcelino}, {Malinen}, {Friesen}, \& {Giuliano}}]{RodriguezBaras2021}
{Rodr{\'\i}guez-Baras}, M., {Fuente}, A., {Rivi{\'e}re-Marichalar}, P., {et~al.} 2021, \aap, 648, A120

\bibitem[{{Ruaud} {et~al.}(2015){Ruaud}, {Loison}, {Hickson}, {Gratier}, {Hersant}, \& {Wakelam}}]{Ruaud2015}
{Ruaud}, M., {Loison}, J.~C., {Hickson}, K.~M., {et~al.} 2015, \mnras, 447, 4004

\bibitem[{{Ruaud} {et~al.}(2016){Ruaud}, {Wakelam}, \& {Hersant}}]{Ruaud2016}
{Ruaud}, M., {Wakelam}, V., \& {Hersant}, F. 2016, \mnras, 459, 3756

\bibitem[{{Ruffle} {et~al.}(1999){Ruffle}, {Hartquist}, {Caselli}, \& {Williams}}]{Ruffle1999}
{Ruffle}, D.~P., {Hartquist}, T.~W., {Caselli}, P., \& {Williams}, D.~A. 1999, \mnras, 306, 691

\bibitem[{{Sch{\"o}ier} {et~al.}(2005){Sch{\"o}ier}, {van der Tak}, {van Dishoeck}, \& {Black}}]{Schoier2005}
{Sch{\"o}ier}, F.~L., {van der Tak}, F.~F.~S., {van Dishoeck}, E.~F., \& {Black}, J.~H. 2005, \aap, 432, 369

\bibitem[{{Scibelli} \& {Shirley}(2020)}]{Scibelli2020}
{Scibelli}, S. \& {Shirley}, Y. 2020, \apj, 891, 73

\bibitem[{{Scibelli} {et~al.}(2024){Scibelli}, {Shirley}, {Meg{\'\i}as}, \& {Jim{\'e}nez-Serra}}]{Scibelli2024}
{Scibelli}, S., {Shirley}, Y., {Meg{\'\i}as}, A., \& {Jim{\'e}nez-Serra}, I. 2024, \mnras, 533, 4104

\bibitem[{{Shen} {et~al.}(2004){Shen}, {Greenberg}, {Schutte}, \& {van Dishoeck}}]{Shen2004}
{Shen}, C.~J., {Greenberg}, J.~M., {Schutte}, W.~A., \& {van Dishoeck}, E.~F. 2004, \aap, 415, 203

\bibitem[{{Shingledecker} {et~al.}(2020){Shingledecker}, {Lamberts}, {Laas}, {Vasyunin}, {Herbst}, {K{\"a}stner}, \& {Caselli}}]{Shingledecker2020}
{Shingledecker}, C.~N., {Lamberts}, T., {Laas}, J.~C., {et~al.} 2020, \apj, 888, 52

\bibitem[{{Sipil{\"a}} {et~al.}(2021){Sipil{\"a}}, {Silsbee}, \& {Caselli}}]{Sipila2021}
{Sipil{\"a}}, O., {Silsbee}, K., \& {Caselli}, P. 2021, \apj, 922, 126

\bibitem[{{Slavicinska} {et~al.}(2024){Slavicinska}, {Boogert}, {Tychoniec}, {van Dishoeck}, {van Gelder}, {Santos}, {Klaassen}, {Kavanagh}, \& {Chuang}}]{Slavicinska2024}
{Slavicinska}, K., {Boogert}, A., {Tychoniec}, {\L}., {et~al.} 2024, arXiv e-prints, arXiv:2410.02860

\bibitem[{{Tafalla} {et~al.}(2006){Tafalla}, {Kumar}, \& {Bachiller}}]{Tafalla2006}
{Tafalla}, M., {Kumar}, M.~S.~N., \& {Bachiller}, R. 2006, \aap, 456, 179

\bibitem[{{Tafalla} {et~al.}(2004){Tafalla}, {Myers}, {Caselli}, \& {Walmsley}}]{Tafalla2004}
{Tafalla}, M., {Myers}, P.~C., {Caselli}, P., \& {Walmsley}, C.~M. 2004, \aap, 416, 191

\bibitem[{{Taillard} {et~al.}(2023){Taillard}, {Wakelam}, {Gratier}, {Dartois}, {Chabot}, {Noble}, {Keane}, {Boogert}, \& {Harsono}}]{Taillard2023}
{Taillard}, A., {Wakelam}, V., {Gratier}, P., {et~al.} 2023, \aap, 670, A141

\bibitem[{{Tinacci} {et~al.}(2022){Tinacci}, {Germain}, {Pantaleone}, {Ferrero}, {Ceccarelli}, \& {Ugliengo}}]{Tinacci2022}
{Tinacci}, L., {Germain}, A., {Pantaleone}, S., {et~al.} 2022, ACS Earth and Space Chemistry, 6, 1514

\bibitem[{{van der Tak} {et~al.}(2007){van der Tak}, {Black}, {Sch{\"o}ier}, {Jansen}, \& {van Dishoeck}}]{vanDerTak2007}
{van der Tak}, F.~F.~S., {Black}, J.~H., {Sch{\"o}ier}, F.~L., {Jansen}, D.~J., \& {van Dishoeck}, E.~F. 2007, \aap, 468, 627

\bibitem[{{Vastel} {et~al.}(2014){Vastel}, {Ceccarelli}, {Lefloch}, \& {Bachiller}}]{Vastel2014}
{Vastel}, C., {Ceccarelli}, C., {Lefloch}, B., \& {Bachiller}, R. 2014, \apjl, 795, L2

\bibitem[{{Vasyunin} {et~al.}(2017){Vasyunin}, {Caselli}, {Dulieu}, \& {Jim{\'e}nez-Serra}}]{Vasyunin2017}
{Vasyunin}, A.~I., {Caselli}, P., {Dulieu}, F., \& {Jim{\'e}nez-Serra}, I. 2017, \apj, 842, 33

\bibitem[{{Vasyunin} \& {Herbst}(2013)}]{Vasyunin2013}
{Vasyunin}, A.~I. \& {Herbst}, E. 2013, \apj, 769, 34

\bibitem[{{Vidal} {et~al.}(2017){Vidal}, {Loison}, {Jaziri}, {Ruaud}, {Gratier}, \& {Wakelam}}]{Vidal2017}
{Vidal}, T. H.~G., {Loison}, J.-C., {Jaziri}, A.~Y., {et~al.} 2017, \mnras, 469, 435

\bibitem[{{Viti} \& {Williams}(1999)}]{Viti1999}
{Viti}, S. \& {Williams}, D.~A. 1999, \mnras, 310, 517

\bibitem[{{Vitorino} {et~al.}(2024){Vitorino}, {Loison}, {Wakelam}, {Congiu}, \& {Dulieu}}]{Vitorino2024}
{Vitorino}, J., {Loison}, J.~C., {Wakelam}, V., {Congiu}, E., \& {Dulieu}, F. 2024, \mnras, 533, 52

\bibitem[{{Wakelam} {et~al.}(2021){Wakelam}, {Dartois}, {Chabot}, {Spezzano}, {Navarro-Almaida}, {Loison}, \& {Fuente}}]{Wakelam2021}
{Wakelam}, V., {Dartois}, E., {Chabot}, M., {et~al.} 2021, \aap, 652, A63

\bibitem[{{Wakelam} {et~al.}(2024){Wakelam}, {Gratier}, {Loison}, {Hickson}, {Penguen}, \& {Mechineau}}]{Wakelam2024}
{Wakelam}, V., {Gratier}, P., {Loison}, J.~C., {et~al.} 2024, arXiv e-prints, arXiv:2407.15958

\bibitem[{{Watanabe} \& {Kouchi}(2002)}]{Watanabe2002}
{Watanabe}, N. \& {Kouchi}, A. 2002, \apjl, 571, L173

\bibitem[{{Wilson}(1999)}]{Wilson1999}
{Wilson}, T.~L. 1999, Reports on Progress in Physics, 62, 143

\bibitem[{{Wirstr{\"o}m} {et~al.}(2011){Wirstr{\"o}m}, {Geppert}, {Hjalmarson}, {Persson}, {Black}, {Bergman}, {Millar}, {Hamberg}, \& {Vigren}}]{Wirstrom2011}
{Wirstr{\"o}m}, E.~S., {Geppert}, W.~D., {Hjalmarson}, {\r{A}}., {et~al.} 2011, \aap, 533, A24

\bibitem[{{Zari} {et~al.}(2016){Zari}, {Lombardi}, {Alves}, {Lada}, \& {Bouy}}]{Zari2016}
{Zari}, E., {Lombardi}, M., {Alves}, J., {Lada}, C.~J., \& {Bouy}, H. 2016, \aap, 587, A106

\bibitem[{{Zucker} {et~al.}(2018){Zucker}, {Schlafly}, {Speagle}, {Green}, {Portillo}, {Finkbeiner}, \& {Goodman}}]{Zucker2018}
{Zucker}, C., {Schlafly}, E.~F., {Speagle}, J.~S., {et~al.} 2018, \apj, 869, 83

\end{thebibliography}
    
    \begin{appendix}
        
        \onecolumn
        
        \section{Molecular spectra - Barnard 1b}\label{sec:spectraBarnard1b}
        	\vfill
            \begin{figure}[h]
                \centering
                \includegraphics[width=0.99\textwidth]{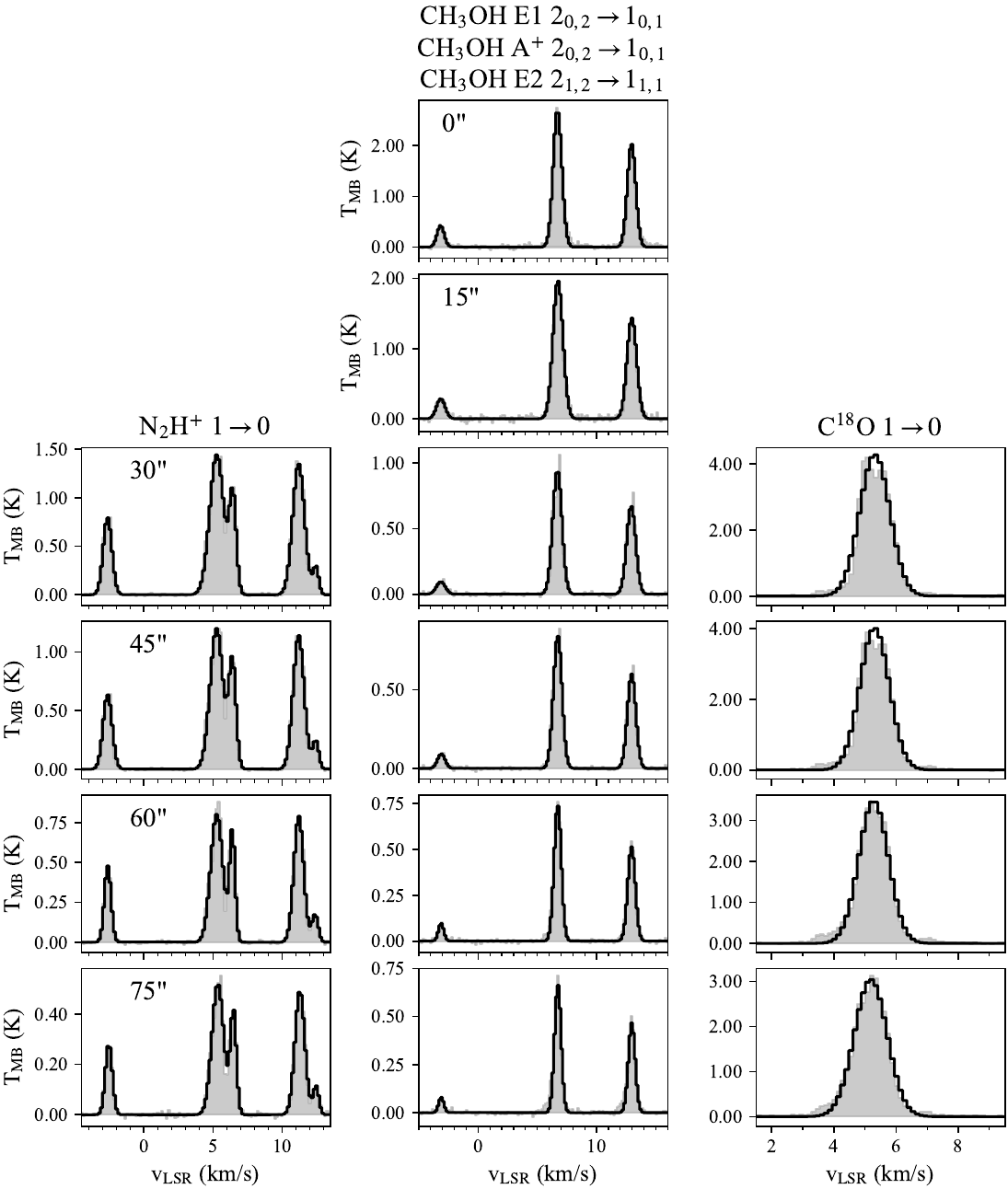}
                \caption{Rotational spectra (gray area) of molecular lines observed at offsets $0''-75''$ with respect to the central position in Barnard 1b (see Table \ref{tab:sources}). The solid black line marks the RADEX fitting of each line while the red solid line corresponds to additional velocity components that were not taken into account in the fitting.}
                \label{fig:spectra_n2hp_ch3oh_30_90_IC348}
            \end{figure}
            \clearpage
            \vfill
            \begin{figure}[h]
                \centering
                \includegraphics[width=0.99\textwidth]{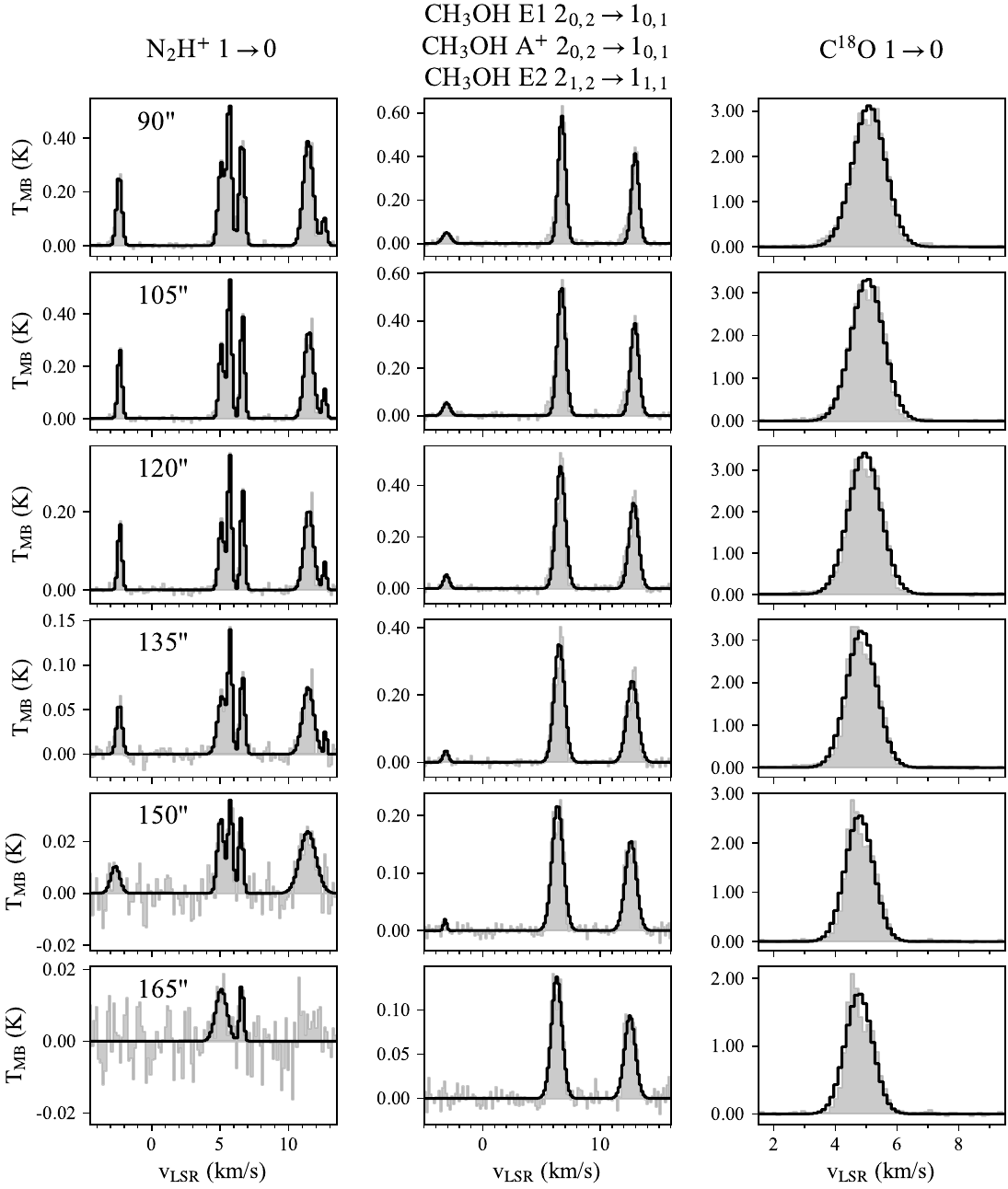}
                \caption{Rotational spectra (gray area) of molecular lines observed at offsets $90''-165''$ with respect to the central position in Barnard 1b (see Table \ref{tab:sources}). The solid black line marks the RADEX fitting of each line while the red solid line corresponds to additional velocity components that were not taken into account in the fitting.}
                \label{fig:spectra_n2hp_ch3oh_30_90_IC348}
            \end{figure}
            \vfill
            \clearpage
            \vfill
            \begin{figure}[h]
                \centering
                \includegraphics[width=\textwidth]{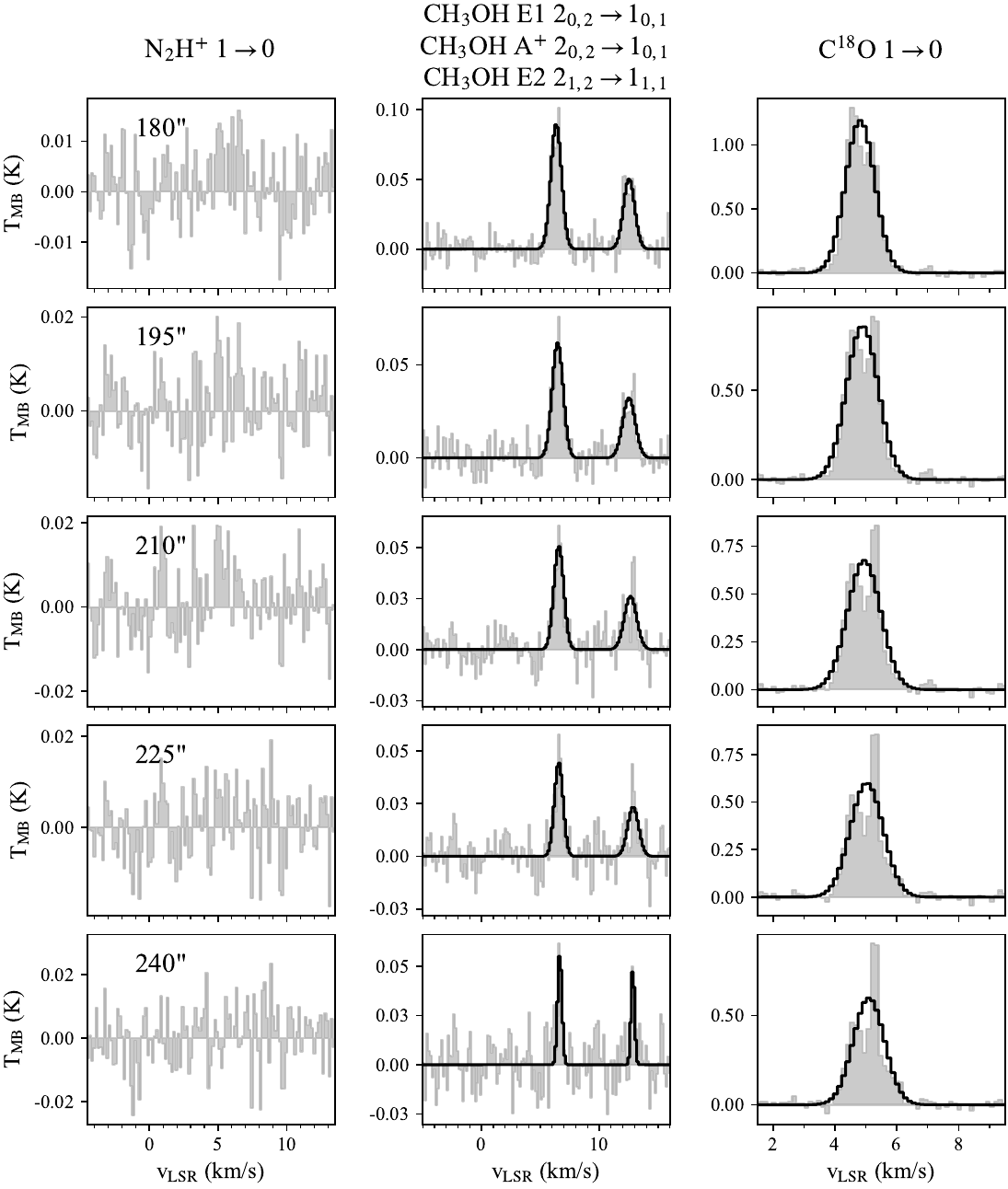}
                \caption{Rotational spectra (gray area) of molecular lines observed at offsets $180''-240''$ with respect to the central position in Barnard 1b (see Table \ref{tab:sources}). The solid black line marks the RADEX fitting of each line while the red solid line corresponds to additional velocity components that were not taken into account in the fitting.}
                \label{fig:spectra_n2hp_ch3oh_30_90_IC348}
            \end{figure}
            \vfill
            \vfill
            \begin{figure}[h]
                \centering
                \includegraphics[width=\textwidth]{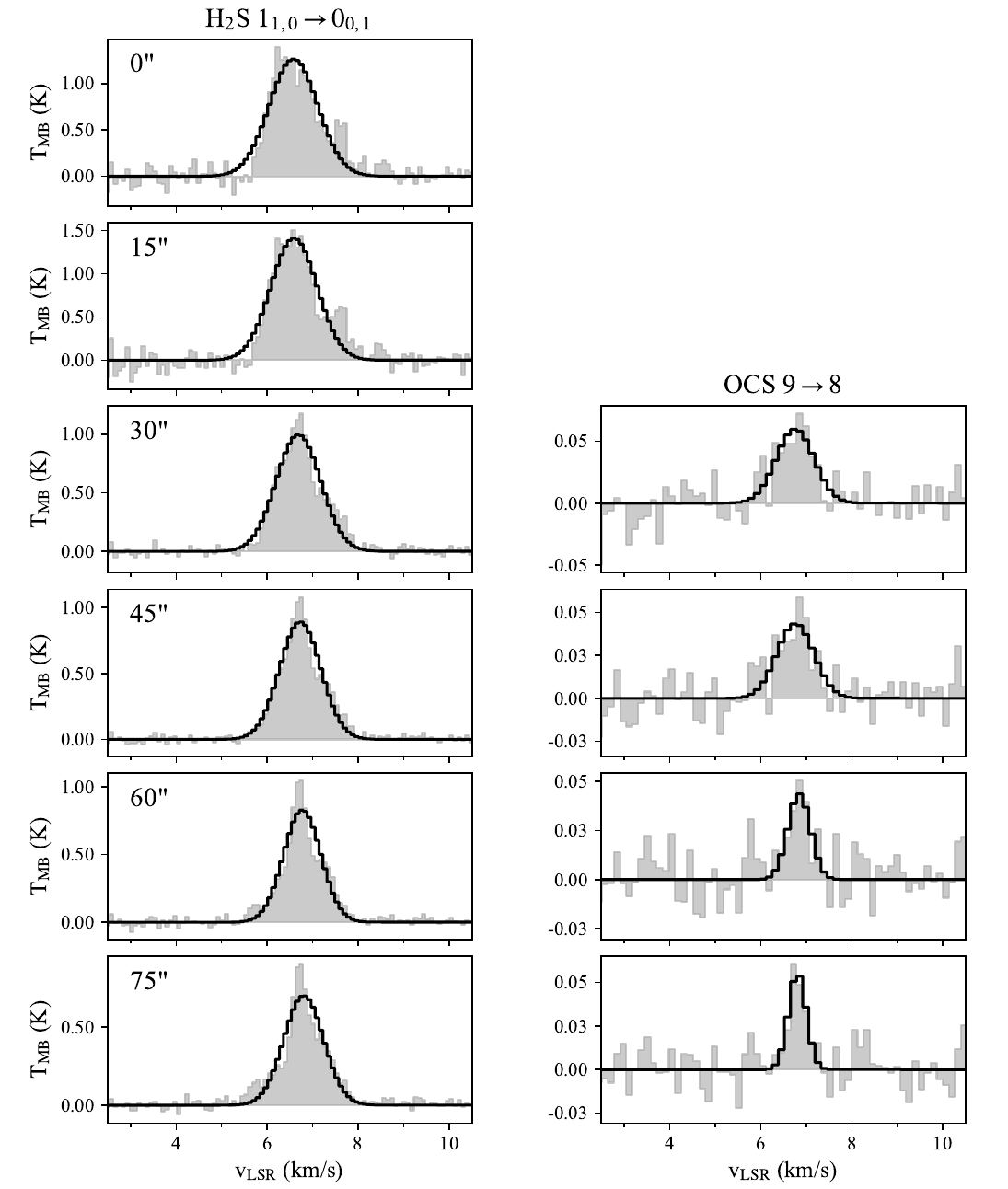}
                \caption{Rotational spectra (gray) of molecular lines observed with the IRAM 30m telescope at offsets $0''-75''$ with respect to the central position in Barnard 1b (see Table \ref{tab:sources}). The CH$_{3}$OH data obtained with the Yebes 40m telescope is shown in the upper right corner in blue. The solid black line marks the RADEX fitting of each line while the red solid line corresponds to additional velocity components that were not taken into account in the fitting.}
                \label{fig:spectra_n2hp_ch3oh_30_90_IC348}
            \end{figure}
            \vfill
            \vfill
            \begin{figure}[h]
                \centering
                \includegraphics[width=\textwidth]{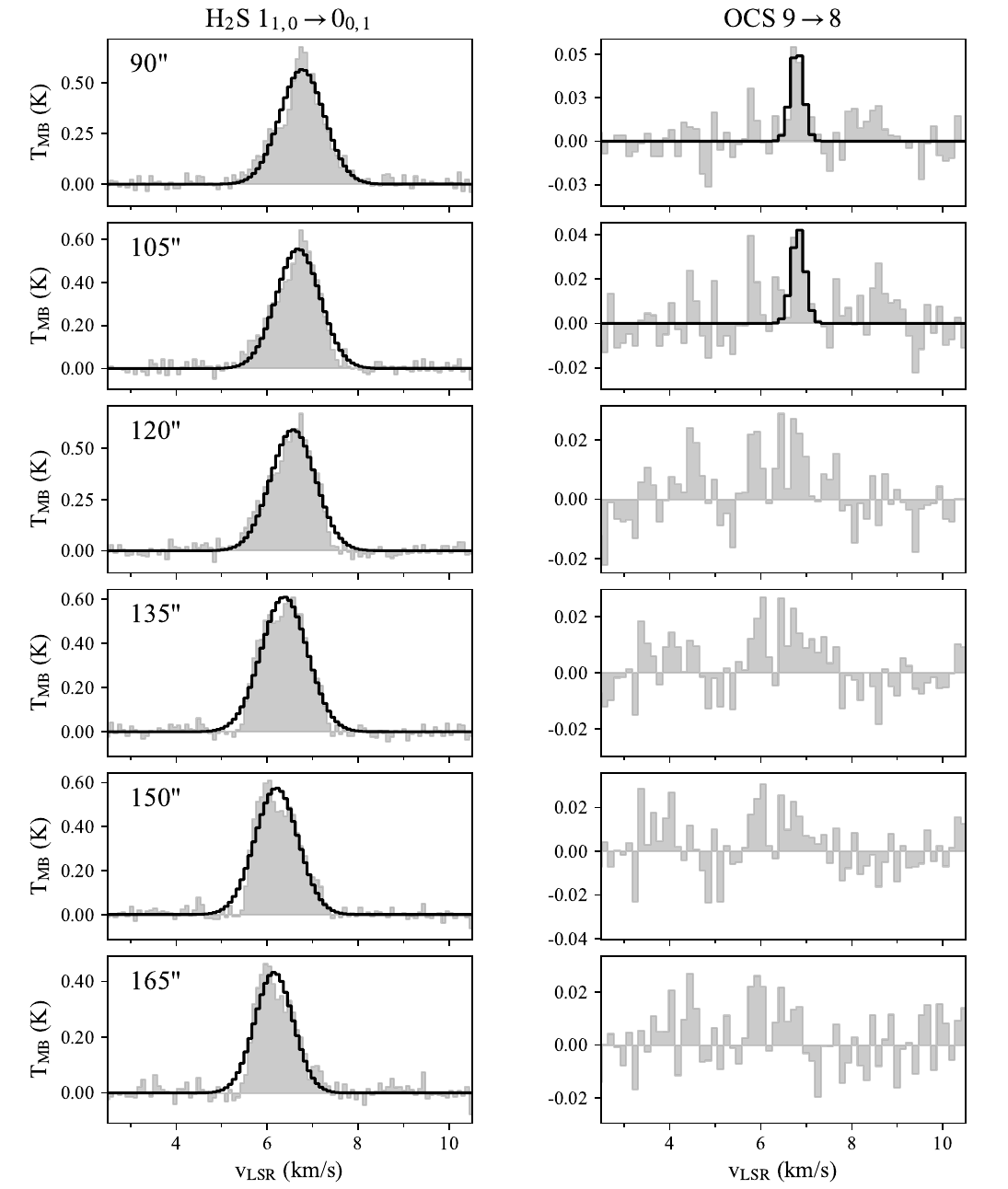}
                \caption{Rotational spectra (gray area) of molecular lines observed at offsets $90''-165''$ with respect to the central position in Barnard 1b (see Table \ref{tab:sources}). The solid black line marks the RADEX fitting of each line while the red solid line corresponds to additional velocity components that were not taken into account in the fitting.}
                \label{fig:spectra_n2hp_ch3oh_30_90_IC348}
            \end{figure}
            \vfill
            \vfill
            \begin{figure}[h]
                \centering
                \includegraphics[width=\textwidth]{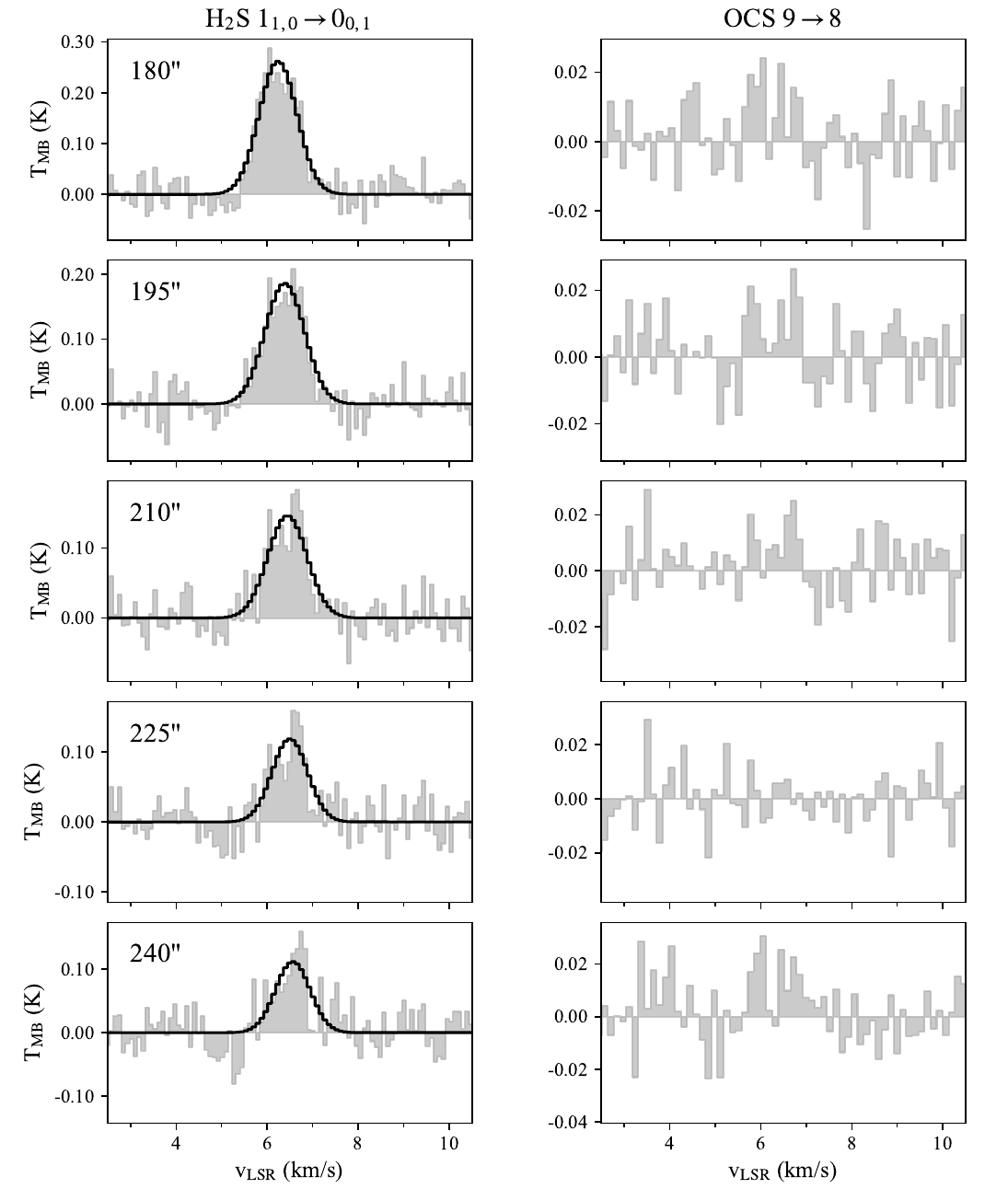}
                \caption{Rotational spectra (gray area) of molecular lines observed at offsets $180''-240''$ with respect to the central position in Barnard 1b (see Table \ref{tab:sources}). The solid black line marks the RADEX fitting of each line while the red solid line corresponds to additional velocity components that were not taken into account in the fitting.}
                \label{fig:spectra_n2hp_ch3oh_30_90_IC348}
            \end{figure}
            \vfill
            \clearpage
        	\newpage
        \section{Line analysis - Barnard 1b}\label{sec:linePropBarnard1b}
        
        \vfill
        
        \begin{table}[h!]
                \centering
                \caption{Properties, main beam temperatures, and integrated intensities of the CH$_{3}$OH rotational transitions at different offsets toward Barnard 1b.}
                \resizebox{\textwidth}{!}{
                \begin{tabular}{lccccrccc}
                    \toprule
				    \multirow{2}{*}{{Offset}} &  \multirow{2}{*}{{Transition}} &  {Frequency} & \multirow{2}{*}{{E$_{\rm up}$} {(K)}} & \multirow{2}{*}{{log(A$_{\rm{ij}}$)}} & \multirow{2}{*}{Peak position (km s$^{-1}$)} & \multirow{2}{*}{Width (km s$^{-1}$)} & \multirow{2}{*}{{T}$_{\rm MB}$ {(K)}} & $\int${T}$_{\rm MB}\ dv$\\
					&  & {(MHz)} & & & & & & {(K km s}$^{{-1}}${)} \\
				    \midrule\midrule
                    \multirow{3}{*}{0"} & E1 $2_{0,2}\rightarrow 1_{0,1}$ & 96744.54  & 20.10 & -5.47 & $-3.17\pm 0.02$ & $0.72\pm 0.05$ & $0.43\pm 0.02$ & $0.33\pm 0.02$\\
                    & A$^{+}$ $2_{0,2}\rightarrow 1_{0,1}$ & 96741.37  & 7.00 & -5.47 & $6.703\pm 0.004$ & $0.859\pm 0.009$ & $2.70\pm 0.02$ & $2.47\pm 0.02$\\
                    & E2 $2_{1,2}\rightarrow 1_{1,1}$ & 96739.36  & 12.50 & -5.59 & $12.946\pm 0.004$ & $0.82\pm 0.01$ & $2.04\pm 0.02$ & $1.79\pm 0.02$ \\
                    \midrule
                    \multirow{3}{*}{15"} & E1 $2_{0,2}\rightarrow 1_{0,1}$ & 96744.54  & 20.10 & -5.47 & $-3.15\pm 0.03$ & $0.86\pm 0.08$ & $0.29\pm 0.02$ & $0.26\pm 0.02$\\
                    & A$^{+}$ $2_{0,2}\rightarrow 1_{0,1}$ & 96741.37  & 7.00 & -5.47 & $6.739\pm 0.005$ & $0.99\pm 0.01$ & $1.97\pm 0.02$ & $2.07\pm 0.02$\\
                    & E2 $2_{1,2}\rightarrow 1_{1,1}$ & 96739.36  & 12.50  & -5.59 & $12.972\pm 0.007$ & $0.95\pm 0.02$ & $1.44\pm 0.02$ & $1.45\pm 0.02$ \\
                    \midrule
                    \multirow{3}{*}{30"} & E1 $2_{0,2}\rightarrow 1_{0,1}$ & 96744.54  & 20.10 & -5.47 & $-3.10\pm 0.03$ & $0.89\pm 0.09$ & $0.10\pm 0.01$ & $0.089\pm 0.007$\\
                    & A$^{+}$ $2_{0,2}\rightarrow 1_{0,1}$ & 96741.37  & 7.00 & -5.47 & $6.739\pm 0.003$ & $0.913\pm 0.008$ & $0.95\pm 0.01$ & $0.918\pm 0.007$\\
                    & E2 $2_{1,2}\rightarrow 1_{1,1}$ & 96739.36  & 12.50  & -5.59 & $12.951\pm 0.005$ & $0.91\pm 0.01$ & $0.67\pm 0.01$ & $0.647\pm 0.007$ \\
                    \midrule
                    \multirow{3}{*}{45"} & E1 $2_{0,2}\rightarrow 1_{0,1}$ & 96744.54  & 20.10 & -5.47 & $-3.10\pm 0.02$ & $0.75\pm 0.06$ & $0.093\pm 0.009$ & $0.074\pm 0.005$\\
                    & A$^{+}$ $2_{0,2}\rightarrow 1_{0,1}$ & 96741.37  & 7.00 & -5.47 & $6.712\pm 0.000$ & $0.852\pm 0.005$ & $0.844\pm 0.009$ & $0.765\pm 0.001$\\
                    & E2 $2_{1,2}\rightarrow 1_{1,1}$ & 96739.36  & 12.50  & -5.59 & $12.949\pm 0.004$ & $0.845\pm 0.009$ & $0.600\pm 0.009$ & $0.539\pm 0.005$ \\
                    \midrule
                    \multirow{3}{*}{60"} & E1 $2_{0,2}\rightarrow 1_{0,1}$ & 96744.54  & 20.10 & -5.47 & $-3.1\pm 0.2$ & $0.5\pm 0.2$ & $0.097\pm 0.009$ & $0.06\pm 0.02$\\
                    & A$^{+}$ $2_{0,2}\rightarrow 1_{0,1}$ & 96741.37  & 7.00 & -5.47 & $6.7\pm 0.2$ & $0.7\pm 0.2$ & $0.742\pm 0.009$ & $0.56\pm 0.02$\\
                    & E2 $2_{1,2}\rightarrow 1_{1,1}$ & 96739.36  & 12.50  & -5.59 & $12.9\pm 0.2$ & $0.8\pm 0.2$ & $0.514\pm 0.009$ & $0.41\pm 0.02$ \\
                    \midrule
                    \multirow{3}{*}{75"} & E1 $2_{0,2}\rightarrow 1_{0,1}$ & 96744.54  & 20.10 & -5.47 & $-3.11\pm 0.03$ & $0.51\pm 0.08$ & $0.080\pm 0.009$ & $0.044\pm 0.005$\\
                    & A$^{+}$ $2_{0,2}\rightarrow 1_{0,1}$ & 96741.37  & 7.00 & -5.47 & $6.723\pm 0.004$ & $0.66\pm 0.01$ & $0.666\pm 0.009$ & $0.466\pm 0.006$\\
                    & E2 $2_{1,2}\rightarrow 1_{1,1}$ & 96739.36  & 12.50  & -5.59 & $12.963\pm 0.006$ & $0.68\pm 0.02$ & $0.467\pm 0.009$ & $0.339\pm 0.006$ \\
                    \midrule
                    \multirow{3}{*}{90"} & E1 $2_{0,2}\rightarrow 1_{0,1}$ & 96744.54  & 20.10 & -5.47 & $-3.08\pm 0.06$ & $0.8\pm 0.2$ & $0.05\pm 0.01$ & $0.044\pm 0.007$\\
                    & A$^{+}$ $2_{0,2}\rightarrow 1_{0,1}$ & 96741.37  & 7.00 & -5.47 & $6.730\pm 0.005$ & $0.73\pm 0.01$ & $0.59\pm 0.01$ & $0.454\pm 0.006$\\
                    & E2 $2_{1,2}\rightarrow 1_{1,1}$ & 96739.36  & 12.50  & -5.59 & $12.970\pm 0.007$ & $0.74\pm 0.02$ & $0.41\pm 0.01$ & $0.327\pm 0.006$\\
                    \midrule
                    \multirow{3}{*}{105"} & E1 $2_{0,2}\rightarrow 1_{0,1}$ & 96744.54  & 20.10 & -5.47 & $-3.09\pm 0.05$ & $0.7\pm 0.1$ & $0.054\pm 0.009$ & $0.041\pm 0.006$\\
                    & A$^{+}$ $2_{0,2}\rightarrow 1_{0,1}$ & 96741.37  & 7.00 & -5.47 & $6.692\pm 0.005$ & $0.84\pm 0.01$ & $0.544\pm 0.009$ & $0.483\pm 0.006$\\
                    & E2 $2_{1,2}\rightarrow 1_{1,1}$ & 96739.36  & 12.50  & -5.59 & $12.928\pm 0.008$ & $0.84\pm 0.02$ & $0.390\pm 0.009$ & $0.346\pm 0.007$\\
                    \midrule
                    \multirow{3}{*}{120"} & E1 $2_{0,2}\rightarrow 1_{0,1}$ & 96744.54  & 20.10 & -5.47 & $-3.11\pm 0.03$ & $0.58\pm 0.09$ & $0.054\pm 0.009$ & $0.034\pm 0.004$\\
                    & A$^{+}$ $2_{0,2}\rightarrow 1_{0,1}$ & 96741.37  & 7.00 & -5.47 & $6.613\pm 0.005$ & $0.95\pm 0.01$ & $0.474\pm 0.009$ & $0.481\pm 0.005$\\
                    & E2 $2_{1,2}\rightarrow 1_{1,1}$ & 96739.36  & 12.50  & -5.59 & $12.844\pm 0.008$ & $0.97\pm 0.02$ & $0.333\pm 0.009$ & $0.342\pm 0.005$\\
                    \midrule
                    \multirow{3}{*}{135"} & E1 $2_{0,2}\rightarrow 1_{0,1}$ & 96744.54  & 20.10 & -5.47 & $-3.2\pm 0.2$ & $0.5\pm 0.2$ & $0.035\pm 0.008$ & $0.02\pm 0.01$\\
                    & A$^{+}$ $2_{0,2}\rightarrow 1_{0,1}$ & 96741.37  & 7.00 & -5.47 & $6.5\pm 0.2$ & $1.0\pm 0.2$ & $0.353\pm 0.008$ & $0.39\pm 0.01$\\
                    & E2 $2_{1,2}\rightarrow 1_{1,1}$ & 96739.36  & 12.50  & -5.59 & $12.7\pm 0.2$ & $1.1\pm 0.2$ & $0.243\pm 0.008$ & $0.30\pm 0.01$\\
                    \midrule
                    \multirow{3}{*}{150"} & E1 $2_{0,2}\rightarrow 1_{0,1}$ & 96744.54  & 20.10 & -5.47 & $-3.22\pm 0.07$ & $0.3\pm 0.1$ & $0.020\pm 0.008$ & $0.007\pm 0.003$\\
                    & A$^{+}$ $2_{0,2}\rightarrow 1_{0,1}$ & 96741.37  & 7.00 & -5.47 & $6.37\pm 0.01$ & $1.12\pm 0.02$ & $0.218\pm 0.008$ & $0.260\pm 0.005$\\
                    & E2 $2_{1,2}\rightarrow 1_{1,1}$ & 96739.36  & 12.50  & -5.59 & $12.597\pm 0.02$ & $1.13\pm 0.03$ & $0.156\pm 0.008$ & $0.188\pm 0.005$\\
                    \midrule
                    \multirow{2}{*}{165"} & A$^{+}$ $2_{0,2}\rightarrow 1_{0,1}$ & 96741.37  & 7.00 & -5.47 & $6.29\pm 0.02$ & $1.06\pm 0.04$ & $0.138\pm 0.009$ & $0.156\pm 0.005$\\
                    & E2 $2_{1,2}\rightarrow 1_{1,1}$ & 96739.36  & 12.50  & -5.59 & $12.52\pm 0.03$ & $1.10\pm 0.05$ & $0.094\pm 0.009$ & $0.109\pm 0.005$\\
                    \midrule
                    \multirow{2}{*}{180"} & A$^{+}$ $2_{0,2}\rightarrow 1_{0,1}$ & 96741.37  & 7.00 & -5.47 & $6.34\pm 0.03$ & $1.06\pm 0.06$ & $0.089\pm 0.009$ & $0.101\pm 0.005$\\
                    & E2 $2_{1,2}\rightarrow 1_{1,1}$ & 96739.36  & 12.50  & -5.59 & $12.53\pm 0.05$ & $1.1\pm 0.1$ & $0.050\pm 0.009$ & $0.061\pm 0.005$\\
                    \midrule
                    \multirow{2}{*}{195"} & A$^{+}$ $2_{0,2}\rightarrow 1_{0,1}$ & 96741.37  & 7.00 & -5.47 & $6.5\pm 0.04$ & $1.1\pm 0.08$ & $0.062\pm 0.008$ & $0.072\pm 0.005$\\
                    & E2 $2_{1,2}\rightarrow 1_{1,1}$ & 96739.36  & 12.50  & -5.59 & $12.5\pm 0.08$ & $1.3\pm 0.2$ & $0.032\pm 0.008$ & $0.044\pm 0.005$\\
                    \midrule
                    \multirow{2}{*}{210"} & A$^{+}$ $2_{0,2}\rightarrow 1_{0,1}$ & 96741.37  & 7.00 & -5.47 & $6.56\pm 0.04$ & $1.0\pm 0.1$ & $0.051\pm 0.008$ & $0.053\pm 0.005$\\
                    & E2 $2_{1,2}\rightarrow 1_{1,1}$ & 96739.36  & 12.50  & -5.59 & $12.7\pm 0.1$ & $1.2\pm 0.2$ & $0.026\pm 0.008$ & $0.035\pm 0.005$\\
                    \midrule
                    \multirow{2}{*}{225"} & A$^{+}$ $2_{0,2}\rightarrow 1_{0,1}$ & 96741.37  & 7.00 & -5.47 & $6.56\pm 0.05$ & $0.9\pm 0.1$ & $0.044\pm 0.009$ & $0.041\pm 0.005$\\
                    & E2 $2_{1,2}\rightarrow 1_{1,1}$ & 96739.36  & 12.50  & -5.59 & $12.9\pm 0.1$ & $1.1\pm 0.4$ & $0.023\pm 0.009$ & $0.028\pm 0.006$\\
                    \midrule
                    \multirow{2}{*}{240"} & A$^{+}$ $2_{0,2}\rightarrow 1_{0,1}$ & 96741.37  & 7.00 & -5.47 & $6.6\pm 0.04$ & $0.4\pm 0.2$ & $0.06\pm 0.01$ & $0.026\pm 0.006$\\
                    & E2 $2_{1,2}\rightarrow 1_{1,1}$ & 96739.36  & 12.50  & -5.59 & $12.850\pm 0.04$ & $0.4\pm 0.1$ & $0.05\pm 0.02$ & $0.019\pm 0.005$\\
    			\bottomrule
			\end{tabular}
            }
			\label{tab:linePropertiesCH3OHB1b}
        \end{table}
        
        \vfill
        
        \begin{table}[h!]
                \centering
                \caption{Properties, main beam temperatures, and integrated intensities of the CH$_{3}$OH rotational transitions observed with the Yebes 40m telescope toward Barnard 1b.}
                \resizebox{0.98\textwidth}{!}{
                \begin{tabular}{lccccrccc}
                    \toprule
				    \multirow{2}{*}{{Offset}} &  \multirow{2}{*}{{Transition}} &  {Frequency} & \multirow{2}{*}{{E$_{\rm up}$} {(K)}} & \multirow{2}{*}{{log(A$_{\rm{ij}}$)}} & \multirow{2}{*}{Peak position (km s$^{-1}$)} & \multirow{2}{*}{Width (km s$^{-1}$)} & \multirow{2}{*}{{T}$_{\rm MB}$ {(K)}} & $\int${T}$_{\rm MB}\ dv$\\
					&  & {(MHz)} & & & & & & {(K km s}$^{{-1}}${)} \\
				    \midrule\midrule
				    $150"$ & A$^{+}$ $1_{0,1}\rightarrow 0_{0,0}$ & 48372.46  & 2.30 & -6.45 & $6.38\pm 0.06$ & $1.4\pm 0.1$ & $0.21\pm 0.02$ & $0.33\pm 0.03$\\
                    \midrule
                    $180"$ & A$^{+}$ $1_{0,1}\rightarrow 0_{0,0}$ & 48372.46  & 2.30 & -6.45 & $6.39\pm 0.08$ & $1.1\pm 0.2$ & $0.15\pm 0.04$ & $0.18\pm 0.04$\\
                    \midrule
                    $240"$ & A$^{+}$ $1_{0,1}\rightarrow 0_{0,0}$ & 48372.46  & 2.30 & -6.45 & \multicolumn{4}{c}{rms $= 0.01$} \\
    			\bottomrule
			\end{tabular}
            }
			\label{tab:linePropertiesCH3OHB1b40m}
        \end{table}
        
        \begin{table}[h!]
    		\centering
            \caption{Properties, main beam temperatures, and integrated intensities of the C$^{18}$O $1\rightarrow0$ and H$_{2}$S $1_{1,0}\rightarrow1_{0,1}$ rotational transitions at different offsets toward Barnard 1b.}
                \resizebox{0.96\textwidth}{!}{
                \begin{tabular}{lccccrccc}
                    \toprule
				    \multirow{2}{*}{{Offset}} &  \multirow{2}{*}{{Transition}} &  {Frequency} & \multirow{2}{*}{{E$_{\rm up}$} {(K)}} & \multirow{2}{*}{{log(A$_{\rm{ij}}$)}} & \multirow{2}{*}{Peak position (km s$^{-1}$)} & \multirow{2}{*}{Width (km s$^{-1}$)} & \multirow{2}{*}{{T}$_{\rm MB}$ {(K)}} & $\int${T}$_{\rm MB}\ dv$\\
					&  & {(MHz)} & & & & & & {(K km s}$^{{-1}}${)} \\
				    \midrule\midrule
				    \multicolumn{9}{c}{C$^{18}$O $1\rightarrow 0$}\\
				    \midrule
                    30" & $1\rightarrow0$ & 109782.17 & 5.27 & -7.20 & $6.614\pm 0.001$ & $1.142\pm 0.002$ & $4.28\pm 0.02$ & $5.21\pm 0.02$ \\
                    \midrule
                    45" & $1\rightarrow0$ & 109782.17 & 5.27 & -7.20 & $6.607\pm 0.002$ & $1.111\pm 0.004$ & $4.02\pm 0.02$ & $4.75\pm 0.02$ \\
                    \midrule
                    60" & $1\rightarrow0$ & 109782.17 & 5.27 & -7.20 & $6.578\pm 0.002$ & $1.122\pm 0.006$ & $3.48\pm 0.02$ & $4.16\pm 0.02$ \\
                    \midrule
                    75" & $1\rightarrow0$ & 109782.17 & 5.27 & -7.20 & $6.486\pm 0.002$ & $1.245\pm 0.004$ & $3.04\pm 0.02$ & $4.03\pm 0.02$ \\
                    \midrule
                    90" & $1\rightarrow0$ & 109782.17 & 5.27 & -7.20 & $6.392\pm 0.002$ & $1.255\pm 0.004$ & $3.12\pm 0.02$ & $4.17\pm 0.01$ \\
                    \midrule
                    105" & $1\rightarrow0$ & 109782.17 & 5.27 & -7.20 & $6.335\pm 0.000$ & $1.244\pm 0.004$ & $3.32\pm 0.02$ & $4.40\pm 0.01$ \\
                    \midrule
                    120" & $1\rightarrow0$ & 109782.17 & 5.27 & -7.20 & $6.247\pm 0.002$ & $1.237\pm 0.004$ & $3.41\pm 0.02$ & $4.50\pm 0.01$ \\
                    \midrule
                    135" & $1\rightarrow0$ & 109782.17 & 5.27 & -7.20 & $6.149\pm 0.000$ & $1.156\pm 0.004$ & $3.22\pm 0.02$ & $3.96\pm 0.01$ \\
                    \midrule
                    150" & $1\rightarrow0$ & 109782.17 & 5.27 & -7.20 & $6.077\pm 0.002$ & $1.075\pm 0.005$ & $2.55\pm 0.02$ & $2.92\pm 0.01$ \\
                    \midrule
                    165" & $1\rightarrow0$ & 109782.17 & 5.27 & -7.20 & $6.060\pm 0.003$ & $1.044\pm 0.006$ & $1.78\pm 0.02$ & $1.98\pm 0.01$ \\
                    \midrule
                    180" & $1\rightarrow0$ & 109782.17 & 5.27 & -7.20 & $6.115\pm 0.004$ & $1.114\pm 0.008$ & $1.19\pm 0.02$ & $1.41\pm 0.01$ \\
                    \midrule
                    195" & $1\rightarrow0$ & 109782.17 & 5.27 & -7.20 & $6.187\pm 0.006$ & $1.21\pm 0.01$ & $0.86\pm 0.02$ & $1.11\pm 0.01$ \\
                    \midrule
                    210" & $1\rightarrow0$ & 109782.17 & 5.27 & -7.20 & $6.243\pm 0.007$ & $1.26\pm 0.02$ & $0.68\pm 0.02$ & $0.91\pm 0.01$ \\
                    \midrule
                    225" & $1\rightarrow0$ & 109782.17 & 5.27 & -7.20 & $6.327\pm 0.008$ & $1.25\pm 0.02$ & $0.60\pm 0.02$ & $0.80\pm 0.01$ \\
                    \midrule
                    240" & $1\rightarrow0$ & 109782.17 & 5.27 & -7.20 & $6.396\pm 0.009$ & $1.19\pm 0.02$ & $0.60\pm 0.02$ & $0.76\pm 0.01$ \\
    			\midrule
    			\multicolumn{9}{c}{H$_{2}$S $1_{1,0}\rightarrow1_{0,1}$}\\
    			\midrule
    			0" & $1_{1,0}\rightarrow1_{0,1}$ & 168762.76 & 27.9 & -4.57 & $6.58\pm 0.02$ & $1.25\pm 0.05$ & $1.26\pm 0.09$ & $1.68\pm 0.05$ \\
                    \midrule
				    15" & $1_{1,0}\rightarrow1_{0,1}$ & 168762.76 & 27.9 & -4.57 & $6.59\pm 0.02$ & $1.19\pm 0.04$ & $1.4\pm 0.1$ & $1.79\pm 0.05$ \\
                    \midrule
                    30" & $1_{1,0}\rightarrow1_{0,1}$ & 168762.76 & 27.9 & -4.57 & $6.684\pm 0.008$ & $1.15\pm 0.02$ & $1.00\pm 0.04$ & $1.22\pm 0.02$ \\
                    \midrule
                    45" & $1_{1,0}\rightarrow1_{0,1}$ & 168762.76 & 27.9 & -4.57 & $6.723\pm 0.007$ & $1.07\pm 0.02$ & $0.89\pm 0.03$ & $1.01\pm 0.01$ \\
                    \midrule
                    60" & $1_{1,0}\rightarrow1_{0,1}$ & 168762.76 & 27.9 & -4.57 & $6.768\pm 0.006$ & $0.97\pm 0.02$ & $0.83\pm 0.03$ & $0.85\pm 0.01$ \\
                    \midrule
                    75" & $1_{1,0}\rightarrow1_{0,1}$ & 168762.76 & 27.9 & -4.57 & $6.801\pm 0.008$ & $1.03\pm 0.02$ & $0.70\pm 0.03$ & $0.77\pm 0.01$ \\
                    \midrule
                    90" & $1_{1,0}\rightarrow1_{0,1}$ & 168762.76 & 27.9 & -4.57 & $6.77\pm 0.01$ & $1.15\pm 0.03$ & $0.57\pm 0.03$ & $0.69\pm 0.01$ \\
                    \midrule
                    105" & $1_{1,0}\rightarrow1_{0,1}$ & 168762.76 & 27.9 & -4.57 & $6.68\pm 0.01$ & $1.13\pm 0.03$ & $0.56\pm 0.03$ & $0.67\pm 0.01$ \\
                    \midrule
                    120" & $1_{1,0}\rightarrow1_{0,1}$ & 168762.76 & 27.9 & -4.57 & $6.57\pm 0.01$ & $1.15\pm 0.02$ & $0.59\pm 0.03$ & $0.73\pm 0.01$ \\
                    \midrule
                    135" & $1_{1,0}\rightarrow1_{0,1}$ & 168762.76 & 27.9 & -4.57 & $6.37\pm 0.01$ & $1.19\pm 0.02$ & $0.61\pm 0.03$ & $0.77\pm 0.01$ \\
                    \midrule
                    150" & $1_{1,0}\rightarrow1_{0,1}$ & 168762.76 & 27.9 & -4.57 & $6.20\pm 0.01$ & $1.11\pm 0.02$ & $0.58\pm 0.03$ & $0.68\pm 0.01$ \\
                    \midrule
                    165" & $1_{1,0}\rightarrow1_{0,1}$ & 168762.76 & 27.9 & -4.57 & $6.15\pm 0.01$ & $0.99\pm 0.03$ & $0.43\pm 0.03$ & $0.45\pm 0.01$ \\
                    \midrule
                    180" & $1_{1,0}\rightarrow1_{0,1}$ & 168762.76 & 27.9 & -4.57 & $6.24\pm 0.02$ & $0.98\pm 0.04$ & $0.26\pm 0.03$ & $0.27\pm 0.01$ \\
                    \midrule
                    195" & $1_{1,0}\rightarrow1_{0,1}$ & 168762.76 & 27.9 & -4.57 & $6.38\pm 0.03$ & $1.02\pm 0.06$ & $0.19\pm 0.03$ & $0.20\pm 0.01$ \\
                    \midrule
                    210" & $1_{1,0}\rightarrow1_{0,1}$ & 168762.76 & 27.9 & -4.57 & $6.44\pm 0.04$ & $1.01\pm 0.09$ & $0.15\pm 0.03$ & $0.16\pm 0.01$ \\
                    \midrule
                    225" & $1_{1,0}\rightarrow1_{0,1}$ & 168762.76 & 27.9 & -4.57 & $6.50\pm 0.05$ & $0.9\pm 0.1$ & $0.12\pm 0.03$ & $0.12\pm 0.01$ \\
                    \midrule
                    240" & $1_{1,0}\rightarrow1_{0,1}$ & 168762.76 & 27.9 & -4.57 & $6.56\pm 0.07$ & $0.9\pm 0.2$ & $0.11\pm 0.04$ & $0.11\pm 0.02$ \\
    			\bottomrule
			\end{tabular}
            }
            
			\label{tab:linePropertiesC18OB1b}
        \end{table}
        
        \begin{table}[h!]
                \centering
                \caption{Properties, main beam temperatures, and integrated intensities of the observed N$_{2}$H$^{+}$ $1\rightarrow0$ rotational transition at different offsets toward Barnard 1b.}
                \resizebox{\textwidth}{!}{
                \begin{tabular}{lccccrcccc}
                    \toprule
				    \multirow{2}{*}{{Offset}} &  \multirow{2}{*}{{Transition}} &  {Frequency} & \multirow{2}{*}{{E$_{\rm up}$} {(K)}} & \multirow{2}{*}{{log(A$_{\rm{ij}}$)}} & \multirow{2}{*}{Peak position (km s$^{-1}$)} & \multirow{2}{*}{Width (km s$^{-1}$)} & \multirow{2}{*}{{T}$_{\rm MB}$ {(K)}} & $\int${T}$_{\rm MB}\ dv$ & \multirow{2}{*}{${\tau_{\rm main}}$}\\
					&  & {(MHz)} & & & & & & {(K km s}$^{{-1}}${)} & \\
				    \midrule\midrule
                    \multirow{5}{*}{30"} & \multirow{5}{*}{$1\rightarrow0$} & \multirow{5}{*}{93173.40} & \multirow{5}{*}{4.47} & \multirow{5}{*}{-4.44} & $-2.6\pm 0.2$ & $0.7\pm 0.2$ & $0.79\pm 0.01$ & $0.63\pm 0.07$ & \multirow{5}{*}{$1.04\pm 0.02$}\\
                        & & & & & $5.3\pm 0.2$ & $1.2\pm 0.2$ & $1.44\pm 0.01$ & $1.86\pm 0.07$ & \\
                        & & & & & $6.4\pm 0.2$ & $0.6\pm 0.2$ & $0.99\pm 0.01$ & $0.62\pm 0.07$ & \\
                        & & & & & $11.2\pm 0.2$ & $1.1\pm 0.2$ & $1.35\pm 0.01$ & $1.46\pm 0.07$ & \\
                        & & & & & $12.4\pm 0.2$ & $0.5\pm 0.2$ & $0.28\pm 0.01$ & $0.14\pm 0.07$ & \\
                    \midrule
                    \multirow{5}{*}{45"} & \multirow{5}{*}{$1\rightarrow0$} & \multirow{5}{*}{93173.40} & \multirow{5}{*}{4.47} & \multirow{5}{*}{-4.44} & $-2.6\pm 0.2$ & $0.8\pm 0.2$ & $0.635\pm 0.007$ & $0.52\pm 0.05$ & \multirow{5}{*}{$0.94\pm 0.03$}\\
                        & & & & & $5.3\pm 0.2$ & $1.1\pm 0.2$ & $1.201\pm 0.007$ & $1.42\pm 0.05$ & \\
                        & & & & & $6.4\pm 0.2$ & $0.6\pm 0.2$ & $0.900\pm 0.007$ & $0.58\pm 0.05$ & \\
                        & & & & & $11.2\pm 0.2$ & $0.9\pm 0.2$ & $1.142\pm 0.007$ & $1.14\pm 0.05$ & \\
                        & & & & & $12.4\pm 0.2$ & $0.5\pm 0.2$ & $0.242\pm 0.007$ & $0.13\pm 0.07$ & \\
                    \midrule
                    \multirow{5}{*}{60"} & \multirow{5}{*}{$1\rightarrow0$} & \multirow{5}{*}{93173.40} & \multirow{5}{*}{4.47} & \multirow{5}{*}{-4.44} & $-2.6\pm 0.2$ & $0.6\pm 0.2$ & $0.479\pm 0.007$ & $0.29\pm 0.04$ & \multirow{5}{*}{$0.71\pm 0.04$}\\
                        & & & & & $5.3\pm 0.2$ & $1.0\pm 0.2$ & $0.802\pm 0.007$ & $0.90\pm 0.04$ & \\
                        & & & & & $6.4\pm 0.2$ & $0.5\pm 0.2$ & $0.683\pm 0.007$ & $0.34\pm 0.04$ & \\
                        & & & & & $11.2\pm 0.2$ & $0.8\pm 0.2$ & $0.792\pm 0.007$ & $0.70\pm 0.04$ & \\
                        & & & & & $12.4\pm 0.2$ & $0.5\pm 0.2$ & $0.174\pm 0.007$ & $0.10\pm 0.04$ & \\
                    \midrule
                    \multirow{5}{*}{75"} & \multirow{5}{*}{$1\rightarrow0$} & \multirow{5}{*}{93173.40} & \multirow{5}{*}{4.47} & \multirow{5}{*}{-4.44} & $-2.530\pm 0.005$ & $0.59\pm 0.01$ & $0.282\pm 0.007$ & $0.177\pm 0.003$ & \multirow{5}{*}{$0.54\pm 0.05$}\\
                        & & & & & $5.353\pm 0.004$ & $1.01\pm 0.01$ & $0.520\pm 0.007$ & $0.560\pm 0.005$ & \\
                        & & & & & $6.476\pm 0.004$ & $0.505\pm 0.008$ & $0.408\pm 0.007$ & $0.219\pm 0.003$ & \\
                        & & & & & $11.286\pm 0.004$ & $0.846\pm 0.009$ & $0.493\pm 0.007$ & $0.443\pm 0.004$ & \\
                        & & & & & $12.46\pm 0.01$ & $0.48\pm 0.03$ & $0.114\pm 0.007$ & $0.059\pm 0.003$ & \\
                    \midrule
                    \multirow{6}{*}{90"} & \multirow{6}{*}{$1\rightarrow0$} & \multirow{6}{*}{93173.40} & \multirow{6}{*}{4.47} & \multirow{6}{*}{-4.44} & $-2.37\pm 0.07$ & $0.5\pm 0.2$ & $0.266\pm 0.007$ & $0.13\pm 0.04$ & \multirow{6}{*}{$0.696\pm 0.006$}\\
                        & & & & & $5.08\pm 0.03$ & $0.53\pm 0.09$ & $0.308\pm 0.007$ & $0.18\pm 0.03$ & \\
                        & & & & & $5.67\pm 0.01$ & $0.40\pm 0.04$ & $0.543\pm 0.007$ & $0.23\pm 0.02$ & \\
                        & & & & & $6.59\pm 0.02$ & $0.45\pm 0.04$ & $0.394\pm 0.007$ & $0.19\pm 0.01$ & \\
                        & & & & & $11.42\pm 0.01$ & $0.83\pm 0.05$ & $0.390\pm 0.007$ & $0.34\pm 0.02$ & \\
                        & & & & & $12.60\pm 0.06$ & $0.4\pm 0.1$ & $0.105\pm 0.007$ & $0.045\pm 0.009$ & \\
                        \midrule
                    \multirow{6}{*}{105"} & \multirow{6}{*}{$1\rightarrow0$} & \multirow{6}{*}{93173.40} & \multirow{6}{*}{4.47} & \multirow{6}{*}{-4.44} & $-2.32\pm 0.05$ & $0.4\pm 0.1$ & $0.270\pm 0.007$ & $0.11\pm 0.03$ & \multirow{6}{*}{$0.68\pm 0.08$}\\
                        & & & & & $5.08\pm 0.02$ & $0.41\pm 0.06$ & $0.285\pm 0.007$ & $0.12\pm 0.01$ & \\
                        & & & & & $5.69\pm 0.01$ & $0.36\pm 0.03$ & $0.548\pm 0.007$ & $0.21\pm 0.01$ & \\
                        & & & & & $6.63\pm 0.01$ & $0.36\pm 0.03$ & $0.403\pm 0.007$ & $0.15\pm 0.01$ & \\
                        & & & & & $11.47\pm 0.02$ & $0.79\pm 0.05$ & $0.334\pm 0.007$ & $0.28\pm 0.02$ & \\
                        & & & & & $12.63\pm 0.04$ & $0.33\pm 0.09$ & $0.114\pm 0.007$ & $0.04\pm 0.01$ & \\
                        \midrule
                    \multirow{6}{*}{120"} & \multirow{6}{*}{$1\rightarrow0$} & \multirow{6}{*}{93173.40} & \multirow{6}{*}{4.47} & \multirow{6}{*}{-4.44} & $-2.32\pm 0.05$ & $0.4\pm 0.1$ & $0.173\pm 0.007$ & $0.06\pm 0.02$ & \multirow{6}{*}{$0.5\pm 0.1$}\\
                        & & & & & $5.09\pm 0.02$ & $0.46\pm 0.07$ & $0.173\pm 0.007$ & $0.09\pm 0.01$ & \\
                        & & & & & $5.69\pm 0.01$ & $0.35\pm 0.03$ & $0.352\pm 0.007$ & $0.130\pm 0.009$ & \\
                        & & & & & $6.64\pm 0.01$ & $0.34\pm 0.03$ & $0.259\pm 0.007$ & $0.094\pm 0.007$ & \\
                        & & & & & $11.47\pm 0.02$ & $0.82\pm 0.05$ & $0.205\pm 0.007$ & $0.18\pm 0.01$ & \\
                        & & & & & $12.64\pm 0.04$ & $0.29\pm 0.08$ & $0.071\pm 0.007$ & $0.022\pm 0.006$ & \\
                        \midrule
%                    \multirow{6}{*}{135"} & \multirow{6}{*}{$1\rightarrow0$} & \multirow{6}{*}{93173.40} & \multirow{6}{*}{4.47} & \multirow{6}{*}{-4.44} & $-2.4\pm 0.2$ & $0.5\pm 0.2$ & $0.057\pm 0.007$ & $0.029\pm 0.004$ & \multirow{6}{*}{$<0.1$}\\
%                        & & & & & $5.1\pm 0.2$ & $0.8\pm 0.2$ & $0.065\pm 0.007$ & $0.054\pm 0.004$ & \\
%                        & & & & & $5.7\pm 0.2$ & $0.3\pm 0.2$ & $0.127\pm 0.007$ & $0.045\pm 0.004$ & \\
%                        & & & & & $6.6\pm 0.2$ & $0.4\pm 0.2$ & $0.090\pm 0.007$ & $0.039\pm 0.004$ & \\
%                        & & & & & $11.4\pm 0.2$ & $1.0\pm 0.2$ & $0.075\pm 0.007$ & $0.08\pm 0.01$ & \\
%                        & & & & & $12.71\pm 0.06$ & $0.2\pm 0.2$ & $0.040\pm 0.007$ & $0.067\pm 0.006$ & \\
%                        \midrule
                        \multirow{6}{*}{135"} & \multirow{6}{*}{$1\rightarrow0$} & \multirow{6}{*}{93173.40} & \multirow{6}{*}{4.47} & \multirow{6}{*}{-4.44} & $-2.4\pm 0.2$ & $0.5\pm 0.2$ & $0.057\pm 0.007$ & $0.029\pm 0.004$ & \multirow{6}{*}{$<0.1$}\\
                        & & & & & $5.1\pm 0.2$ & $0.8\pm 0.2$ & $0.065\pm 0.007$ & $0.054\pm 0.004$ & \\
                        & & & & & $5.7\pm 0.2$ & $0.3\pm 0.2$ & $0.127\pm 0.007$ & $0.045\pm 0.004$ & \\
                        & & & & & $6.6\pm 0.2$ & $0.4\pm 0.2$ & $0.090\pm 0.007$ & $0.039\pm 0.004$ & \\
                        & & & & & $11.4\pm 0.2$ & $1.0\pm 0.2$ & $0.075\pm 0.007$ & $0.08\pm 0.01$ & \\
                        & & & & & $12.71\pm 0.06$ & $0.2\pm 0.2$ & $0.040\pm 0.007$ & $0.067\pm 0.006$ & \\
                        \midrule
                        \multirow{5}{*}{150"} & \multirow{5}{*}{$1\rightarrow0$} & \multirow{5}{*}{93173.40} & \multirow{5}{*}{4.47} & \multirow{5}{*}{-4.44} & $-2.7\pm 0.2$ & $0.8\pm 0.2$ & $0.011\pm 0.007$ & $0.009\pm 0.003$ & \multirow{5}{*}{$<0.1$}\\
                        & & & & & $5.04\pm 0.06$ & $0.6\pm 0.2$ & $0.029\pm 0.007$ & $0.017\pm 0.004$ & \\
                        & & & & & $5.75\pm 0.04$ & $0.40\pm 0.09$ & $0.036\pm 0.007$ & $0.015\pm 0.003$ & \\
                        & & & & & $6.51\pm 0.04$ & $0.36\pm 0.06$ & $0.029\pm 0.007$ & $0.011\pm 0.002$ & \\
                        & & & & & $11.40\pm 0.09$ & $1.4\pm 0.2$ & $0.024\pm 0.007$ & $0.035\pm 0.004$ & \\
                        \midrule
                        \multirow{2}{*}{165"} & \multirow{2}{*}{$1\rightarrow0$} & \multirow{2}{*}{93173.40} & \multirow{2}{*}{4.47} & \multirow{2}{*}{-4.44} & $5.1\pm 0.1$ & $1.0\pm 0.4$ & $0.015\pm 0.007$ & $0.015\pm 0.004$ & \multirow{2}{*}{$<0.1$}\\
                        & & & & & $6.56\pm 0.07$ & $0.3\pm 0.1$ & $0.016\pm 0.007$ & $0.006\pm 0.002$ & \\
                        \midrule
                        $180"$ & $1\rightarrow0$ & 93173.40 & 4.47 & -4.44 & & & rms $= 0.006$ & & \\
                        \midrule
                        $195"$ & $1\rightarrow0$ & 93173.40 & 4.47 & -4.44 & & & rms $= 0.006$ & & \\
                        \midrule
                        $210"$ & $1\rightarrow0$ & 93173.40 & 4.47 & -4.44 & & & rms $= 0.007$ & & \\
                        \midrule
                        $225"$ & $1\rightarrow0$ & 93173.40 & 4.47 & -4.44 & & & rms $= 0.007$ & & \\
                        \midrule
                        $240"$ & $1\rightarrow0$ & 93173.40 & 4.47 & -4.44 & & & rms $= 0.01$ & & \\
    			\bottomrule
			\end{tabular}
            }
			\label{tab:linePropertiesN2H+B1b}
        \end{table}
        
        \vfill
        
        \begin{table}[h!]
                \centering
                \caption{Properties, main beam temperatures, and integrated intensities of the OCS $9\rightarrow8$ rotational transition at different offsets toward Barnard 1b.}
                \resizebox{\textwidth}{!}{
                \begin{tabular}{lccccrccc}
                    \toprule
				    \multirow{2}{*}{{Offset}} &  \multirow{2}{*}{{Transition}} &  {Frequency} & \multirow{2}{*}{{E$_{\rm up}$} {(K)}} & \multirow{2}{*}{{log(A$_{\rm{ij}}$)}} & \multirow{2}{*}{Peak position (km s$^{-1}$)} & \multirow{2}{*}{Width (km s$^{-1}$)} & \multirow{2}{*}{{T}$_{\rm MB}$ {(K)}} & $\int${T}$_{\rm MB}\ dv$\\
					&  & {(MHz)} & & & & & & {(K km s}$^{{-1}}${)} \\
				    \midrule\midrule
                    30" & $9\rightarrow8$ & 109463.06  & 26.27 & -5.43 & $6.75\pm 0.06$ & $1.0\pm 0.2$ & $0.06\pm 0.02$ & $0.062\pm 0.008$\\
                    \midrule
                    45" & $9\rightarrow8$ & 109463.06  & 26.27 & -5.43 & $6.76\pm 0.07$ & $1.0\pm 0.2$ & $0.04\pm 0.01$ & $0.044\pm 0.007$\\
                    \midrule
                    60" & $9\rightarrow8$ & 109463.06  & 26.27 & -5.43 & $6.84\pm 0.05$ & $0.6\pm 0.1$ & $0.04\pm 0.01$ & $0.027\pm 0.005$\\
                    \midrule
                    75" & $9\rightarrow8$ & 109463.06  & 26.27 & -5.43 & $6.80\pm 0.03$ & $0.47\pm 0.09$ & $0.06\pm 0.01$ & $0.027\pm 0.004$\\
                    \midrule
                    90" & $9\rightarrow8$ & 109463.06  & 26.27 & -5.43 & $6.79\pm 0.03$ & $0.3\pm 0.1$ & $0.05\pm 0.01$ & $0.019\pm 0.004$\\
                    \midrule
                    105" & $9\rightarrow8$ & 109463.06  & 26.27 & -5.43 & $6.82\pm 0.04$ & $0.4\pm 0.1$ & $0.04\pm 0.01$ & $0.016\pm 0.004$\\
                    \midrule
                    120" & $9\rightarrow8$ & 109463.06  & 26.27 & -5.43 & \multicolumn{4}{c}{rms $=0.01$} \\
                    \midrule
                    135" & $9\rightarrow8$ & 109463.06  & 26.27 & -5.43 & \multicolumn{4}{c}{rms $=0.01$} \\
                    \midrule
                    150" & $9\rightarrow8$ & 109463.06  & 26.27 & -5.43 & \multicolumn{4}{c}{rms $=0.01$} \\
                    \midrule
                    165" & $9\rightarrow8$ & 109463.06  & 26.27 & -5.43 & \multicolumn{4}{c}{rms $=0.01$} \\
                    \midrule
                    180" & $9\rightarrow8$ & 109463.06  & 26.27 & -5.43 & \multicolumn{4}{c}{rms $=0.01$}\\
                    \midrule
                    195" & $9\rightarrow8$ & 109463.06  & 26.27 & -5.43 & \multicolumn{4}{c}{rms $=0.01$} \\
                    \midrule
                    210" & $9\rightarrow8$ & 109463.06  & 26.27 & -5.43 & \multicolumn{4}{c}{rms $=0.01$} \\
                    \midrule
                    225" & $9\rightarrow8$ & 109463.06  & 26.27 & -5.43 & \multicolumn{4}{c}{rms $=0.01$} \\
                    \midrule
                    240" & $9\rightarrow8$ & 109463.06  & 26.27 & -5.43 & \multicolumn{4}{c}{rms $=0.02$}\\
    			\bottomrule
			\end{tabular}
            }
			\label{tab:linePropertiesOCSB1b}
        \end{table}
      
      \vfill
      
      	\newpage
      	\clearpage
        \section{Molecular spectra - IC348}\label{sec:spectraIC348}
        
            \vfill
            \begin{figure}[h]
                \centering
                \includegraphics[width=0.99\textwidth]{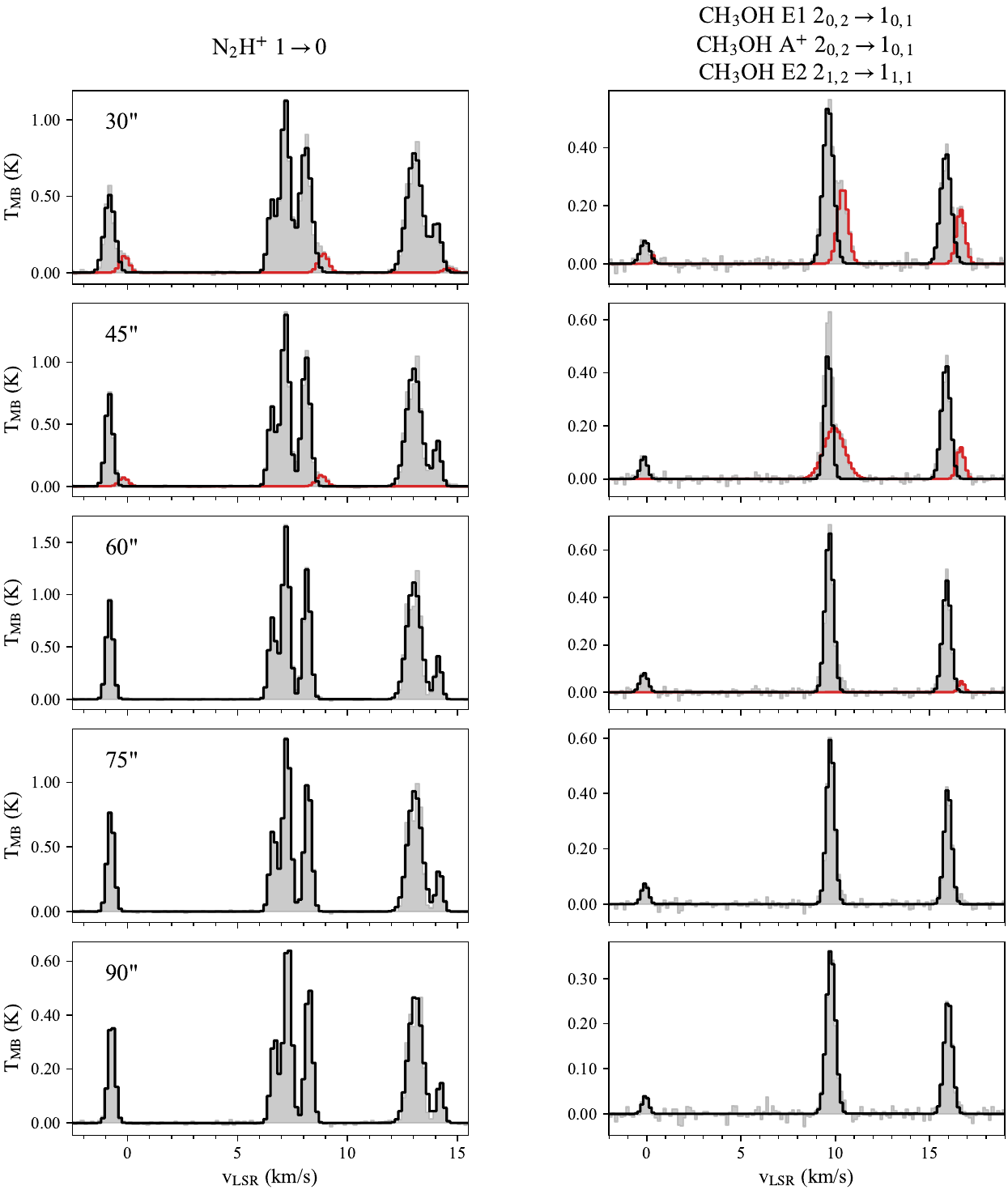}
                \caption{Rotational spectra (gray area) of molecular lines observed at offsets $30''-90''$ with respect to the central position in IC348 (see Table \ref{tab:sources}). The solid black line marks the RADEX fitting of each line while the red solid line corresponds to additional velocity components that were not taken into account in the fitting.}
                \label{fig:spectra_n2hp_ch3oh_30_90_IC348}
            \end{figure}
            \clearpage
            \begin{figure}[h]
                \centering
                \includegraphics[width=\textwidth]{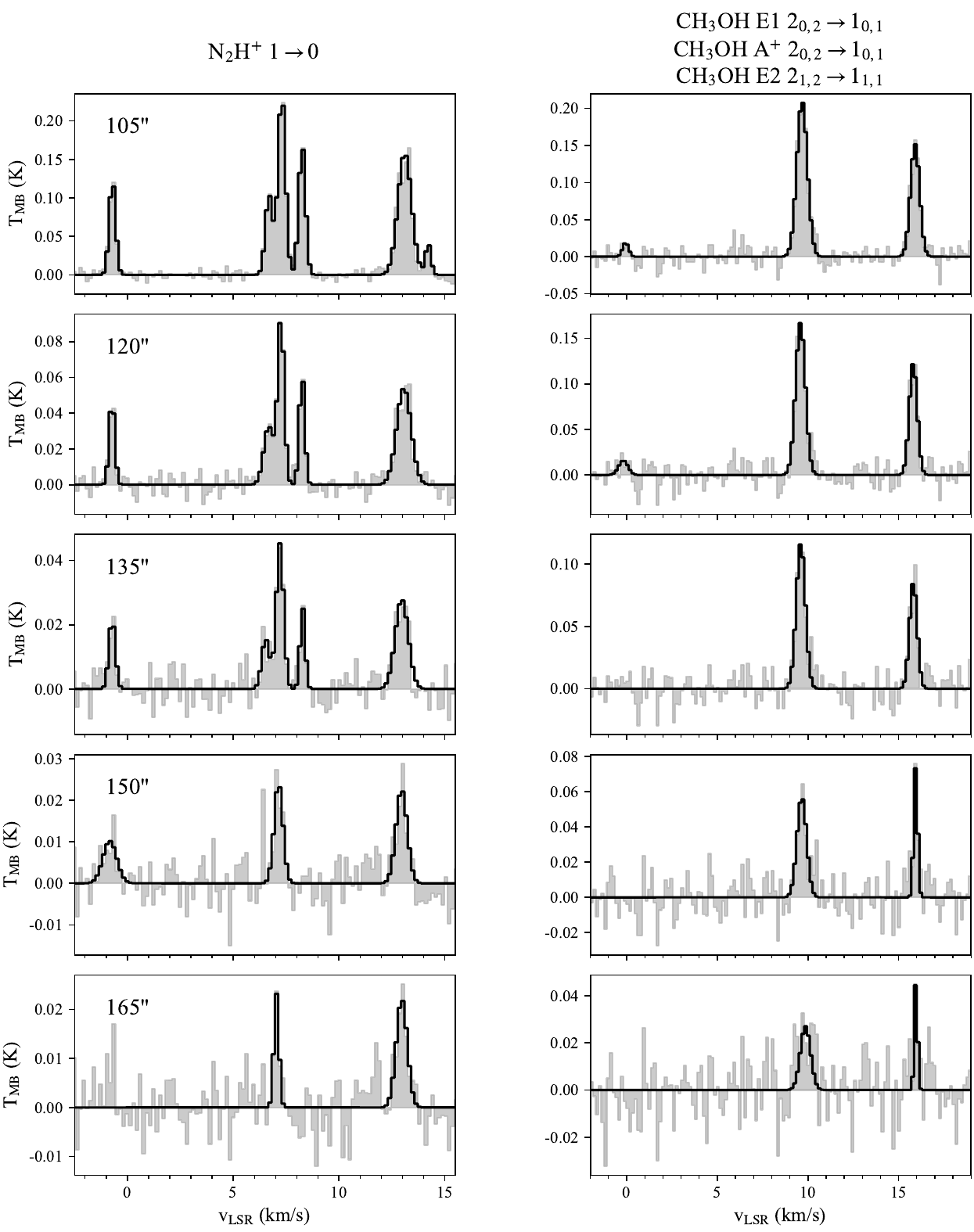}
                \caption{Rotational spectra (gray area) of the molecular lines observed at offsets $105''-165''$ with respect to the central position in IC348 (see Table \ref{tab:sources}). The solid black line marks the RADEX fitting of each line.}
                \label{fig:spectra_n2hp_ch3oh_105_165_IC348}
            \end{figure}
            \vfill
            \clearpage
            \begin{figure}[h]
                \centering
                \includegraphics[width=\textwidth]{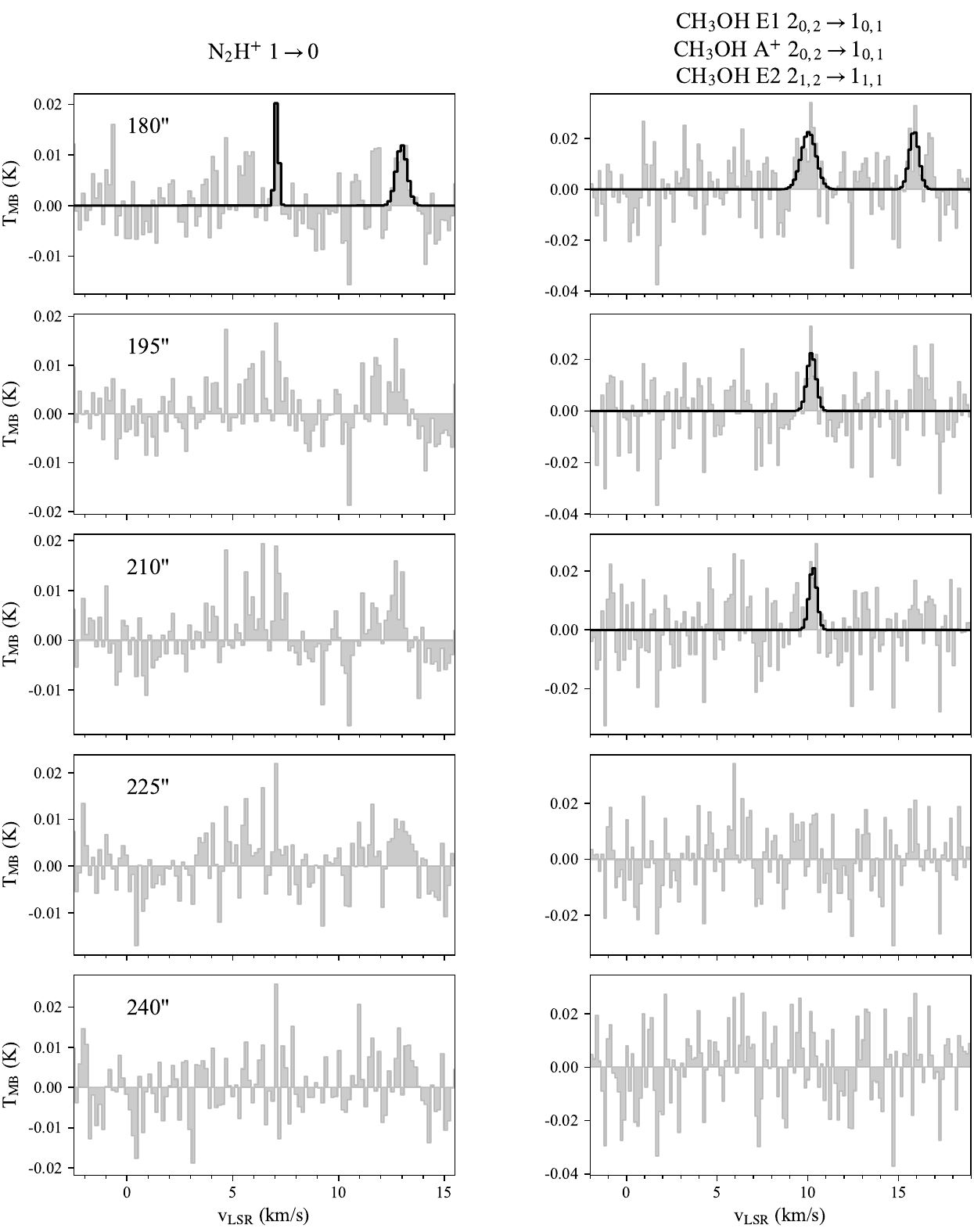}
                \caption{Rotational spectra (gray area) of the molecular lines observed at offsets $180''-240''$ with respect to the central position in IC348 (see Table \ref{tab:sources}). The solid black line marks the RADEX fitting of each line.}
                \label{fig:spectra_n2hp_ch3oh_180_240_IC348}
            \end{figure}
            \vfill
            \clearpage
            \vfill
            \begin{figure}[h]
                \centering
                \includegraphics[width=\textwidth]{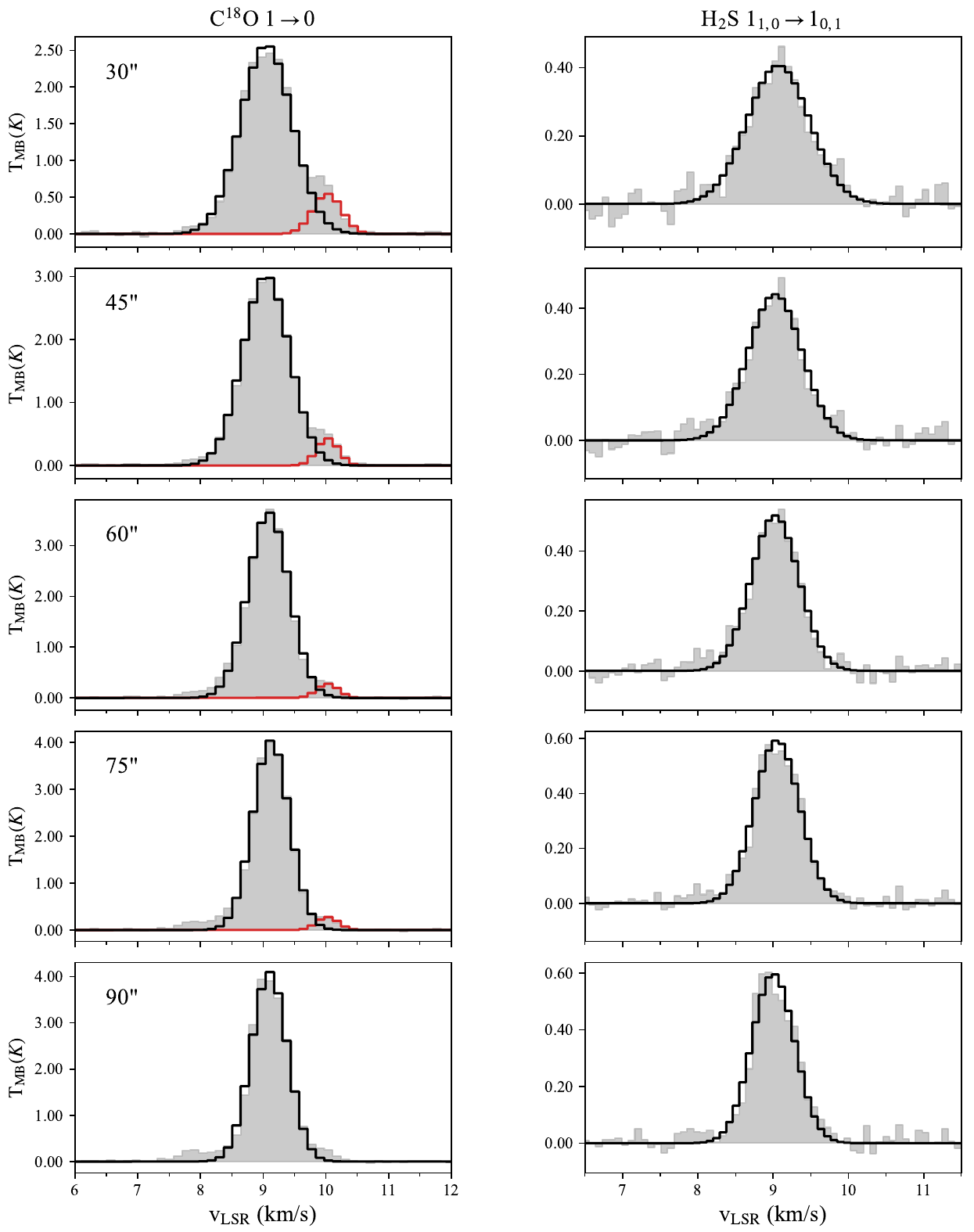}
                \caption{Rotational spectra (gray area) of the molecular lines observed at offsets $30''-90''$ with respect to the central position in IC348 (see Table \ref{tab:sources}). The solid black line marks the RADEX fitting of each line while the red solid line corresponds to additional velocity components that were not taken into account in the fitting.}
                \label{fig:spectra_c18o_h2s_30_90_IC348}
            \end{figure}
            \vfill
            \clearpage
            \vfill
            \begin{figure}[h]
                \centering
                \includegraphics[width=\textwidth]{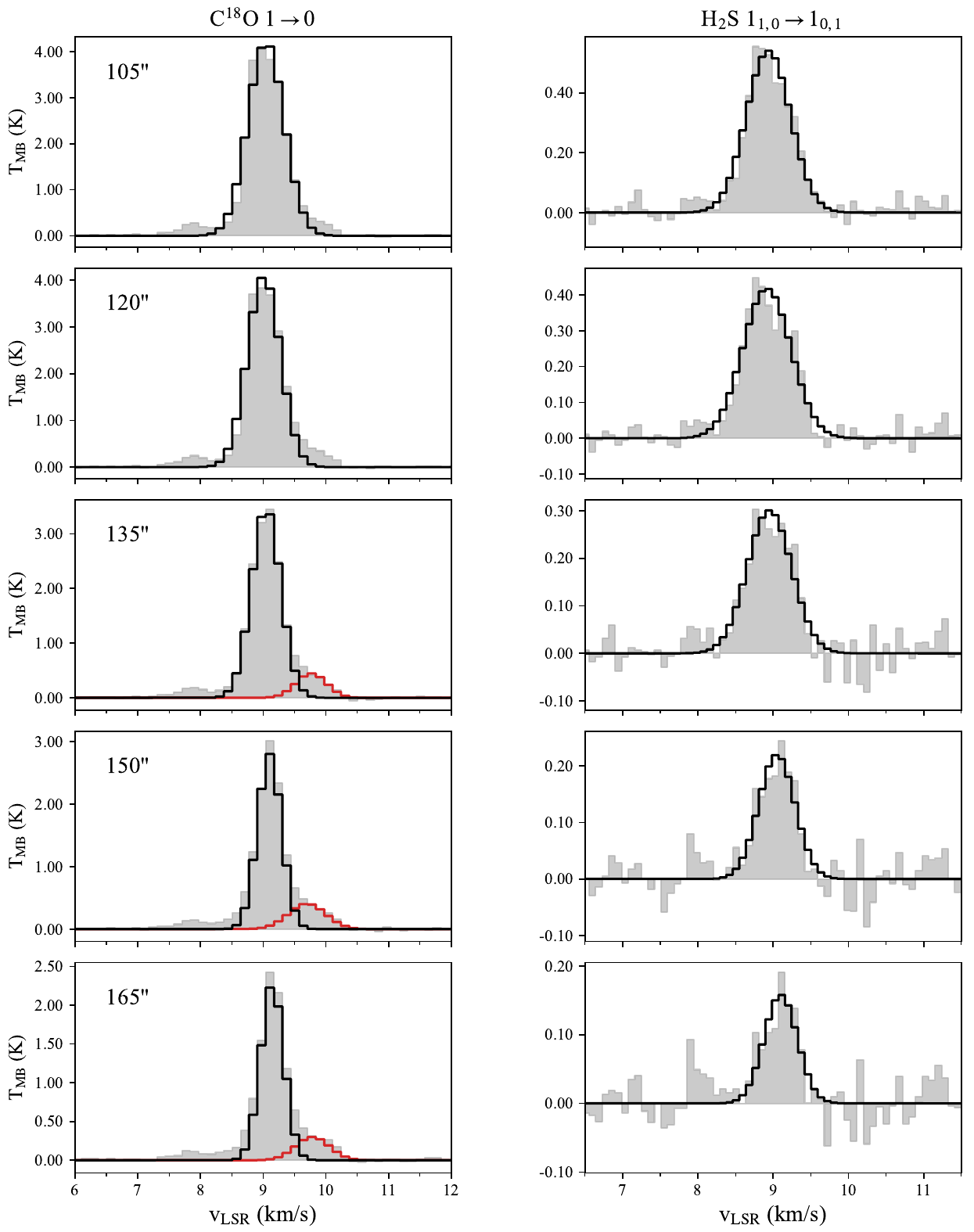}
                \caption{Rotational spectra (gray area) of the molecular lines observed at offsets $105''-165''$ with respect to the central position in IC348 (see Table \ref{tab:sources}). The solid black line marks the RADEX fitting of each line while the red solid line corresponds to additional velocity components that were not taken into account in the fitting.}
                \label{fig:spectra_c18o_h2s_105_165_IC348}
            \end{figure}
            \vfill
            \clearpage
            \begin{figure}[h]
                \centering
                \includegraphics[width=\textwidth]{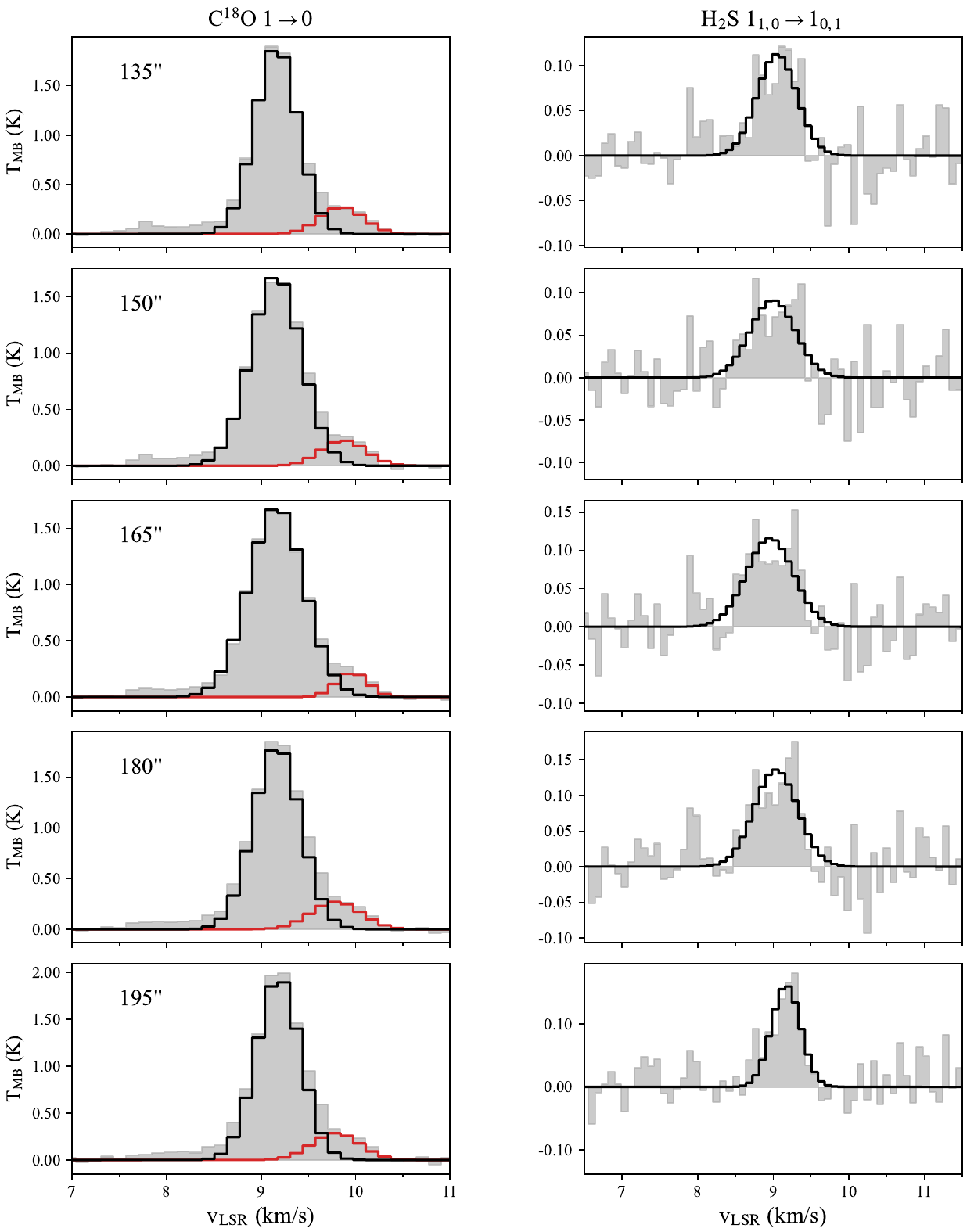}
                \caption{Rotational spectra (gray area) of the molecular lines observed at offsets $180''-240''$ with respect to the central position in IC348 (see Table \ref{tab:sources}). The solid black line marks the RADEX fitting of each line while the red solid line corresponds to additional velocity components that were not taken into account in the fitting.}
                \label{fig:spectra_c18o_h2s_180_240_IC348}
            \end{figure}
            \vfill
        \clearpage
        
        \newpage
      
      \section{Line analysis - IC348}\label{sec:linePropIC348}

		\begin{table}[h!]
                \centering
                \caption{Properties, main beam temperatures, and integrated intensities of the CH$_{3}$OH rotational transitions at different offsets toward IC348.}
                \resizebox{0.98\textwidth}{!}{
                \begin{tabular}{lccccrccc}
                    \toprule
				    \multirow{2}{*}{{Offset}} &  \multirow{2}{*}{{Transition}} &  {Frequency} & \multirow{2}{*}{{E$_{\rm up}$} {(K)}} & \multirow{2}{*}{{log(A$_{\rm{ij}}$)}} & \multirow{2}{*}{Peak position (km s$^{-1}$)} & \multirow{2}{*}{Width (km s$^{-1}$)} & \multirow{2}{*}{{T}$_{\rm MB}$ {(K)}} & $\int${T}$_{\rm MB}\ dv$\\
					&  & {(MHz)} & & & & & & {(K km s}$^{{-1}}${)} \\
				    \midrule\midrule
                    \multirow{6}{*}{30"} & E1 $2_{0,2}\rightarrow 1_{0,1}$ & 96744.54  & 20.10 & -5.47 & $-0.09\pm 0.05$ & $0.6\pm 0.1$ & $0.08\pm 0.02$ & $0.050\pm 0.008$\\ \cmidrule{6-9}
                    & \multirow{2}{*}{A$^{+}$ $2_{0,2}\rightarrow 1_{0,1}$} & \multirow{2}{*}{96741.37}  & \multirow{2}{*}{7.00} & \multirow{2}{*}{-5.47} & $9.628\pm 0.003$ & $0.65\pm 0.02$ & $0.55\pm 0.02$ & $0.38\pm 0.01$\\
                    &  &  & & & $10.40\pm 0.02$ & $0.66\pm 0.04$ & $0.26\pm 0.02$ & $0.18\pm 0.01$\\
                    \cmidrule{6-9}
                    & \multirow{2}{*}{E2 $2_{1,2}\rightarrow 1_{1,1}$} & \multirow{2}{*}{96739.36}  & \multirow{2}{*}{12.50}  & \multirow{2}{*}{-5.59}  & $16.65\pm 0.02$ & $0.55\pm 0.05$ & $0.19\pm 0.02$ & $0.11\pm 0.01$ \\
                    & & & & & $15.87\pm 0.01$ & $0.66\pm 0.03$ & $0.39\pm 0.02$ & $0.27\pm 0.01$ \\
                    \midrule
                    \multirow{6}{*}{45"} & E1 $2_{0,2}\rightarrow 1_{0,1}$ & 96744.54  & 20.10 & -5.47 & $-0.16\pm 0.03$ & $0.44\pm 0.08$ & $0.09\pm 0.01$ & $0.040\pm 0.005$\\ \cmidrule{6-9}
                    & \multirow{2}{*}{A$^{+}$ $2_{0,2}\rightarrow 1_{0,1}$} & \multirow{2}{*}{96741.37}  & \multirow{2}{*}{7.00} & \multirow{2}{*}{-5.47} & $9.618\pm 0.006$ & $0.45\pm 0.02$ & $0.48\pm 0.01$ & $0.23\pm 0.02$\\
                    &  &  & & & $9.94\pm 0.04$ & $1.30\pm 0.05$ & $0.19\pm 0.01$ & $0.27\pm 0.02$\\
                    \cmidrule{6-9}
                    & \multirow{2}{*}{E2 $2_{1,2}\rightarrow 1_{1,1}$} & \multirow{2}{*}{96739.36}  & \multirow{2}{*}{12.50}  & \multirow{2}{*}{-5.59}  & $15.869\pm 0.005$ & $0.57\pm 0.02$ & $0.44\pm 0.01$ & $0.264\pm 0.006$ \\
                    & & & & & $16.65\pm 0.02$ & $0.48\pm 0.05$ & $0.12\pm 0.01$ & $0.062\pm 0.006$ \\
                    \midrule

                    \multirow{5}{*}{60"} & E1 $2_{0,2}\rightarrow 1_{0,1}$ & 96744.54  & 20.10 & -5.47 & $-0.2\pm 0.2$ & $0.5\pm 0.2$ & $0.08\pm 0.01$ & $0.04\pm 0.02$\\ \cmidrule{6-9}
                    & A$^{+}$ $2_{0,2}\rightarrow 1_{0,1}$ & 96741.37  & 7.00 & -5.47 & $9.7\pm 0.2$ & $0.5\pm 0.2$ & $0.68\pm 0.01$ & $0.38\pm 0.02$\\
                    \cmidrule{6-9}
                    & \multirow{2}{*}{E2 $2_{1,2}\rightarrow 1_{1,1}$} & \multirow{2}{*}{96739.36}  & \multirow{2}{*}{12.50}  & \multirow{2}{*}{-5.59}  & $15.9\pm 0.2$ & $0.5\pm 0.2$ & $0.47\pm 0.01$ & $0.27\pm 0.02$ \\
                    & & & & & $16.7\pm 0.2$ & $0.4\pm 0.2$ & $0.05\pm 0.01$ & $0.02\pm 0.02$ \\
                    \midrule

                    \multirow{3}{*}{75"} & E1 $2_{0,2}\rightarrow 1_{0,1}$ & 96744.54  & 20.10 & -5.47 & $-0.10\pm 0.03$ & $0.44\pm 0.06$ & $0.08\pm 0.01$ & $0.035\pm 0.005$\\
                    & A$^{+}$ $2_{0,2}\rightarrow 1_{0,1}$ & 96741.37  & 7.00 & -5.47 & $9.748\pm 0.004$ & $0.51\pm 0.01$ & $0.60\pm 0.01$ & $0.324\pm 0.005$\\
                    & E2 $2_{1,2}\rightarrow 1_{1,1}$ & 96739.36  & 12.50  & -5.59 & $15.973\pm 0.005$ & $0.48\pm 0.01$ & $0.42\pm 0.01$ & $0.218\pm 0.005$ \\
                    \midrule

                    \multirow{3}{*}{90"} & E1 $2_{0,2}\rightarrow 1_{0,1}$ & 96744.54  & 20.10 & -5.47 & $-0.07\pm 0.05$ & $0.41\pm 0.09$ & $0.04\pm 0.01$ & $0.018\pm 0.004$\\
                    & A$^{+}$ $2_{0,2}\rightarrow 1_{0,1}$ & 96741.37  & 7.00 & -5.47 & $9.763\pm 0.006$ & $0.57\pm 0.02$ & $0.37\pm 0.01$ & $0.222\pm 0.005$\\
                    & E2 $2_{1,2}\rightarrow 1_{1,1}$ & 96739.36  & 12.50  & -5.59 & $15.992\pm 0.009$ & $0.53\pm 0.02$ & $0.26\pm 0.01$ & $0.144\pm 0.005$ \\
                    \midrule

                    \multirow{3}{*}{105"} & E1 $2_{0,2}\rightarrow 1_{0,1}$ & 96744.54  & 20.10 & -5.47 & $-0.1\pm 0.1$ & $0.4\pm 0.2$ & $0.02\pm 0.01$ & $0.008\pm 0.004$\\
                    & A$^{+}$ $2_{0,2}\rightarrow 1_{0,1}$ & 96741.37  & 7.00 & -5.47 & $9.68\pm 0.01$ & $0.69\pm 0.03$ & $0.21\pm 0.01$ & $0.153\pm 0.006$\\
                    & E2 $2_{1,2}\rightarrow 1_{1,1}$ & 96739.36  & 12.50  & -5.59 & $15.91\pm 0.02$ & $0.59\pm 0.04$ & $0.15\pm 0.01$ & $0.096\pm 0.006$ \\
                    \midrule

                    \multirow{3}{*}{120"} & E1 $2_{0,2}\rightarrow 1_{0,1}$ & 96744.54  & 20.10 & -5.47 & $-0.2\pm 0.2$ & $0.6\pm 0.3$ & $0.02\pm 0.01$ & $0.006\pm 0.005$\\
                    & A$^{+}$ $2_{0,2}\rightarrow 1_{0,1}$ & 96741.37  & 7.00 & -5.47 & $9.59\pm 0.02$ & $0.63\pm 0.04$ & $0.17\pm 0.01$ & $0.112\pm 0.006$\\
                    & E2 $2_{1,2}\rightarrow 1_{1,1}$ & 96739.36  & 12.50  & -5.59 & $15.81\pm 0.02$ & $0.52\pm 0.04$ & $0.12\pm 0.01$ & $0.068\pm 0.005$ \\
                    \midrule

                    \multirow{3}{*}{135"} & E1 $2_{0,2}\rightarrow 1_{0,1}$ & 96744.54  & 20.10 & -5.47 & \multicolumn{4}{c}{rms $= 0.01$} \\
                    & A$^{+}$ $2_{0,2}\rightarrow 1_{0,1}$ & 96741.37  & 7.00 & -5.47 & $9.60\pm 0.02$ & $0.59\pm 0.06$ & $0.12\pm 0.01$ & $0.074\pm 0.006$\\
                    & E2 $2_{1,2}\rightarrow 1_{1,1}$ & 96739.36  & 12.50  & -5.59 & $15.81\pm 0.02$ & $0.50\pm 0.06$ & $0.09\pm 0.01$ & $0.046\pm 0.005$ \\
                    \midrule

                    \multirow{3}{*}{150"} & E1 $2_{0,2}\rightarrow 1_{0,1}$ & 96744.54  & 20.10 & -5.47 & \multicolumn{4}{c}{rms $= 0.01$} \\
                    & A$^{+}$ $2_{0,2}\rightarrow 1_{0,1}$ & 96741.37  & 7.00 & -5.47 & $9.66\pm 0.05$ & $0.6\pm 0.2$ & $0.06\pm 0.01$ & $0.038\pm 0.007$\\
                    & E2 $2_{1,2}\rightarrow 1_{1,1}$ & 96739.36  & 12.50  & -5.59 & $15.94\pm 0.02$ & $0.26\pm 0.07$ & $0.08\pm 0.01$ & $0.021\pm 0.004$ \\
                    \midrule

                    \multirow{3}{*}{165"} & E1 $2_{0,2}\rightarrow 1_{0,1}$ & 96744.54  & 20.10 & -5.47 & \multicolumn{4}{c}{rms $= 0.01$} \\
                    & A$^{+}$ $2_{0,2}\rightarrow 1_{0,1}$ & 96741.37  & 7.00 & -5.47 & $9.9\pm 0.1$ & $1.1\pm 0.2$ & $0.03\pm 0.01$ & $0.031\pm 0.007$\\
                    & E2 $2_{1,2}\rightarrow 1_{1,1}$ & 96739.36  & 12.50  & -5.59 & $15.96\pm 0.04$ & $0.2\pm 0.1$ & $0.05\pm 0.01$ & $0.011\pm 0.004$ \\
                    \midrule
                    
                    \multirow{3}{*}{180"} & E1 $2_{0,2}\rightarrow 1_{0,1}$ & 96744.54  & 20.10 & -5.47 & \multicolumn{4}{c}{rms $= 0.01$} \\
                    & A$^{+}$ $2_{0,2}\rightarrow 1_{0,1}$ & 96741.37  & 7.00 & -5.47 & $10.0\pm 0.2$ & $1.0\pm 0.3$ & $0.02\pm 0.01$ & $0.023\pm 0.007$\\
                    & E2 $2_{1,2}\rightarrow 1_{1,1}$ & 96739.36  & 12.50  & -5.59 & $15.9\pm 0.1$ & $0.6\pm 0.3$ & $0.02\pm 0.01$ & $0.016\pm 0.006$ \\
                    \midrule

                    \multirow{3}{*}{195"} & E1 $2_{0,2}\rightarrow 1_{0,1}$ & 96744.54  & 20.10 & -5.47 & \multicolumn{4}{c}{rms $= 0.01$}\\
                    & A$^{+}$ $2_{0,2}\rightarrow 1_{0,1}$ & 96741.37  & 7.00 & -5.47 & $10.2\pm 0.1$ & $0.6\pm 0.3$ & $0.02\pm 0.01$ & $0.014\pm 0.005$ \\
                    & E2 $2_{1,2}\rightarrow 1_{1,1}$ & 96739.36  & 12.50  & -5.59 & \multicolumn{4}{c}{rms $= 0.01$}\\
                    \midrule

                    \multirow{3}{*}{210"} & E1 $2_{0,2}\rightarrow 1_{0,1}$ & 96744.54  & 20.10 & -5.47 & \multicolumn{4}{c}{rms $= 0.01$}\\
                    & A$^{+}$ $2_{0,2}\rightarrow 1_{0,1}$ & 96741.37  & 7.00 & -5.47 & $10.3\pm 0.1$ & $0.5\pm 0.2$ & $0.02\pm 0.01$ & $0.012\pm 0.004$ \\
                    & E2 $2_{1,2}\rightarrow 1_{1,1}$ & 96739.36  & 12.50  & -5.59 & \multicolumn{4}{c}{rms $= 0.01$}\\
                    \midrule

                    \multirow{3}{*}{225"} & E1 $2_{0,2}\rightarrow 1_{0,1}$ & 96744.54  & 20.10 & -5.47 & \\
                    & A$^{+}$ $2_{0,2}\rightarrow 1_{0,1}$ & 96741.37  & 7.00 & -5.47 & \multicolumn{4}{c}{rms $= 0.01$}\\
                    & E2 $2_{1,2}\rightarrow 1_{1,1}$ & 96739.36  & 12.50  & -5.59 & \\
                    \midrule

                    \multirow{3}{*}{240"} & E1 $2_{0,2}\rightarrow 1_{0,1}$ & 96744.54  & 20.10 & -5.47 & \\
                    & A$^{+}$ $2_{0,2}\rightarrow 1_{0,1}$ & 96741.37  & 7.00 & -5.47 & \multicolumn{4}{c}{rms $= 0.02$}\\
                    & E2 $2_{1,2}\rightarrow 1_{1,1}$ & 96739.36  & 12.50  & -5.59 & \\
    			\bottomrule
			\end{tabular}
            }
			\label{tab:linePropertiesCH3OHIC348}
        \end{table}

\begin{table}[h!]
    		\centering
            \caption{Properties, main beam temperatures, and integrated intensities of the C$^{18}$O $1\rightarrow0$ rotational transition at different offsets toward IC348.}
                \resizebox{\textwidth}{!}{
                \begin{tabular}{lcccccccc}
                    \toprule
				    \multirow{2}{*}{{Offset}} &  \multirow{2}{*}{{Transition}} &  {Frequency} & \multirow{2}{*}{{E$_{\rm up}$} {(K)}} & \multirow{2}{*}{{log(A$_{\rm{ij}}$)}} & \multirow{2}{*}{Peak position (km s$^{-1}$)} & \multirow{2}{*}{Width (km s$^{-1}$)} & \multirow{2}{*}{{T}$_{\rm MB}$ {(K)}} & $\int${T}$_{\rm MB}\ dv$\\
					&  & {(MHz)} & & & & & & {(K km s}$^{{-1}}${)} \\
				    \midrule\midrule
                    30" & $1\rightarrow0$ & 109782.17 & 5.27 & -7.20 & $9.050\pm 0.002$ & $0.974\pm 0.006$ & $2.58\pm 0.02$ & $2.67\pm 0.01$ \\
                    \midrule
                    45" & $1\rightarrow0$ & 109782.17 & 5.27 & -7.20 & $9.046\pm 0.001$ & $0.877\pm 0.004$ & $3.02\pm 0.02$ & $2.82\pm 0.01$ \\
                    \midrule
                    60" & $1\rightarrow0$ & 109782.17 & 5.27 & -7.20 & $9.085\pm 0.001$ & $0.774\pm 0.003$ & $3.65\pm 0.02$ & $3.01\pm 0.01$ \\
                    \midrule
                    75" & $1\rightarrow0$ & 109782.17 & 5.27 & -7.20 & $9.124\pm 0.002$ & $0.688\pm 0.004$ & $4.04\pm 0.02$ & $2.96\pm 0.02$ \\
                    \midrule
                    90" & $1\rightarrow0$ & 109782.17 & 5.27 & -7.20 & $9.099\pm 0.001$ & $0.680\pm 0.003$ & $4.10\pm 0.02$ & $2.97\pm 0.01$ \\
                    \midrule
                    105" & $1\rightarrow0$ & 109782.17 & 5.27 & -7.20 & $9.044\pm 0.000$ & $0.681\pm 0.002$ & $4.21\pm 0.02$ & $3.05\pm 0.01$ \\
                    \midrule
                    120" & $1\rightarrow0$ & 109782.17 & 5.27 & -7.20 & $9.0\pm 0.1$ & $0.6\pm 0.1$ & $4.09\pm 0.02$ & $2.70\pm 0.08$ \\
                    \midrule
                    135" & $1\rightarrow0$ & 109782.17 & 5.27 & -7.20 & $9.046\pm 0.000$ & $0.550\pm 0.003$ & $3.47\pm 0.02$ & $2.03\pm 0.01$ \\
                    \midrule
                    150" & $1\rightarrow0$ & 109782.17 & 5.27 & -7.20 & $9.101\pm 0.001$ & $0.451\pm 0.003$ & $2.80\pm 0.02$ & $1.35\pm 0.01$ \\
                    \midrule
                    165" & $1\rightarrow0$ & 109782.17 & 5.27 & -7.20 & $9.144\pm 0.000$ & $0.435\pm 0.005$ & $2.27\pm 0.02$ & $1.05\pm 0.02$ \\
                    \midrule
                    180" & $1\rightarrow0$ & 109782.17 & 5.27 & -7.20 & $9.161\pm 0.003$ & $0.54\pm 0.01$ & $1.90\pm 0.02$ & $1.09\pm 0.03$ \\
                    \midrule
                    195" & $1\rightarrow0$ & 109782.17 & 5.27 & -7.20 & $9.2\pm 0.1$ & $0.6\pm 0.1$ & $1.70\pm 0.02$ & $1.14\pm 0.02$ \\
                    \midrule
                    210" & $1\rightarrow0$ & 109782.17 & 5.27 & -7.20 & $9.163\pm 0.003$ & $0.691\pm 0.008$ & $1.70\pm 0.02$ & $1.25\pm 0.02$ \\
                    \midrule
                    225" & $1\rightarrow0$ & 109782.17 & 5.27 & -7.20 & $9.166\pm 0.003$ & $0.587\pm 0.006$ & $1.81\pm 0.02$ & $1.13\pm 0.01$ \\
                    \midrule
                    240" & $1\rightarrow0$ & 109782.17 & 5.27 & -7.20 & $9.183\pm 0.005$ & $0.550\pm 0.005$ & $1.95\pm 0.02$ & $1.14\pm 0.04$ \\
    			\bottomrule
			\end{tabular}
            }
            
			\label{tab:linePropertiesC18OIC348}
        \end{table}

		\vspace*{2 cm}
		
		\begin{table}[h!]
    		\centering
            \caption{Properties, main beam temperatures, and integrated intensities of the H$_{2}$S $1_{1,0}\rightarrow 1_{0,1}$ rotational transition at different offsets toward IC348.}
                \resizebox{\textwidth}{!}{
                \begin{tabular}{lccccrccc}
                    \toprule
				    \multirow{2}{*}{{Offset}} &  \multirow{2}{*}{{Transition}} &  {Frequency} & \multirow{2}{*}{{E$_{\rm up}$} {(K)}} & \multirow{2}{*}{{log(A$_{\rm{ij}}$)}} & \multirow{2}{*}{Peak position (km s$^{-1}$)} & \multirow{2}{*}{Width (km s$^{-1}$)} & \multirow{2}{*}{{T}$_{\rm MB}$ {(K)}} & $\int${T}$_{\rm MB}\ dv$\\
					&  & {(MHz)} & & & & & & {(K km s}$^{{-1}}${)} \\
				    \midrule\midrule
                    30" & $1_{1,0}\rightarrow1_{0,1}$ & 168762.76 & 27.9 & -4.57 & $9.07\pm 0.02$ & $0.98\pm 0.05$ & $0.41\pm 0.04$ & $0.42\pm 0.02$ \\
                    \midrule
                    45" & $1_{1,0}\rightarrow1_{0,1}$ & 168762.76 & 27.9 & -4.57 & $9.02\pm 0.01$ & $0.85\pm 0.04$ & $0.44\pm 0.03$ & $0.40\pm 0.01$ \\
                    \midrule
                    60" & $1_{1,0}\rightarrow1_{0,1}$ & 168762.76 & 27.9 & -4.57 & $9.02\pm 0.01$ & $0.75\pm 0.03$ & $0.52\pm 0.03$ & $0.41\pm 0.01$ \\
                    \midrule
                    75" & $1_{1,0}\rightarrow1_{0,1}$ & 168762.76 & 27.9 & -4.57 & $9.05\pm 0.01$ & $0.71\pm 0.02$ & $0.59\pm 0.03$ & $0.45\pm 0.01$ \\
                    \midrule
                    90" & $1_{1,0}\rightarrow1_{0,1}$ & 168762.76 & 27.9 & -4.57 & $9.001\pm 0.009$ & $0.67\pm 0.02$ & $0.60\pm 0.03$ & $0.43\pm 0.01$ \\
                    \midrule
                    105" & $1_{1,0}\rightarrow1_{0,1}$ & 168762.76 & 27.9 & -4.57 & $8.94\pm 0.01$ & $0.69\pm 0.02$ & $0.54\pm 0.03$ & $0.40\pm 0.01$ \\
                    \midrule
                    120" & $1_{1,0}\rightarrow1_{0,1}$ & 168762.76 & 27.9 & -4.57 & $8.92\pm 0.01$ & $0.75\pm 0.03$ & $0.42\pm 0.03$ & $0.33\pm 0.01$ \\
                    \midrule
                    135" & $1_{1,0}\rightarrow1_{0,1}$ & 168762.76 & 27.9 & -4.57 & $9.05\pm 0.02$ & $0.69\pm 0.04$ & $0.30\pm 0.03$ & $0.22\pm 0.01$ \\
                    \midrule
                    150" & $1_{1,0}\rightarrow1_{0,1}$ & 168762.76 & 27.9 & -4.57 & $9.04\pm 0.02$ & $0.57\pm 0.05$ & $0.22\pm 0.03$ & $0.13\pm 0.01$ \\
                    \midrule
                    165" & $1_{1,0}\rightarrow1_{0,1}$ & 168762.76 & 27.9 & -4.57 & $9.10\pm 0.03$ & $0.52\pm 0.07$ & $0.16\pm 0.03$ & $0.09\pm 0.01$ \\
                    \midrule
                    180" & $1_{1,0}\rightarrow1_{0,1}$ & 168762.76 & 27.9 & -4.57 & $9.05\pm 0.05$ & $0.62\pm 0.09$ & $0.11\pm 0.03$ & $0.07\pm 0.01$ \\
                    \midrule
                    195" & $1_{1,0}\rightarrow1_{0,1}$ & 168762.76 & 27.9 & -4.57 & $8.99\pm 0.07$ & $0.7\pm 0.1$ & $0.09\pm 0.03$ & $0.07\pm 0.01$ \\
                    \midrule
                    210" & $1_{1,0}\rightarrow1_{0,1}$ & 168762.76 & 27.9 & -4.57 & $8.96\pm 0.05$ & $0.74\pm 0.09$ & $0.12\pm 0.03$ & $0.09\pm 0.01$ \\
                    \midrule
                    225" & $1_{1,0}\rightarrow1_{0,1}$ & 168762.76 & 27.9 & -4.57 & $9.04\pm 0.04$ & $0.68\pm 0.08$ & $0.14\pm 0.03$ & $0.10\pm 0.01$ \\
                    \midrule
                    240" & $1_{1,0}\rightarrow1_{0,1}$ & 168762.76 & 27.9 & -4.57 & $9.17\pm 0.04$ & $0.46\pm 0.09$ & $0.16\pm 0.04$ & $0.08\pm 0.01$ \\
    			\bottomrule
			\end{tabular}
            }
            
			\label{tab:linePropertiesH2SIC348}
        \end{table}

            \begin{table}[h!]
                \centering
                \caption{Properties, main beam temperatures, and integrated intensities of the N$_{2}$H$^{+}$ $1\rightarrow0$ rotational transition at different offsets toward IC348. We include the opacities of the resolved hyperfine structure components.}
                \resizebox{\textwidth}{!}{
                \begin{tabular}{lccccrcccc}
                    \toprule
				    \multirow{2}{*}{{Offset}} &  \multirow{2}{*}{{Transition}} &  {Frequency} & \multirow{2}{*}{{E$_{\rm up}$} {(K)}} & \multirow{2}{*}{{log(A$_{\rm{ij}}$)}} & \multirow{2}{*}{Peak position (km s$^{-1}$)} & \multirow{2}{*}{Width (km s$^{-1}$)} & \multirow{2}{*}{{T}$_{\rm MB}$ {(K)}} & $\int${T}$_{\rm MB}\ dv$ & \multirow{2}{*}{${\tau_{\rm main}}$}\\
					&  & {(MHz)} & & & & & & {(K km s}$^{{-1}}${)} & \\
				    \midrule\midrule
					 \multirow{6}{*}{30"} & \multirow{6}{*}{$1\rightarrow0$} & \multirow{6}{*}{93173.40} & \multirow{6}{*}{4.47} & \multirow{6}{*}{-4.44} & $-0.8\pm 0.2$ & $0.6\pm 0.2$ & $0.509\pm 0.006$ & $0.31\pm 0.04$ & \multirow{6}{*}{$0.5\pm 0.1$}\\
                        & & & & & $6.55\pm 0.04$ & $0.41\pm 0.09$ & $0.471\pm 0.006$ & $0.20\pm 0.04$ & \\
                        & & & & & $7.18\pm 0.02$ & $0.47\pm 0.05$ & $1.133\pm 0.006$ & $0.57\pm 0.05$ & \\
                        & & & & & $8.10\pm 0.02$ & $0.63\pm 0.07$ & $0.830\pm 0.006$ & $0.56\pm 0.05$ & \\
                        & & & & & $13.05\pm 0.03$ & $0.84\pm 0.07$ & $0.784\pm 0.006$ & $0.70\pm 0.05$ & \\
                        & & & & & $14.06\pm 0.05$ & $0.5\pm 0.1$ & $0.322\pm 0.006$ & $0.17\pm 0.04$ &\\
                    \midrule
                    \multirow{6}{*}{45"} & \multirow{6}{*}{$1\rightarrow0$} & \multirow{6}{*}{93173.40} & \multirow{6}{*}{4.47} & \multirow{6}{*}{-4.44} & $-0.84\pm 0.06$ & $0.4\pm 0.2$ & $0.753\pm 0.005$ & $0.3\pm 0.1$ & \multirow{6}{*}{$0.64\pm 0.07$}\\
                        & & & & & $6.56\pm 0.03$ & $0.39\pm 0.07$ & $0.640\pm 0.005$ & $0.27\pm 0.04$ & \\
                        & & & & & $7.18\pm 0.01$ & $0.43\pm 0.04$ & $1.396\pm 0.005$ & $0.64\pm 0.04$ & \\
                        & & & & & $8.12\pm 0.02$ & $0.47\pm 0.05$ & $1.049\pm 0.005$ & $0.52\pm 0.04$ & \\
                        & & & & & $13.01\pm 0.02$ & $0.79\pm 0.05$ & $0.466\pm 0.005$ & $0.80\pm 0.04$ & \\
                        & & & & & $14.09\pm 0.05$ & $0.4\pm 0.1$ & $0.366\pm 0.005$ & $0.16\pm 0.04$ &\\
                    \midrule
                     \multirow{6}{*}{60"} & \multirow{6}{*}{$1\rightarrow0$} & \multirow{6}{*}{93173.40} & \multirow{6}{*}{4.47} & \multirow{6}{*}{-4.44} & $-0.81\pm 0.06$ & $0.4\pm 0.1$ & $0.944\pm 0.005$ & $0.4\pm 0.1$ & \multirow{6}{*}{$0.84\pm 0.02$}\\
                        & & & & & $6.59\pm 0.03$ & $0.39\pm 0.07$ & $0.782\pm 0.005$ & $0.32\pm 0.05$ & \\
                        & & & & & $7.20\pm 0.01$ & $0.40\pm 0.03$ & $1.648\pm 0.005$ & $0.70\pm 0.05$ & \\
                        & & & & & $8.16\pm 0.02$ & $0.40\pm 0.04$ & $1.241\pm 0.005$ & $0.52\pm 0.04$ & \\
                        & & & & & $13.02\pm 0.02$ & $0.76\pm 0.05$ & $1.115\pm 0.005$ & $0.90\pm 0.06$ & \\
                        & & & & & $14.134\pm 0.04$ & $0.34\pm 0.09$ & $0.411\pm 0.005$ & $0.15\pm 0.04$ &\\
                    \midrule
                     \multirow{6}{*}{75"} & \multirow{6}{*}{$1\rightarrow0$} & \multirow{6}{*}{93173.40} & \multirow{6}{*}{4.47} & \multirow{6}{*}{-4.44} & $-0.77\pm 0.05$ & $0.4\pm 0.1$ & $0.791\pm 0.005$ & $0.32\pm 0.09$ & \multirow{6}{*}{$0.912\pm 0.006$}\\
                        & & & & & $6.63\pm 0.03$ & $0.38\pm 0.07$ & $0.647\pm 0.005$ & $0.26\pm 0.04$ & \\
                        & & & & & $7.25\pm 0.01$ & $0.41\pm 0.03$ & $1.378\pm 0.005$ & $0.60\pm 0.04$ & \\
                        & & & & & $8.20\pm 0.02$ & $0.39\pm 0.04$ & $1.034\pm 0.005$ & $0.43\pm 0.03$ & \\
                        & & & & & $13.06\pm 0.02$ & $0.76\pm 0.05$ & $0.938\pm 0.005$ & $0.75\pm 0.05$ & \\
                        & & & & & $14.18\pm 0.05$ & $0.3\pm 0.1$ & $0.334\pm 0.005$ & $0.12\pm 0.03$ &\\
                    \midrule
                     \multirow{6}{*}{90"} & \multirow{6}{*}{$1\rightarrow0$} & \multirow{6}{*}{93173.40} & \multirow{6}{*}{4.47} & \multirow{6}{*}{-4.44} & $-0.73\pm 0.06$ & $0.4\pm 0.1$ & $0.394\pm 0.005$ & $0.16\pm 0.05$ & \multirow{6}{*}{$0.84\pm 0.04$}\\
                        & & & & & $6.67\pm 0.03$ & $0.38\pm 0.07$ & $0.326\pm 0.005$ & $0.13\pm 0.02$ & \\
                        & & & & & $7.29\pm 0.01$ & $0.41\pm 0.03$ & $0.702\pm 0.005$ & $0.31\pm 0.02$ & \\
                        & & & & & $8.24\pm 0.02$ & $0.39\pm 0.04$ & $0.524\pm 0.005$ & $0.22\pm 0.02$ & \\
                        & & & & & $13.09\pm 0.02$ & $0.76\pm 0.05$ & $0.478\pm 0.005$ & $0.39\pm 0.02$ & \\
                        & & & & & $14.22\pm 0.05$ & $0.3\pm 0.1$ & $0.158\pm 0.005$ & $0.06\pm 0.02$ &\\
                    \midrule
                    \multirow{6}{*}{105"} & \multirow{6}{*}{$1\rightarrow0$} & \multirow{6}{*}{93173.40} & \multirow{6}{*}{4.47} & \multirow{6}{*}{-4.44} & $-0.71\pm 0.06$ & $0.4\pm 0.1$ & $0.124\pm 0.005$ & $0.05\pm 0.02$ & \multirow{6}{*}{$0.3\pm 0.1$}\\
                        & & & & & $6.69\pm 0.03$ & $0.41\pm 0.08$ & $0.104\pm 0.005$ & $0.045\pm 0.007$ & \\
                        & & & & & $7.30\pm 0.01$ & $0.42\pm 0.03$ & $0.237\pm 0.005$ & $0.104\pm 0.007$ & \\
                        & & & & & $8.26\pm 0.02$ & $0.38\pm 0.04$ & $0.169\pm 0.005$ & $0.068\pm 0.005$ & \\
                        & & & & & $13.11\pm 0.02$ & $0.76\pm 0.05$ & $0.158\pm 0.005$ & $0.128\pm 0.007$ & \\
                        & & & & & $14.22\pm 0.06$ & $0.3\pm 0.1$ & $0.041\pm 0.005$ & $0.013\pm 0.005$ &\\
                    \midrule
                    \multirow{5}{*}{120"} & \multirow{5}{*}{$1\rightarrow0$} & \multirow{5}{*}{93173.40} & \multirow{5}{*}{4.47} & \multirow{5}{*}{-4.44} & $-0.73\pm 0.05$ & $0.3\pm 0.1$ & $0.048\pm 0.005$ & $0.016\pm 0.005$ & \multirow{5}{*}{$<0.1$}\\
                        & & & & & $6.68\pm 0.07$ & $0.6\pm 0.2$ & $0.033\pm 0.005$ & $0.020\pm 0.005$ & \\
                        & & & & & $7.26\pm 0.02$ & $0.36\pm 0.04$ & $0.093\pm 0.005$ & $0.036\pm 0.004$ & \\
                        & & & & & $8.25\pm 0.02$ & $0.31\pm 0.04$ & $0.062\pm 0.005$ & $0.020\pm 0.002$ & \\
                        & & & & & $13.07\pm 0.03$ & $0.81\pm 0.07$ & $0.054\pm 0.005$ & $0.046\pm 0.003$ & \\
                    \midrule
                    \multirow{5}{*}{135"} & \multirow{5}{*}{$1\rightarrow0$} & \multirow{5}{*}{93173.40} & \multirow{5}{*}{4.47} & \multirow{5}{*}{-4.44} & $-0.73\pm 0.04$ & $0.36\pm 0.11$ & $0.022\pm 0.005$ & $0.008\pm 0.002$ & \multirow{5}{*}{$<0.1$}\\
                        & & & & & $6.53\pm 0.07$ & $0.4\pm 0.1$ & $0.016\pm 0.005$ & $0.007\pm 0.002$ & \\
                        & & & & & $7.21\pm 0.02$ & $0.41\pm 0.05$ & $0.045\pm 0.005$ & $0.020\pm 0.002$ & \\
                        & & & & & $8.29\pm 0.03$ & $0.29\pm 0.08$ & $0.025\pm 0.005$ & $0.008\pm 0.002$ & \\
                        & & & & & $12.97\pm 0.04$ & $0.73\pm 0.08$ & $0.028\pm 0.005$ & $0.022\pm 0.002$ & \\
                    \midrule
                    \multirow{3}{*}{150"} & \multirow{3}{*}{$1\rightarrow0$} & \multirow{3}{*}{93173.40} & \multirow{3}{*}{4.47} & \multirow{3}{*}{-4.44} & $-0.8\pm 0.2$ & $0.8\pm 0.1$ & $0.010\pm 0.005$ & $0.008\pm 0.004$ & \multirow{3}{*}{$<0.1$}\\
                        & & & & & $7.14\pm 0.05$ & $0.5\pm 0.1$ & $0.024\pm 0.005$ & $0.013\pm 0.003$ & \\
                        & & & & & $12.96\pm 0.06$ & $0.62\pm 0.16$ & $0.023\pm 0.005$ & $0.015\pm 0.003$ & \\
                    \midrule
                    \multirow{2}{*}{165"} & \multirow{2}{*}{$1\rightarrow0$} & \multirow{2}{*}{93173.40} & \multirow{2}{*}{4.47} & \multirow{2}{*}{-4.44} & $7.04\pm 0.03$ & $0.29\pm 0.08$ & $0.023\pm 0.005$ & $0.007\pm 0.002$ & \multirow{2}{*}{$<0.1$}\\
                        & & & & & $12.97\pm 0.05$ & $0.62\pm 0.13$ & $0.022\pm 0.005$ & $0.015\pm 0.002$ & \\
                    \midrule
                    \multirow{2}{*}{180"} & \multirow{2}{*}{$1\rightarrow0$} & \multirow{2}{*}{93173.40} & \multirow{2}{*}{4.47} & \multirow{2}{*}{-4.44} & $7.08\pm 0.04$ & $0.22\pm 0.08$ & $0.021\pm 0.005$ & $0.008\pm 0.003$ & \multirow{2}{*}{$<0.1$}\\
                        & & & & & $13.0\pm 0.1$ & $0.6\pm 0.2$ & $0.012\pm 0.005$ & $0.008\pm 0.003$ & \\
                    \midrule
                    $195"$ & $1\rightarrow0$ & 93173.40 & 4.47 & -4.44 & & & rms $= 0.005$ & & \\
                    \midrule
                    $210"$ & $1\rightarrow0$ & 93173.40 & 4.47 & -4.44 & & & rms $= 0.006$ & & \\
                    \midrule
                    $225"$ & $1\rightarrow0$ & 93173.40 & 4.47 & -4.44 & & & rms $= 0.006$ & & \\
                    \midrule
                    240" & $1\rightarrow0$ & 93173.40 & 4.47 & -4.44 & & & rms $= 0.007$ & & \\
    			\bottomrule
			\end{tabular}
            }
            
			\label{tab:linePropertiesN2H+IC348}
        \end{table}

		\newpage
		\begin{landscape}
			\section{Binding energy models with varying initial sulfur abundances}\label{sec:all Models}
			\flushleft
			In this section we show the comparison between observed molecular abundances and all the possible models combining the binding energies of each row in Table \ref{tab:molpeceresModels} with the initial sulfur abundances listed in Table \ref{tab:initialSulfurAb}.
			\subsection{H$_{2}$S abundances and models toward Barnard 1b}
			\begin{figure}[h]
                \centering
                \includegraphics[width=1.3\textwidth]{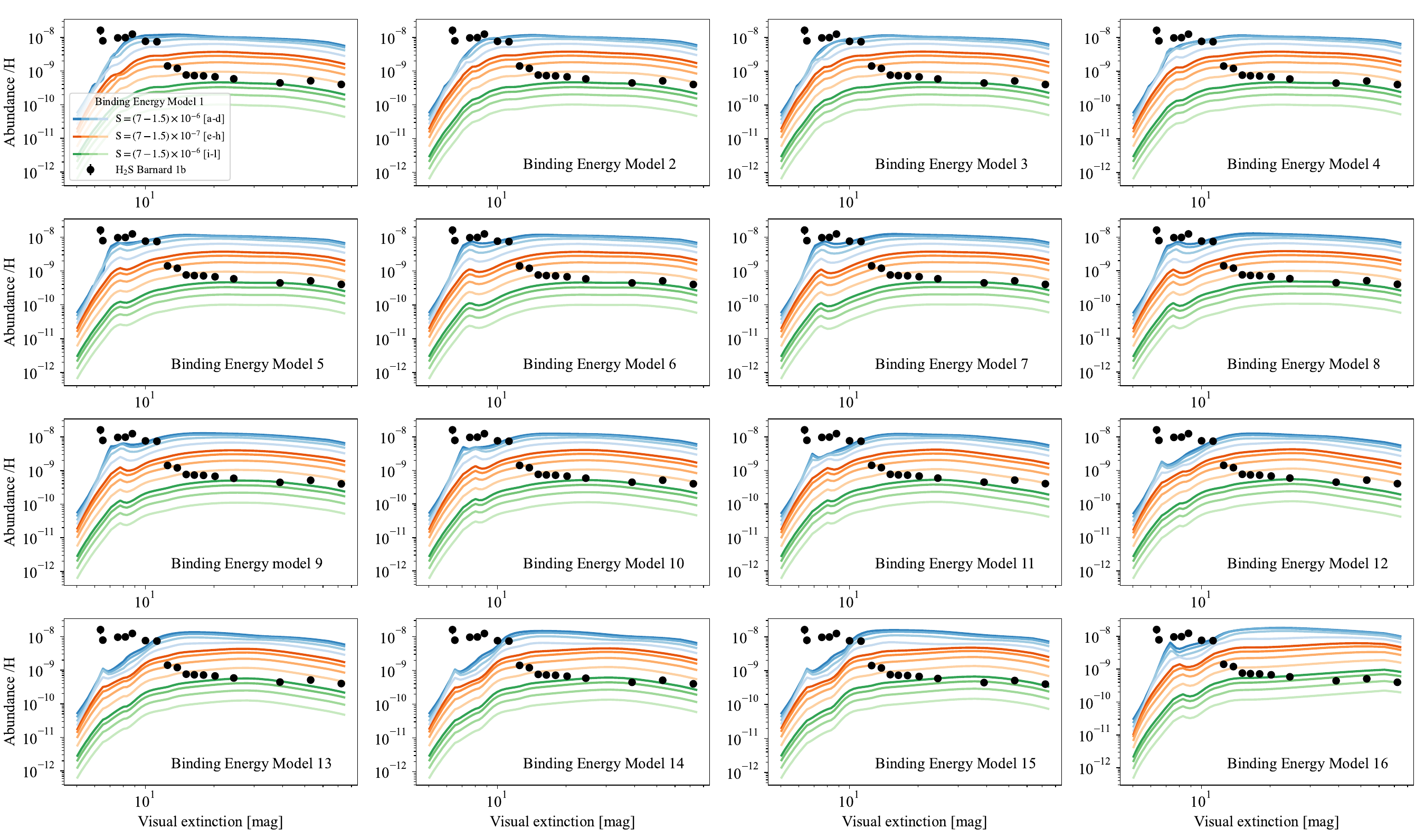}
                \caption{}
                \label{fig:allModelsH2SB1b}
            \end{figure}
            \subsection{H$_{2}$S abundances and models toward IC 348}
            \vfill
			\begin{figure}[h]
                \centering
                \includegraphics[width=1.3\textwidth]{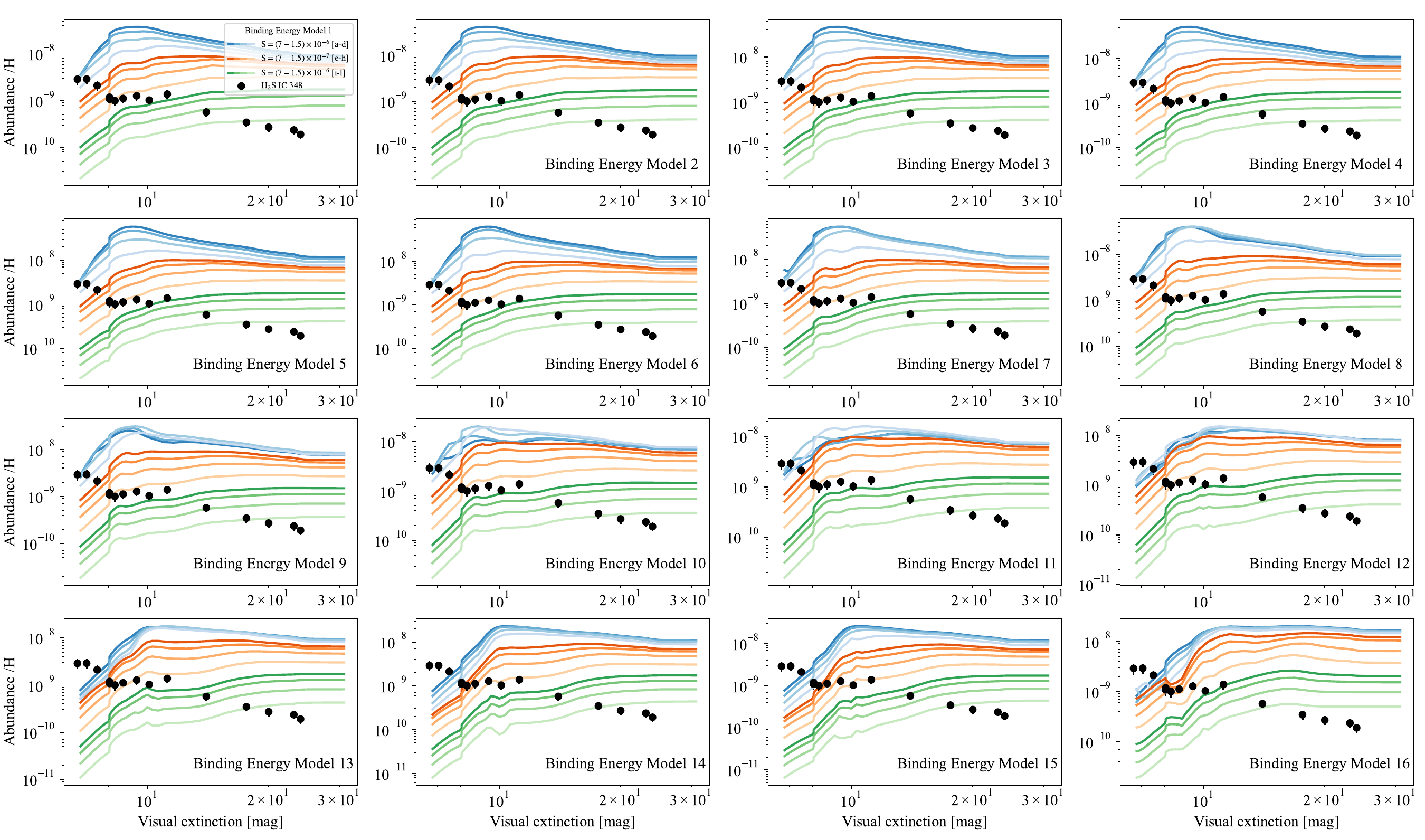}
                \caption{}
                \label{fig:allModelsH2SIC348}
            \end{figure}
            \vfill
            \newpage
            \subsection{CH$_{3}$OH abundances and models toward Barnard 1b}
            \vfill
			\begin{figure}[h]
                \centering
                \includegraphics[width=1.3\textwidth]{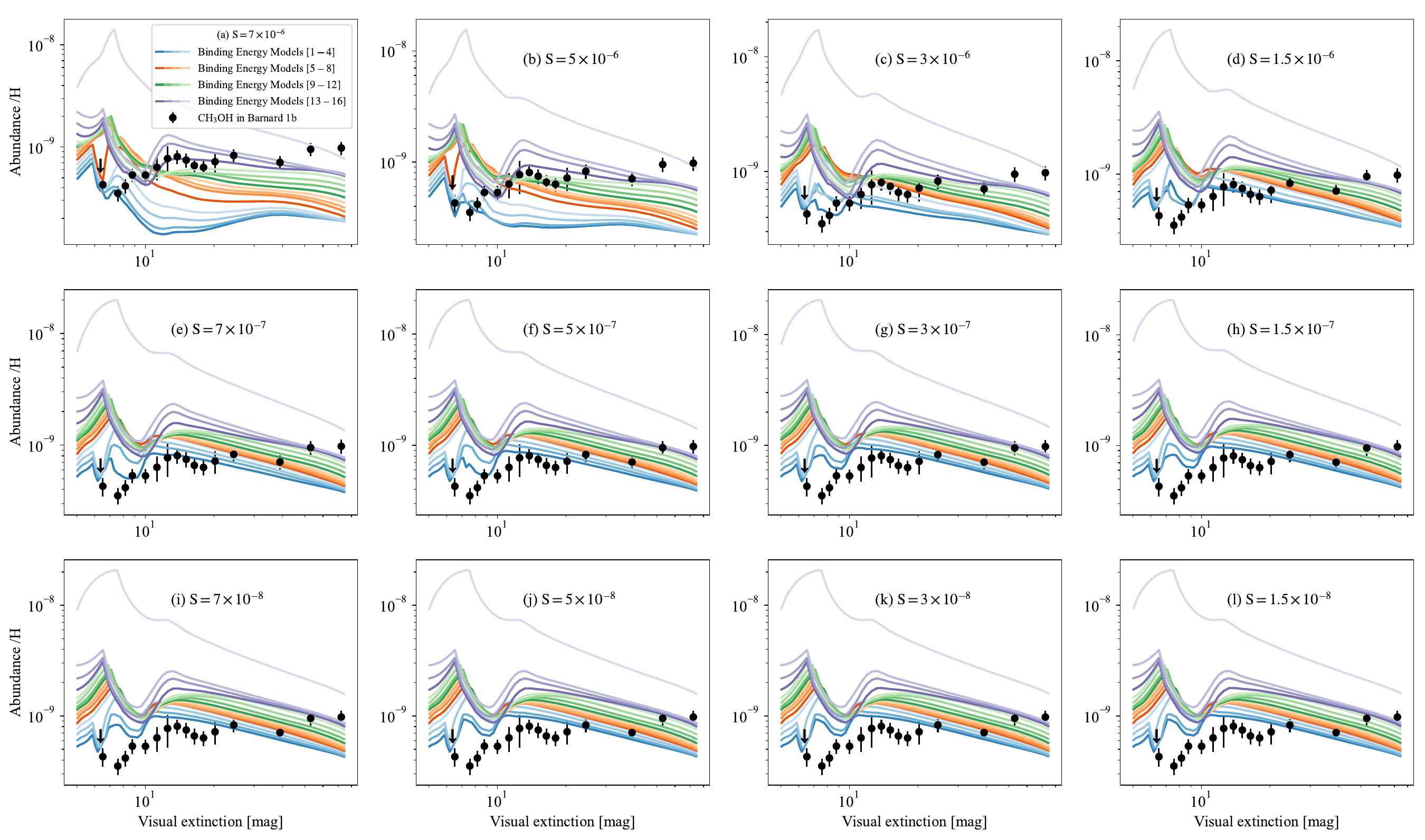}
                \caption{}
                \label{fig:allModelsCH3OHB1b}
            \end{figure}
            \vfill
            \newpage
            \subsection{CH$_{3}$OH abundances and models toward IC 348}
            \vfill
			\begin{figure}[h]
                \centering
                \includegraphics[width=1.3\textwidth]{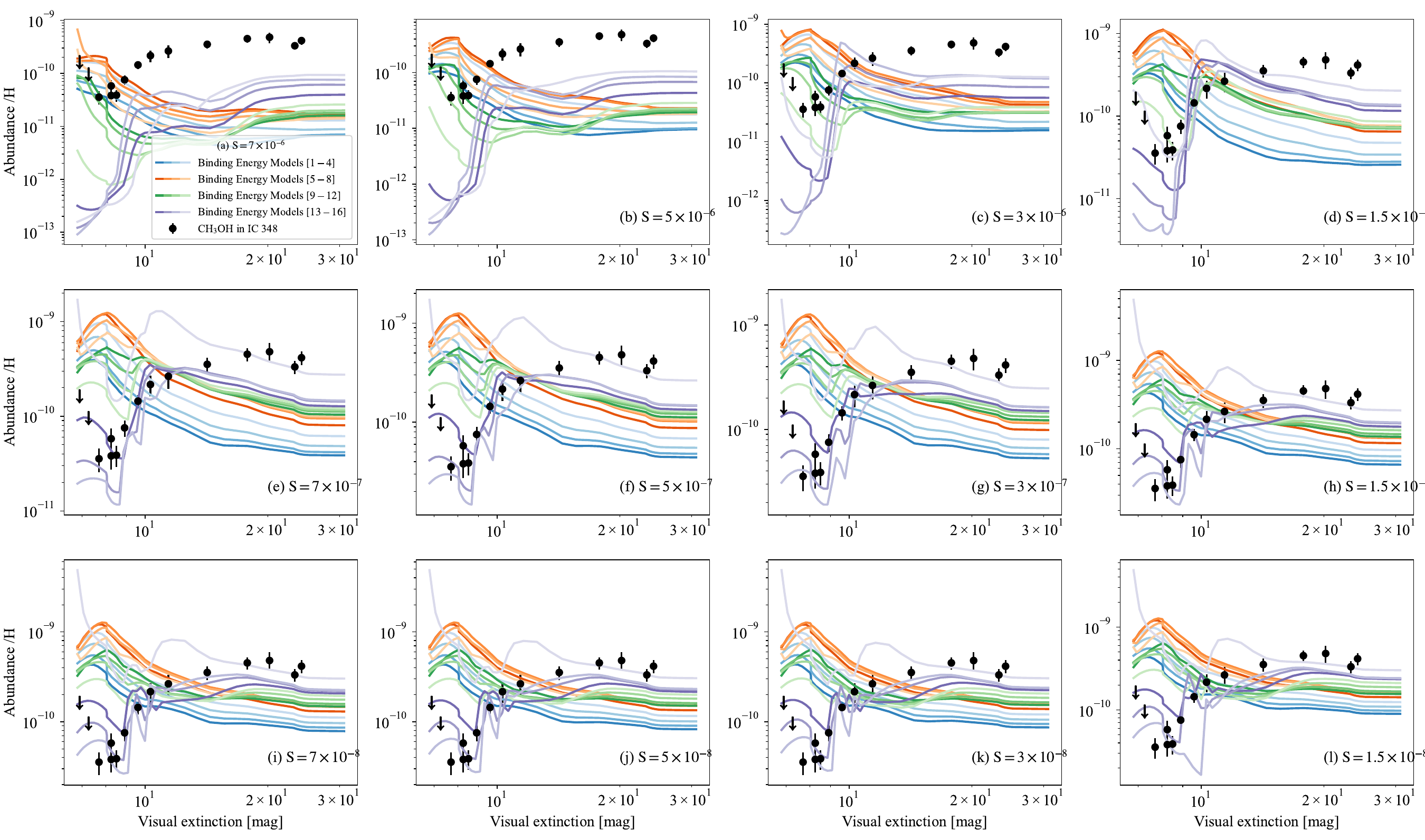}
                \caption{}
                \label{fig:allModelsCH3OHIC348}
            \end{figure}
            \vfill
            \newpage
            \subsection{OCS abundances and models toward Barnard 1b}
            \vfill
			\begin{figure}[h]
                \centering
                \includegraphics[width=1.3\textwidth]{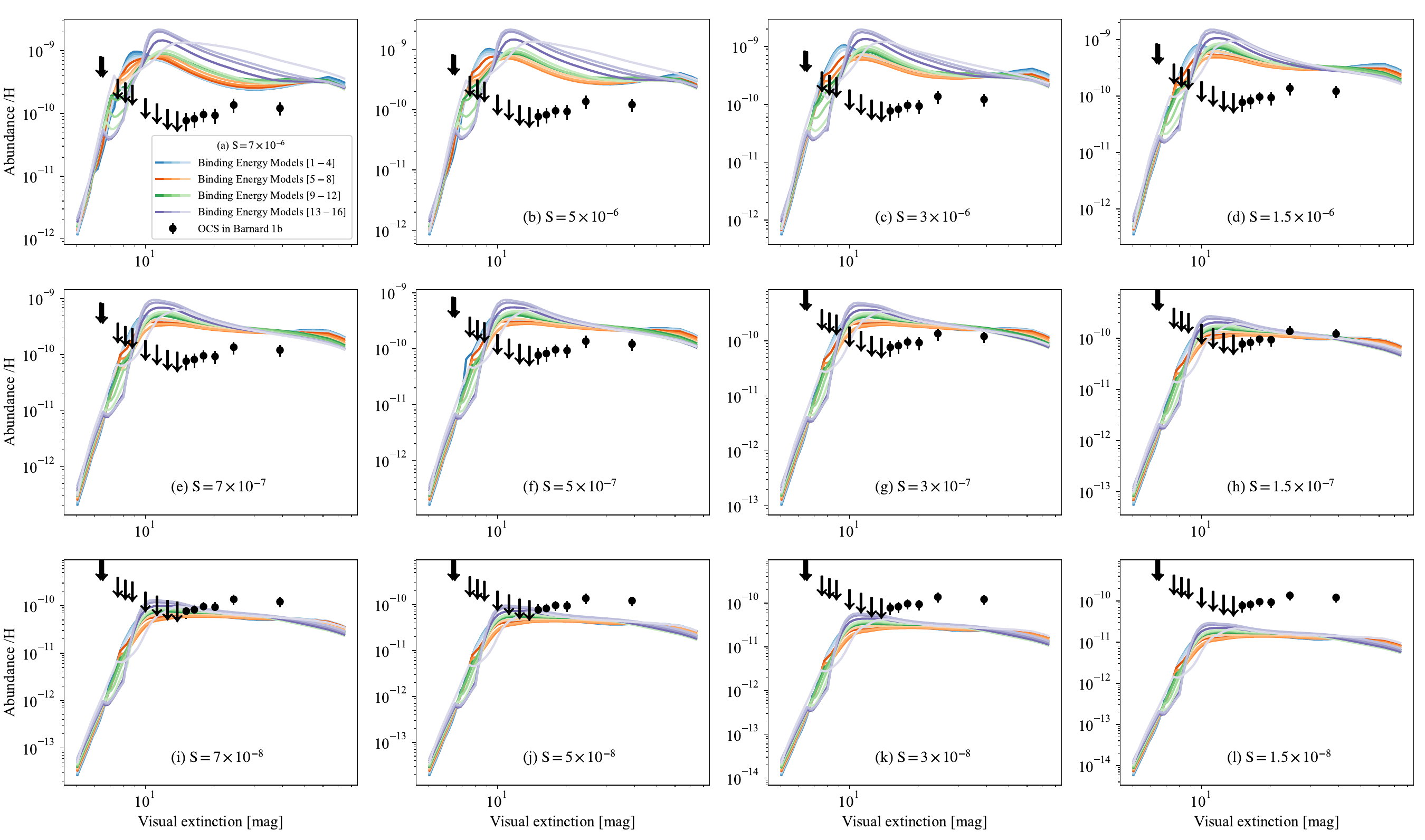}
                \caption{}
                \label{fig:allModelsOCSB1b}
            \end{figure}
            \vfill
            \newpage
            \subsection{OCS abundances and models toward IC 348}
            \vfill
			\begin{figure}[h]
                \centering
                \includegraphics[width=1.3\textwidth]{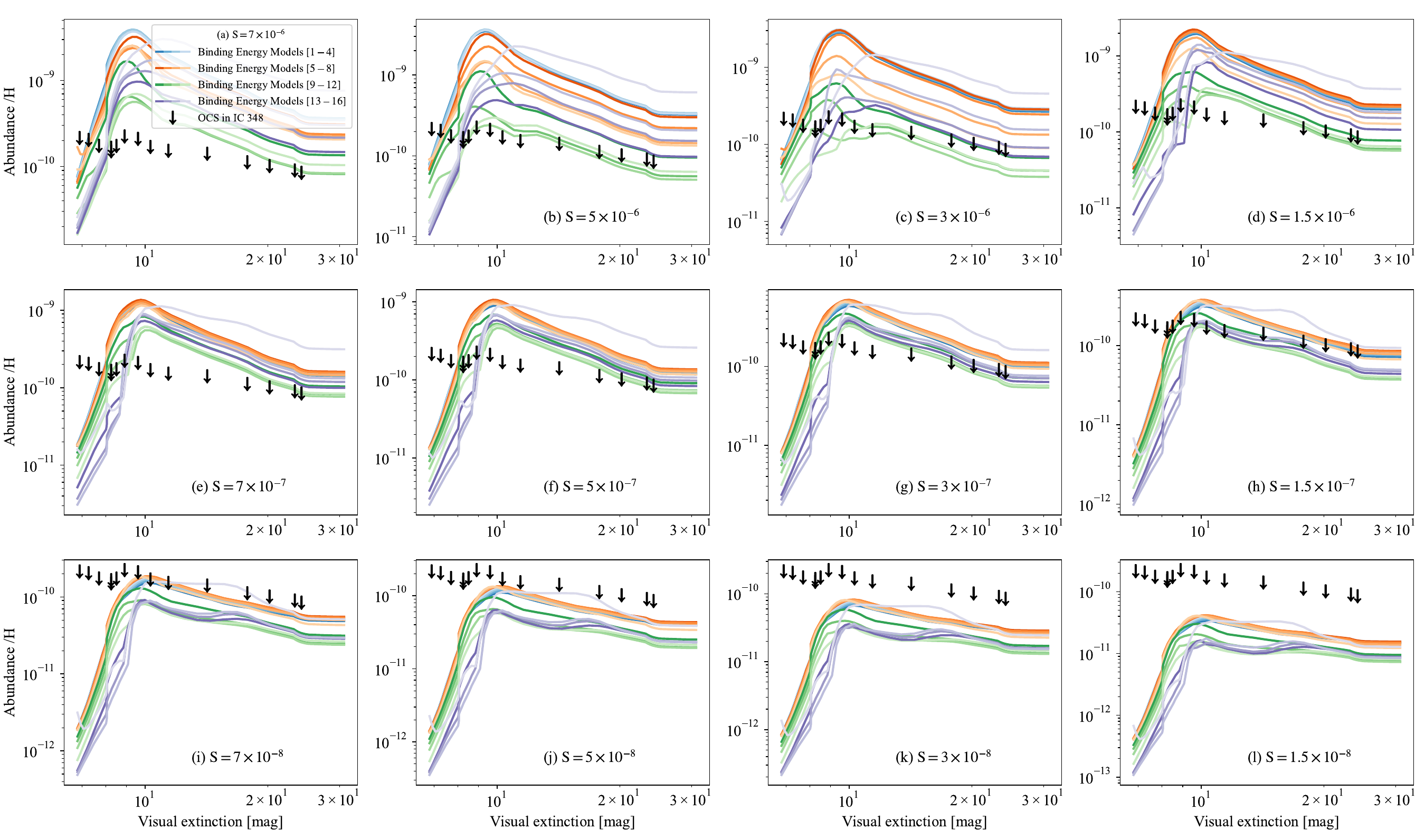}
                \caption{}
                \label{fig:allModelsOCSIC348}
            \end{figure}
            \vfill
            \newpage
            \subsection{CO abundances and models toward Barnard 1b}
            \vfill
			\begin{figure}[h]
                \centering
                \includegraphics[width=1.3\textwidth]{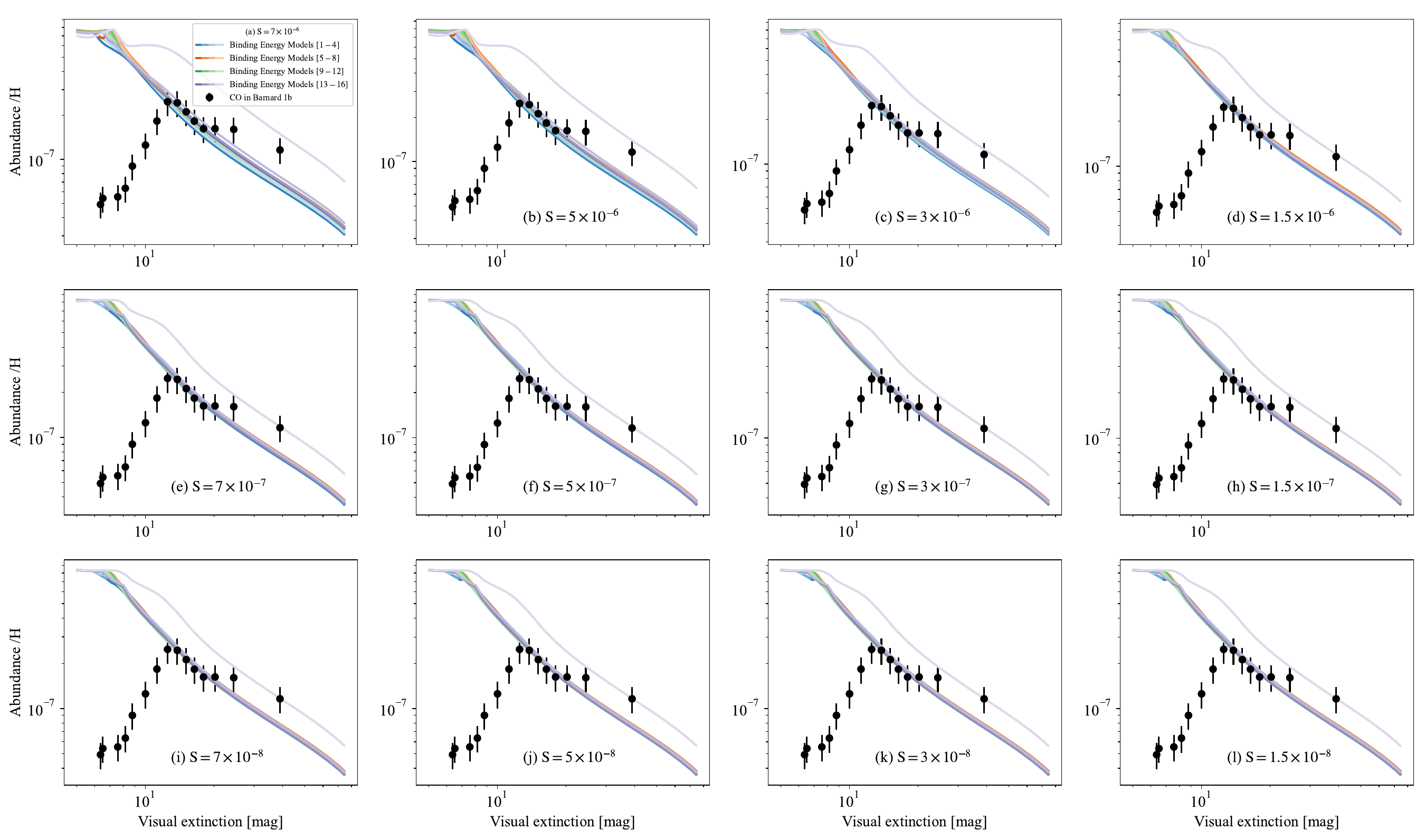}
                \caption{}
                \label{fig:allModelsCOB1b}
            \end{figure}
            \vfill
            \newpage
            \subsection{CO abundances and models toward IC 348}
            \vfill
			\begin{figure}[h]
                \centering
                \includegraphics[width=1.3\textwidth]{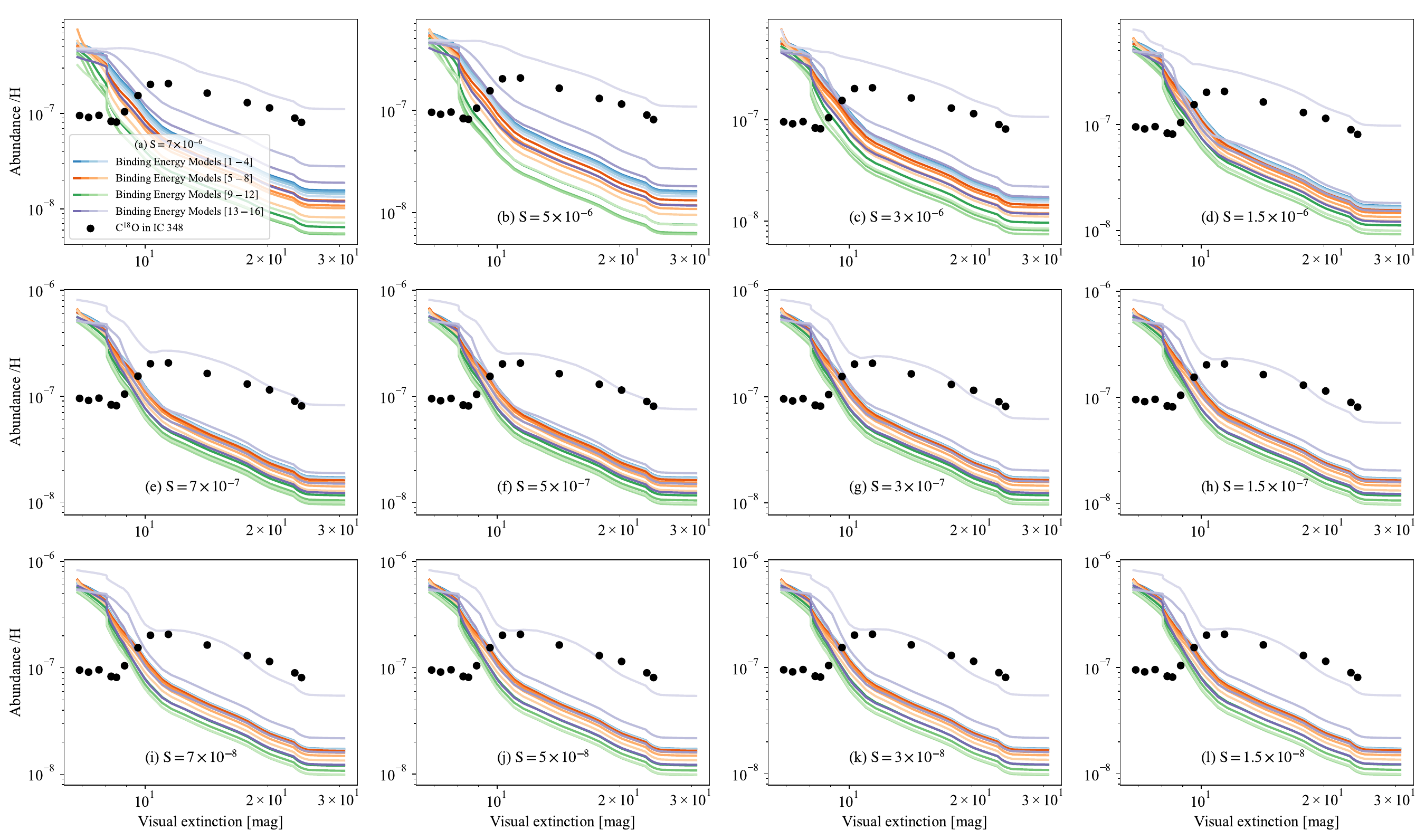}
                \caption{}
                \label{fig:allModelsCOIC348}
            \end{figure}
            \vfill
		\end{landscape}

    \end{appendix}

\end{document}